\pdfoutput=1
\documentclass[10pt]{iopart}
\usepackage{graphicx}
\usepackage{latexsym}

\usepackage{iopams}  
\begin{document}
\def\sun{\hbox{$\odot$}}
\def\aap{A\&A\,  }
\def\aaps{A\&AS  }
\def\acp{Anal. Cell. Pathol. } 
\def\aj{AJ  }
\def\aplett{Astrophys. Lett.\,  }
\def\apj{ApJ\,  }
\def\apjl{ApJ\,  }
\def\apjs{ApJS  }
\def\apss{Astrophysics and Space Science  }
\def\araa{ARA\&A  }
\def\azh{AZh}
\def\bain{BAN  }
\def\cjaa{Chinese J. Astron. Astrophys.  }
\def\eup{Europhys. Lett.  }
\def\iaucirc{IAU circ.  } 
\def\icarus{Icarus} 
\def\jaa{J. Astrophys. Astr.  }
\def\jpc{J. Phys. C  } 
\def\JPG{J. Phys. G\,  }
\def\jsp{J. Stat. Phys  } 
\def\jcp{J. Comput. Phys.  } 
\def\jcpp{Journal of Chemical  Physics  } 
\def\jrasc{JRASC  } 
\def\mnras{MNRAS\,  }
\def\oe{Optic  Express }
\def\nat{Nature\,  }
\def\npb{Nuc. Phys. B   }
\def\pasj{PASJ\,  }
\def\solphys{Sol. Phys.\,  }
\def\planss{Planet. Space Sci.  }
\def\pasp{PASP  }
\def\pasa{PASA  }
\def\POF{Physics of Fluids  }
\def\physrep{Phys. Rep.\,  }
\def\pla{Phys. Lett. A   }
\def\pra{Phys. Rev. A   }
\def\prb{Phys. Rev. B   }
\def\prd{Phys. Rev. D   }
\def\pre{Phys. Rev. E   }
\def\prl{Phys. Rev. Lett.    }
\def\physa{Phys. A    }
\def\rmp{Rev. Mod. Phys.  }
\def\rpp{Rep.Prog.Phys.   }
\def\za{Z. Astrophys.  } 
\title
{
Photometric Effects and Voronoi-diagrams as a  mixed model
for the spatial  distribution of galaxies
}

\author     {L. Zaninetti}
\address    {Dipartimento  di Fisica ,
 via P.Giuria 1,\\ I-10125 Turin,Italy }
\ead {zaninetti@ph.unito.it}

\begin {abstract}
We review the  model of  the  Voronoi Diagrams which  
allows to reproduce 
the   large-scale structures  of our universe as given by the astronomical
catalogs.
The  observed number of galaxies in a given  solid angle  
with a chosen   flux/magnitude  
versus the redshift presents 
a maximum  that is a function of the flux/magnitude  ;  
it can be explained by a detailed analysis of the standard luminosity
function  for galaxies as well by two  
new luminosity function for galaxies.
The current status of the research on the statistics  of the  
Voronoi Diagrams is reviewed.
\end  {abstract}
\vspace{2pc}

\maketitle

\section{Introduction}

\begin{figure}
 \begin{center}
\includegraphics[width=7cm]{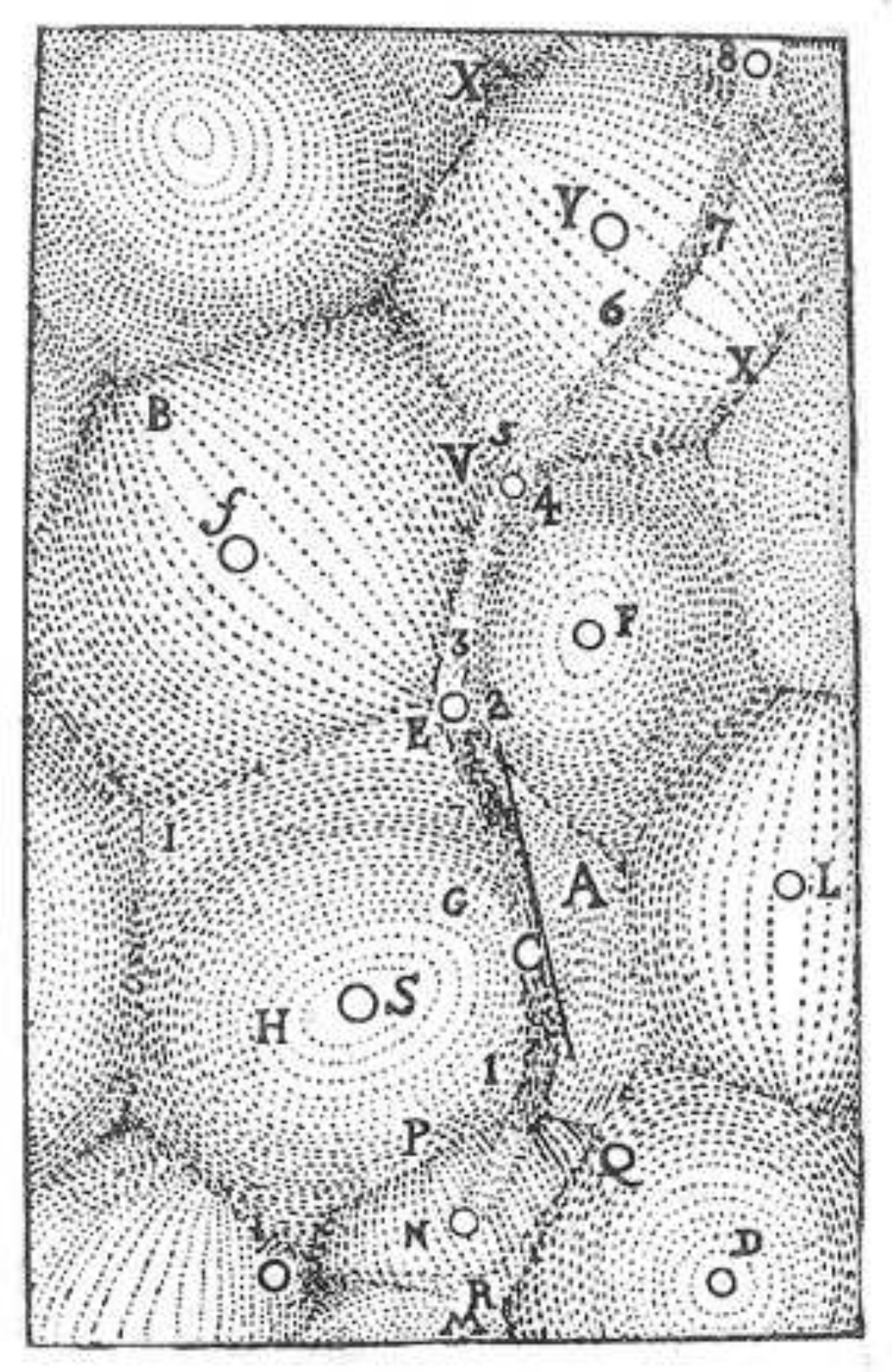}
\end{center}
\caption
{
" God first partitioned the plenum into equal-sized portions, and then placed
these bodies into various circular motions that, ultimately, formed the
three elements of matter and the vortex systems" ,extracted  
from  \cite{Descartes1644}, volume III , article 46.   
}
          \label{descartes}%
    \end{figure}

During the last thirty years the spatial distribution of 
galaxies has been   investigated  from 
the point of view of 
geometrical  and  physical theories.
One first target was to reproduce the two-point 
correlation function $\xi (r)$ for galaxies which 
on  average scales  as $\approx (\frac {r} {5.7Mpc})^{-1.8}$,
see 
\cite{Jones2005,Gallagher2000}.
The statistical theories  
of spatial galaxy distribution
can be classified as 
\begin{itemize}
\item {\bf Levy flights}: the random walk with a variable 
 step  length    can lead to a correlation function in  agreement
with the observed data, see 
\cite{Mandelbrot1975b,
Soneira1977,
Soneira1978,
Peebles1980}.
\item {\bf Percolation}: the theory of 
          primordial explosions
         can lead to the formation of structures, see 
\cite{Charlton1986,ferraro}. 
          Percolation is also used 
         as a tool to organize : (i) the mass and 
         galaxy distributions obtained in 3D
         simulations of  cold dark matter (CDM)
         and  hot dark matter (HDM), see 
\cite{Klypin1993}, 
         (ii)
         the galaxy groups  and clusters in 
         volume-limited samples of the
         Sloan Digital Sky Survey (SDSS), see 
         \cite{Berlind2006}
\item{\bf Statistical approach}
The statistics of the voids  between  galaxies
was   analyzed
in the Center for Astrophysics Redshift survey,
see  \cite{Vogeley1991,Vogeley1994} ,
in the
IRAS Point Source Catalog Redshift Survey (PSCz)
and Updated Zwicky Catalog of Galaxies (UZC),
see \cite{Vogeley2002}
and  in the Sloan Digital Sky Survey,
see  \cite{Vogeley2011}.

\end{itemize}
The geometrical models are well represented
by the concept of Voronoi Diagrams, after
the two historical records by \cite{voronoi_1907,voronoi}.
The concept of Voronoi Diagrams
dates back to the vortex theory
applied to the solar system as developed in the
17th century, see  Fig. \ref{descartes} extracted from   
\cite{Descartes1644,Aurenhammer2000}.
The  first application of the Voronoi Diagrams  to the 
astrophysics is due to  \cite{kiang}.
         The applications  of  the Voronoi Diagrams to the galaxies 
         started with
         \cite{icke1987}, 
         where a sequential clustering
         process was adopted in order to insert the initial seeds,
         and they continued 
         with 
         \cite{Weygaert1989,pierre1990,barrow1990,Coles1990a,Coles1990b,
          Weygaert1991a,Weygaert1991b,zaninettig,Ikeuchi1991,
           Subba1992,Weygaert1994,Goldwirth1995,
           Weygaert2002,Weygaert2003,Zaninetti2006}.
         An updated review of the 3D Voronoi Diagrams 
         applied to  cosmology can be found in 
          \cite{Weygaert2002,Weygaert2003}.
         The 3D Voronoi tessellation was first
        applied to identify groups of galaxies in
         the structure of a super-cluster, 
         see
         \cite{Ebeling1993,Bernardeau1996,Schaap2000,
         Marinoni2002,
         Melnyk2006,
         Schaap2009, 
         Elyiv2009}.
The  physical models 
that produce the observed properties of galaxies
are  intimately  related, 
for example through  the Lagrangian approximation, 
and can be 
approximately classified as 
\begin{itemize}
\item {\bf Cosmological N-body}: Through N-body experiments 
       by 
      \cite{Aarseth1978} 
      it  is possible to simulate 
      groups which are   analogous to the studies of groups 
      among bright Zwicky-catalog galaxies, see
      \cite{Gott1979a} 
      or 
       covariance functions in  simulations of galaxy
       clustering in an expanding universe 
       which are  found to be power laws in the nonlinear regime 
       with slopes centered on 1.9
       \cite{Gott1979b}.
       Using gigaparticle N-body simulations to study galaxy 
       cluster populations in Hubble volumes,
       \cite{Evrard2002} 
       created
       mock sky surveys of dark matter structure to z~=1.4 
       over $10000^{\circ}~sq.~deg$ and to
       z~=0.5 over two full spheres.
       In short,
       N-body calculations seek to model
       the full nonlinear system by making discrete the matter 
       distribution 
       and following its evolution in a
       Lagrangian fashion,
       while N-body simulations are usually understood 
       to concern gravity only.

\item {\bf Dynamical Models}: Starting from a power law 
      of primordial inhomogeneities it is possible 
      to obtain a two-point correlation function 
      for galaxies with an exponent similar to that 
      observed, see
      \cite{Peebles1974a,Peebles1974b,Gott1975}.

        Another line of work is to assume that
        the velocity field is of a potential type; 
        this assumption is 
        called the Zel'dovich  approximation, 
         see 
         \cite{ZelDovich1970,ZelDovich1989,Coles1995}.
        The Zel'dovich formalism
        is    a Lagrangian approximation
        of the fully nonlinear set of equations. 
        In this sense it is ``gravity'' only and does not include
        a pressure term.

\item   {\bf The halo models}:
         The halo model
         describes nonlinear structures as virialized dark-matter
         halos of different mass, placing them in space according
         to the linear large-scale density field which is completely
         described by the initial power spectrum,
         see  
         \cite{Neyman1952,Scherrer1991,Cooray2002}.
         Figure 19 in 
         \cite{Jones2005}, for example,
         reports the exact nonlinear model matter distribution   
         compared with its halo-model representation.

\end{itemize}

The absence of clear information on the 3D displacement
of the physical results as a function of the redshift 
and the selected magnitude characterize  
the cosmological N-body,
the dynamical and the hydrodynamical models.
This  absence  of detailed information 
leads to   the analysis of the following questions:
\begin {itemize}
\item Is it possible to
      compare the theoretical and observational 
      number of galaxies  as a  function of the redshift
      for a fixed flux/magnitude ?
\item  What is the role of the Malmquist bias 
      when theoretical and observed numbers  of
      galaxies versus the redshift are compared? 

\item Is it possible to find an algorithm which  describes
      the intersection between a slice 
       that  starts from the center of the box 
       and   
      the faces of  irregular  Poissonian Voronoi Polyhedrons?
\item Is it possible to model the intersection between a sphere
      of a given redshift and the faces of 
       irregular  Poissonian Voronoi Polyhedrons?
\item Does the developed theory  match   the observed 
      slices of galaxies as given, for example,
      by the 
      2dF Galaxy Redshift Survey?
\item Does the developed algorithm explain
      the voids  appearance 
      in all sky surveys such as the RC3?

\item Is it possible to integrate the usual
      probability density functions (PDFs)
      which characterize the main parameters of 2D and 3D 
      Poisson Voronoi tessellation (PVT)
      in order to obtain an analytical expression
      for the survival function?
\item
     Is it possible to model the normalized areas
     of $V_p(2,3)$  with the known PDFs?
\item
     Can we transform the normalized volumes and areas
     into equivalent radius distributions?

\item Is it possible to derive the probability density function
for the  
radii
of 2D sections in the 
Poissonian Voronoi
Tessellation (PVT)
and  in  a well selected case of   non Poissonian Voronoi
Tessellation (NPVT)?

\item Is it possible to evaluate the probability 
      of  having  a super-void once the averaged void's 
      diameter is fixed?

\item Is it possible to compute the correlation 
      function  for galaxies by introducing the 
      concept of  thick faces  
      of  irregular Voronoi  polyhedrons?      

\item Is it possible to find the acoustic oscillations
      of the correlation function at $\approx~100Mpc$ 
      in   simulated  slices 
      of  the Voronoi  diagrams. 

\end{itemize}
In order  to answer these questions,  
Section~\ref{sec_formulary} briefly reviews 
the elementary  formulas adopted
and Section~\ref{sec_lfs}  reviews 
the standard 
luminosity function for galaxies as well  two new ones.
The adopted catalogs as well the 
Malmquist bias  were presented  in Section \ref{sec_test}.
An accurate  test of the number of galaxies as a  function
of the redshift is performed on 
the 2dF Galaxy Redshift Survey (2dFGRS),
see Section~\ref{sec_maximum}.
Section~\ref{sec_voronoi}   reports the technique 
which allows us  to extract  the galaxies belonging to 
the Voronoi polyhedron.
Section \ref{sec_statistics} reports the
apparent distribution in effective radius
of the 3D PVT as well as their associated
survival functions,
 the fit
of the normalized area distribution
of the sectional PVT with the Kiang
function and the exponential distribution.
Section  \ref{secstereology}
reviews 
the
probability of a plane intersecting a given 
sphere , the stereological approach,  
and then insert in the fundamental 
integral of the stereology the 
cell's radius   of  the new PDF.
Section~\ref{sec_cellular}   
simulates the redshift dependence of  the  2dFGRS
as well as the overall  
Third Reference Catalog of Bright Galaxies (RC3).
Section  \ref{sec_corr} reports the simulation of the 
correlation function computed on the thick faces of 
the Voronoi polyhedron.

\section{Useful   formulas}

\label{sec_formulary}
This Section  reviews  
the elementary cosmology adopted
and  
the system of   magnitudes.

\label{formulary}

\subsection{Basic formulas}

Starting from  
\cite{Hubble1929} 
the suggested correlation
between expansion velocity  and distance is
\begin {equation}
V= H_0 D  = c_l \, z  
\quad  ,
\label {clz}
\end{equation}
where $H_0$ is the Hubble constant 
 $H_0 = 100 h \mathrm{\ km\ s}^{-1}\mathrm{\ Mpc}^{-1}$, with $h=1$
when  $h$ is not specified,
$D$ is the distance in $Mpc$ ,
$c_l$ is  the  light velocity  and
$z$   
is the redshift defined
as
\begin{equation}
z = \frac { \lambda_{obs} - \lambda_{em} } { \lambda_{em}} \quad ,
\end{equation}
with $\lambda_{obs}$  and  
$\lambda_{em}$ denoting respectively
 the
wavelengths of the observed and emitted lines
as  determined from the lab source, the so called Doppler
effect .
Concerning   the value of  $H_0$ we will adopt
a recent value as obtained by
the 
Cepheid-calibrated luminosity of Type Ia supernovae ,
see \cite{Sandage2006},
\begin{equation}
H_0 =(62.3 \pm 5 ) \mathrm{\ km\ s}^{-1}\mathrm{\ Mpc}^{-1}
\quad .
\end {equation}

The quantity $c_lz$ , a velocity , or $z$ , a number ,
characterizes the catalog of galaxies.

We recall that 
the galaxies have peculiar velocities, 
making the measured redshifts a combination of
cosmological redshift plus a contribution 
on behalf of the peculiar velocity.

The  maximum redshift here considered is $z\approx 0.1 $
meaning a maximum velocity of expansion of $\approx $
30000 $\frac{Km}{s}$ ;   up 
to that value  the  space is assumed to be Euclidean.
We now evaluate the error  connected with 
the use of the simplest cosmological  model.
For a zero cosmological  constant  , $\Lambda$,  we have the 
following expression  for the luminosity distance $D_L$ 
\begin{eqnarray}
D_L=
\frac{c_l\,z}{H_0} 
\left  \{ 
1 + \frac { z ( 1-q_0)}{\sqrt {1+2q_0z } +1 +q_0z  }      
\right  \}  
\quad  for ~ \Lambda=0 \quad ,  
\end{eqnarray}
where $q_0$ is  the deceleration parameter,
see \cite{Mattig1958,lang}.

\subsection{Magnitude System}

The  absolute magnitude of a galaxy ,$M$,  is connected
to  the  apparent magnitude $m$ through the 
relationship 
\begin{equation} 
M = m - 5 Log (\frac {c_lz}{H_0}) - 25   
\quad .
\label{absolute}
\end{equation}
In an Euclidean ,non-relativistic 
and homogeneous universe 
the flux of radiation,
$ f$  ,  expressed in $ \frac {L_{\sun}}{Mpc^2}$ units,
where $L_{\sun}$ represents the luminosity of the sun
,  is 
\begin{equation}
f  = \frac{L}{4 \pi D^2}  
\quad ,
\end{equation}
where $D$   represents the distance of the galaxy 
expressed in $Mpc$,
and  
\begin{equation}
D=\frac{c_l z}{H_0} 
\quad  .
\end{equation}

The relationship connecting the absolute magnitude, $M$ ,
 of a
galaxy  to  its luminosity is
\begin{equation}
\frac {L}{L_{\sun}} =
10^{0.4(M_{\sun} - M)}
\quad ,
\label{mlrelation}
\end {equation}
where $M_{\sun}$ is the reference magnitude 
of the sun at the considered bandpass.

The flux   expressed in $ \frac {L_{\sun}}{Mpc^2}$ units 
as  a function of the  apparent magnitude is
\begin{eqnarray}
f=
7.957 \times 10^8 \,{e^{ 0.921\,{\it M_{\sun}}- 0.921\,{\it
m}}}
\quad    \frac {L_{\sun}}{Mpc^2} \quad , 
\label{damaf}
\end {eqnarray}
and  the inverse relationship is 
\begin{eqnarray}
m=
M_{\sun}- 1.0857\,\ln  \left(  0.1256 \times 10^{-8} f \right) 
\quad . 
\label{dafam} 
\end {eqnarray}
The equations in  this section will be used 
in the numerical code which allows us to simulate
the large scale structures as a function of $z$ 
and the selected magnitude .
\section{Luminosity function for galaxies}

\label{sec_lfs}
This Section  reviews  
the  standard luminosity function  
for galaxies,
and two  new
luminosity  function  for galaxies in the following $LF$.

\subsection{The  Schechter function }

The  Schechter function , introduced by 
\cite{schechter}
,
provides a useful fit  for the 
$LF$  of galaxies 
\begin{equation}
\Phi (L) dL  = (\frac {\Phi^*}{L^*}) (\frac {L}{L^*})^{\alpha}
\exp \bigl ( {-  \frac {L}{L^*}} \bigr ) dL \quad  ,
\label{equation_schechter}
\end {equation}
here $\alpha$ sets the slope for low values of $L$ , $L^*$ is the
characteristic luminosity and $\Phi^*$ is the normalization.
The equivalent distribution in absolute magnitude is 
\begin{equation}
\Phi (M)dM=(0.4  ln 10) \Phi^* 10^{0.4(\alpha +1 ) (M^*-M)}  
\exp \bigl ({- 10^{0.4(M^*-M)}} \bigr)  dM \, ,
\label{equation_schechter_M}
\end {equation}
where $M^*$ is the characteristic magnitude as derived from the
data.
The scaling with  $h$ is  $M^* - 5\log_{10}h$ and
$\Phi^* ~h^3~[Mpc^{-3}]$.
The joint distribution in {\it z}  and {\it f}  for galaxies ,
see formula~(1.104) in
 \cite{pad} 
or formula~(1.117) 
in 
\cite{Padmanabhan_III_2002} 
,
 is
\begin{equation}
\frac{dN}{d\Omega dz df} =  
4 \pi  \bigl ( \frac {c_l}{H_0} \bigr )^5    z^4 \Phi (\frac{z^2}{z_{crit}^2})
\label{nfunctionz}  
\quad ,
\end {equation}
where $d\Omega$ , $dz$ and  $ df $ represent the differential of
the solid angle , the redshift and the flux respectively.
The critical value of z ,   $z_{crit}$ , is 
\begin{equation}
 z_{crit}^2 = \frac {H_0^2  L^* } {4 \pi f c_l^2}
\quad .
\end{equation} 
The number of galaxies , $N_S(z,f_{min},f_{max})$  
comprised between a minimum value of flux,
 $f_{min}$,  and  maximum value of flux $f_{max}$ ,
can be computed  through  the following integral 
\begin{equation}
N_S (z) = \int_{f_{min}} ^{f_{max}}
4 \pi  \bigl ( \frac {c_l}{H_0} \bigr )^5    z^4 \Phi (\frac{z^2}{z_{crit}^2})
df
\quad .
\label{integrale_schechter} 
\end {equation}
This integral does not  have  an analytical solution 
and we  must   perform 
a numerical integration.

The number of galaxies in {\it z} and {\it f} as given by 
formula~(\ref{nfunctionz})  has a maximum  at  $z=z_{pos-max}$ ,
where 
\begin{equation}
 z_{pos-max} = z_{crit}  \sqrt {\alpha +2 }
\quad ,
\end{equation} 
which  can be re-expressed   as
\begin{equation}
 z_{pos-max} =
\frac
{
\sqrt {2+\alpha}\sqrt {{10}^{ 0.4\,{\it M_{\sun}}- 0.4\,{\it M^*}}}{
\it H_0}
}
{
2\,\sqrt {\pi }\sqrt {f}{\it c_l}
}
\quad  .
\label{zmax_sch}
\end{equation}

\subsection{The  mass-luminosity relationship }

A new  $LF$ for galaxies as derived 
in \cite{Zaninetti2008} is 
\begin{equation}
\Psi (L) dL  =  (\frac{1}{a \Gamma(c) } ) (\frac {\Psi^*}{L^*})
\left (\frac {L}{L^*} \right  )^{\frac{c-a}{a}} 
\exp \left  ( {-
\left ( \frac {L}{L^*}\right  )^{\frac{1}{a}}} \right   ) dL \quad
, \label{equation_schechter_mia}
\end {equation}
where $\Psi^*$  is  a normalization factor which defines the
overall density of galaxies , a  number  per cubic $Mpc$,
 $1/a$ is  an exponent  which  connects the mass to the
luminosity
and $c$  is connected  with the dimensionality 
of the fragmentation, $c=2d$ , where $d$ represents 
the dimensionality  of the considered space : 1,2,3 . 
The scaling with  $h$ is  $M^* - 5\log_{10}h$ and
$\Psi^* ~h^3~[Mpc^{-3}]$.
The  distribution 
in  absolute
magnitude is 
\begin{eqnarray}
\Psi (M) dM  =  \nonumber \\ (0.4  ln 10 \frac {1}{a \Gamma(c)}) \Psi^*
10^{0.4(\frac{c}{a}) (M^*-M)} 
 \exp \bigl ({-
10^{0.4(M^*-M)(\frac{1}{a})}} \bigr)  dM \, .
\label{equation_mia}
\end {eqnarray}
This function contains the  parameters $M^*$ ,{\it a},
{\it c}
and  $\Psi^*$ which  are   derived from the operation of fitting
the observational data.

The joint distribution in $z$ and $f$ ,
in presence of the ${\mathcal M}-L$ $LF$  
 (equation~(\ref{equation_schechter_mia})) is 
\begin{equation}
\frac{dN}{d\Omega dz df} =  
4 \pi  \bigl ( \frac {c_l}{H_0} \bigr )^5    z^4 \Psi (\frac{z^2}{z_{crit}^2})
\label{nfunctionz_mia}  
\quad .
\end {equation}
The number of galaxies , $N_{{\mathcal M}-L}(z,f_{min},f_{max})$  
comprised between 
 $f_{min}$  and   $f_{max}$ ,
can be computed  through  the following integral 
\begin{equation}
N_{{\mathcal M}-L} (z) =
 \int_{f_{min}} ^{f_{max}}
4 \pi  \bigl ( \frac {c_l}{H_0} \bigr )^5    z^4 \Psi (\frac{z^2}{z_{crit}^2})
df 
\quad ,
\label{integrale_mia} 
\end {equation}
and also in this case  
a numerical integration must be performed.

The number of galaxies as  given 
by formula~(\ref{nfunctionz_mia}) has a maximum at 
$z_{pos-max}$  where 
\begin{equation}
 z_{pos-max} = z_{crit} 
\left( {\it c}+a \right) ^{a/2}
\quad ,
\end{equation} 
which can be re-expressed as
\begin{equation}
 z_{pos-max} =  
\frac
{
\left( a+{\it c} \right) ^{1/2\,a}\sqrt {{10}^{ 0.4\,{\it M_{\sun}}-
0.4\,{\it M^*}}}{\it H_0}
}
{
2\,\sqrt {\pi }\sqrt {f}{\it c_l}
}
\quad .
\label{zmax_mia}
\end{equation} 

\subsection{The generalized gamma distribution with four parameters}

The starting point is the 
probability density function 
(in the following PDF)
named generalized gamma 
that we report exactly as in 
\cite{evans}:
\begin{equation}
G(x;a,b,c,k)=
\frac 
{
{\it k}\, \left( {\frac {x-a}{b}} \right) ^{c{\it k}-1}{{\rm e}^{-
 \left( {\frac {x-a}{b}} \right) ^{{\it k}}}}
}
{
b\Gamma \left( c \right) 
}
\label{base5}
\quad ,
\end{equation}
where 
$\Gamma$ is the gamma function, 
$a$ is the location parameter, 
$b$ is the scale parameter,
$c$ and $k$ are two shape parameters.
A  $LF$  can be  derived 
inserting  $a=0$, $x=L$ and $b=L^*$: 
\begin{equation}
\Psi(L;L^*,c,k,\Psi^*)=
\Psi^*\frac 
{
{\it k}\, \left( {\frac {L}{L^*}} \right) ^{c{\it k}-1}{{\rm e}^{-
 \left( {\frac {L}{L^*}} \right) ^{{\it k}}}}
}
{
L^*\Gamma \left( c \right) 
}
\quad .
\label{lf4}
\end{equation}

The mathematical range of existence is $ 0 \leq L < \infty $
and the number of parameters is four because $a=0$ and
$\Psi^*$ have been added. 
The averaged luminosity is 
\begin{equation}
{ \langle L \rangle } 
=
\frac
{
{\it L^*}\,\Gamma \left( {\frac {1+ck}{k}} \right) 
}
{
\Gamma \left( c \right) 
}
\quad ,
\label{lmedio4}
\end{equation}
and the mode is
at 
\begin{equation}
L=
\left( {\frac {ck-1}{k}} \right) ^{\frac{1}{k}} {\it L^*}
\quad . 
\label{mode4} 
\end{equation}
The magnitude version of the $LF$ is
\begin{equation}
\Psi (M) dM =
\frac
{
{ \Psi^*}0.4\ln \left( 10 \right)k{10}^{- 0.4ck \left( M- { M^*}
 \right) }{{\rm e}^{- {10}^{- 0.4 \left( M- { 
M^*} \right) k}}} 
}
{
\Gamma \left( c \right) 
} 
\nonumber   \\
 dM \, .
\label{pdf4magni}
\end {equation}
The mode when expressed in magnitude is 
at 
\begin{equation}
M = -\frac 
{
 1.0857\,\ln \left( {\frac {c{\it k}- 1}{{\it k}}} \right) 
}
{
k
} 
+ {\it M^*}
\quad .
\label{mode4magni}
\end{equation}

This function contains the four parameters $c$,
$k$, $M^*$
and $\Psi^*$,
more details  as well  other two new $LF$  are 
reported in \cite{Zaninetti2010f}.

The joint distribution in $z$ and $f$ 
of the generalized gamma $LF$ 
is 
\begin{equation}
\frac{dN}{d\Omega dz df} = 
4 \pi \bigl ( \frac {c_l}{H_0} \bigr )^5 z^4 \Psi (\frac{z^2}{z_{crit}^2})
\label{nfunctionz_gammagene} 
\quad.
\end {equation}
The number of galaxies, $N_{LF4}(z,f_{min},f_{max})$ 
of the  generalized gamma $LF$ 
comprised between 
 $f_{min}$ and $f_{max}$,
can be computed through 
the following integral: 
\begin{equation}
N_{LF4} (z) = \int_{f_{min}} ^{f_{max}}
4 \pi \bigl ( \frac {c_l}{H_0} \bigr )^5 z^4 \Psi (\frac{z^2}{z_{crit}^2})
df 
\quad,
\label{integrale_gammagene} 
\end {equation}
and in this case  
a numerical integration must be performed.

The number of galaxies
of the generalized gamma LF
 as given 
by formula~(\ref{nfunctionz_gammagene}) 
has a maximum at 
$z_{pos-max}$ where 
\begin{equation}
 z_{pos-max} = 
{e^{1/2\,{\frac {\ln \left( 1+c{\it k} \right) -\ln \left( {\it k}
 \right) }{{\it k}}}}}{\it z_{crit}}
\quad,
\end{equation} 
which can be re-expressed as
\begin{equation}
 z_{pos-max} = 
\frac
{
{e^{1/2\,{\frac {\ln \left( 1+{\it c}\,{\it k} \right) -\ln 
 \left( {\it k} \right) }{{\it k}}}}}\sqrt {{10}^{ 0.4\,{\it 
{\it M_{\sun}}}
- 0.4\,{\it M^*}}}{\it H_0k}
}
{
2\,\sqrt {\pi }\sqrt {f}{\it c_l}
}
\quad.
\label{zmax_gammagene}
\end{equation}

\section{The adopted catalogs}

\label{sec_test}
We now introduce the  processed 
catalogs   
of galaxies,
the statistics of  1024  observed cosmic voids 
as well  the Malmquist bias.

\subsection{The astronomical catalogs}
\label{catalogs}
A first example  is  the 2dFGRS data release 
available on   the  web site: \newline
 http://msowww.anu.edu.au/2dFGRS/.
In particular we added together the file parent.ngp.txt which  
contains 145652 entries for NGP strip sources and 
the file parent.sgp.txt which 
contains 204490 entries for SGP strip sources.
Once the   heliocentric redshift  was  selected 
we  processed 219107 galaxies with 
$0.001 \leq z \leq 0.3$.

A second example is  the catalog  RC3, see  \cite{RC31991}, 
which  is available  at the following address 
http://vizier.u-strasbg.fr/viz-bin/VizieR?-source=VII/155.

\begin{figure}\begin{center}
\includegraphics[width=7cm]{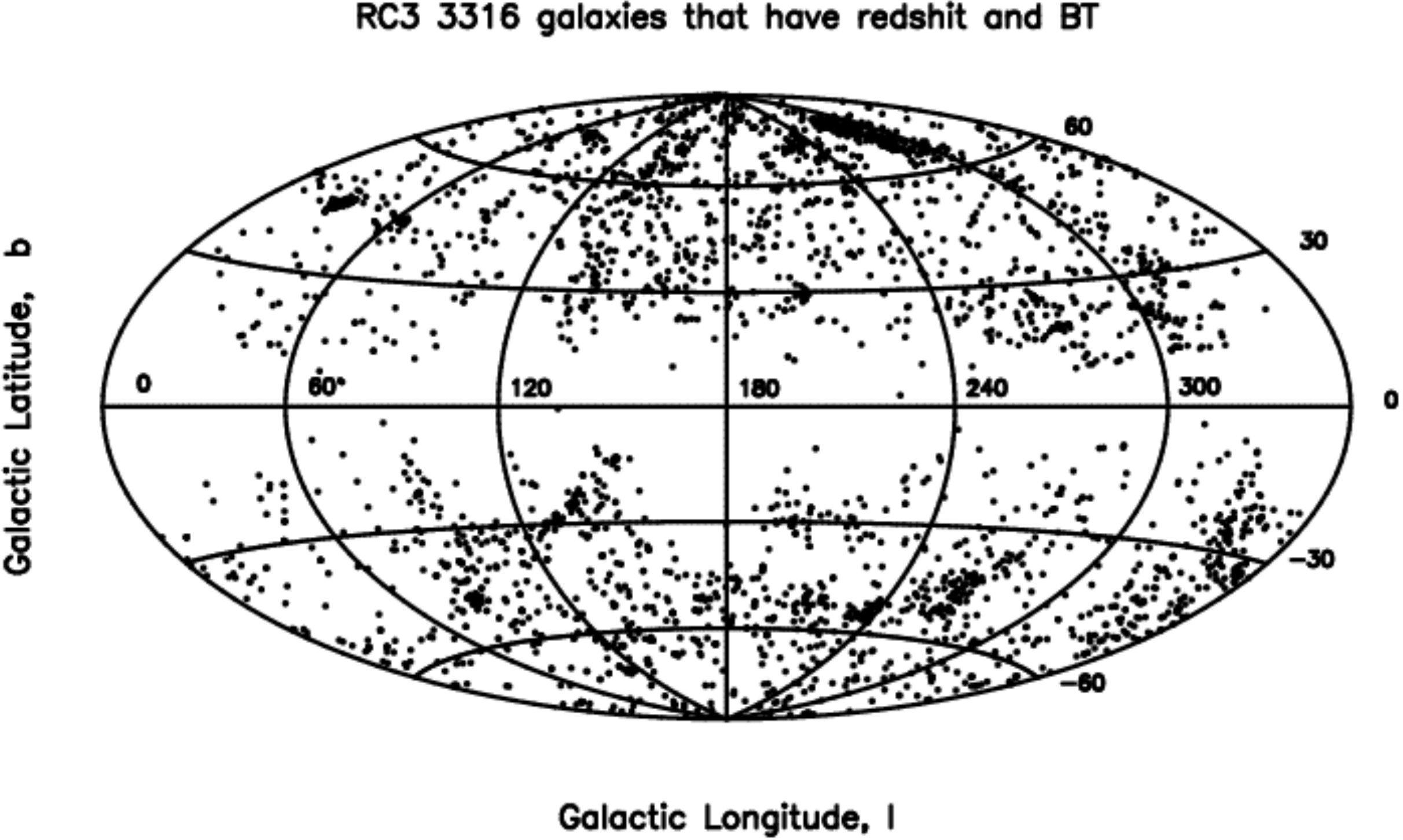}
\end{center}\caption{
  Hammer-Aitoff  projection in galactic coordinates 
  of 3316  galaxies  
  in the  RC3  which have BT and redshift.
  The ZOA is due to our own galaxy. 
}
          \label{rc3_all}%
    \end{figure}

This catalog    attempts to be reasonably complete for galaxies
having apparent diameters larger than 1 arcmin at the D25
isophotal level and total B-band magnitudes BT,
brighter than about 15.5, 
with a redshift not in excess of 15000 km/s.
All the galaxies in  the
RC3 catalog  which have redshift and BT  are reported
in Fig. \ref{rc3_all}.
In the case of RC3 the covered area 
 is $4\pi$ steradians with the 
exclusion of the {\it Zone of Avoidance (ZOA)}

\begin{figure}\begin{center}
\includegraphics[width=7cm]{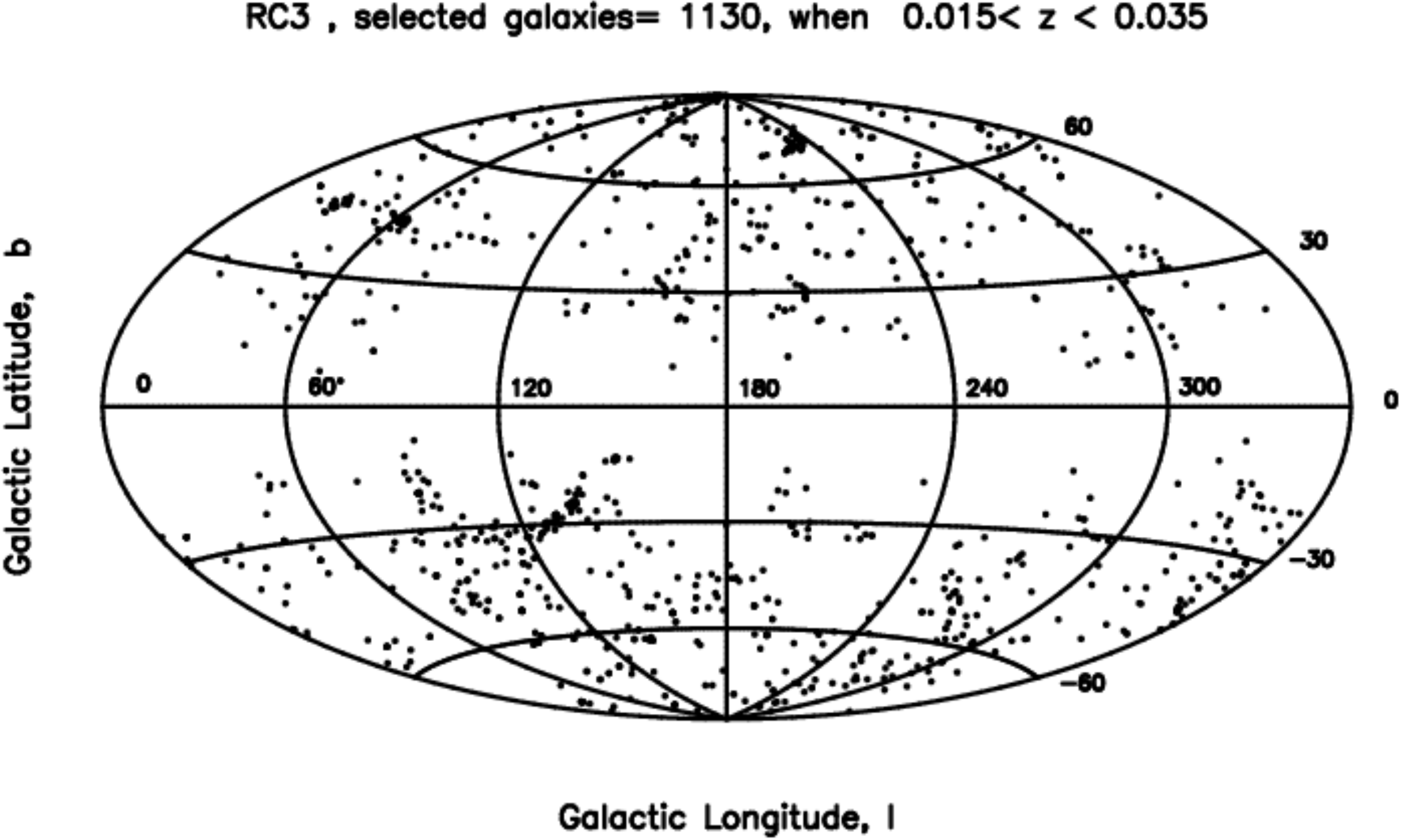}
\end{center}\caption{
Hammer-Aitoff  projection  
in galactic coordinates 
(observational counterpart 
of  $V_s(2,3)$ ) 
  of 1130  galaxies  
 in the RC3  which have BT and  $0.015 < z < 0.035$.
}
          \label{rc3_z}%
    \end{figure}

\begin{figure}\begin{center}
\includegraphics[width=7cm]{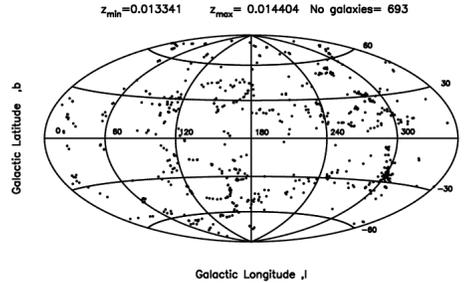}
\end{center}\caption
{
Hammer-Aitoff  projection in galactic coordinates
of a spherical cut of the Large Super Cluster
data (LSC)
at
$0.0133  \leq z \leq 0.0144$
or
$56.9 Mpc  \leq D  \leq 61.67$
.
}
          \label{2mrs_data}%
    \end{figure}

Fig. \ref{rc3_z}  reports the RC3 galaxies
in a given window in $z$.
A third  example  is 
the Two-Micron All Sky Survey (2MASS) 
 which has instruments in the infrared
(1-2.2 $\mu m$ ) and therefore detects  the galaxies
in the so called "Zone of Avoidance" ,
see \cite{Jarrett2004,Huchra2007}.
Fig. \ref{2mrs_data} reports a spherical cut
at a given radius
of the Local Super-cluster (LSC) according to
2MASS Redshift Survey (2MRS), which
is available  online  at  
https://www.cfa.harvard.edu/~huchra\,.
In the case of  2MRS the covered area 
 is $4\pi$ steradians.

A fourth  example is 
the second CFA2 redshift   Survey , which started in 1984,
and produced slices showing that the spatial 
distribution of galaxies
is not random but distributed on filaments that represent the 2D
projection of 3D bubbles. 
We recall that a slice comprises all the
galaxies with magnitude $m_b~\leq~16.5$ in a strip of $6^{\circ}$
wide and about $130^{\circ}$ long. 
One of such slice (the so
called first CFA strip) is visible at the following address
http://cfa-www.harvard.edu/~huchra/zcat/ ; more details can be
found in~\cite{geller}. 
The already mentioned slice can be
down-loaded from http://cfa-www.harvard.edu/~huchra/zcat/n30.dat/
. 

\subsection{Statistics of the voids}
\label{secstat}
The distribution  of the effective radius  
and 
the radius of the maximal enclosed sphere
between galaxies
of the Sloan Digital Sky Survey
Data Release 7 (SDSS DR7)
has been reported  in
\cite{Vogeley2011}.
This catalog  contains   1054  voids:
Table~\ref{statvoids} shows the basic
statistical  parameters of the 
effective radius.
\begin{table}
 \caption[]
{
The statistical  parameters
of the effective radius in  SDSS DR7.
}
 \label{statvoids}
 \[
 \begin{array}{lc}
 \hline
 \hline
 \noalign{\smallskip}
parameter                  &   value                          \\ \noalign{\smallskip}
elements                   &  1024               \\ \noalign{\smallskip}
mean                       &  18.23h^{-1}~ Mpc   \\ \noalign{\smallskip}
variance                   &  23.32h^{-2}~ Mpc^2 \\ \noalign{\smallskip}
standard~ deviation        &  4.82h^{-1} ~ Mpc   \\ \noalign{\smallskip}
skewness                   &  0.51         \\ \noalign{\smallskip}
kurtosis                   &  0.038        \\ \noalign{\smallskip}
maximum ~value             &  34.12h^{-1}~ Mpc   \\ \noalign{\smallskip}
minimum ~value             &  9.9h^{-1}~   Mpc   \\ \noalign{\smallskip} \hline
 \hline
 \end{array}
 \]
 \end {table}


\subsection{Malmquist bias}

\begin{figure}\begin{center}
\includegraphics[width=7cm]{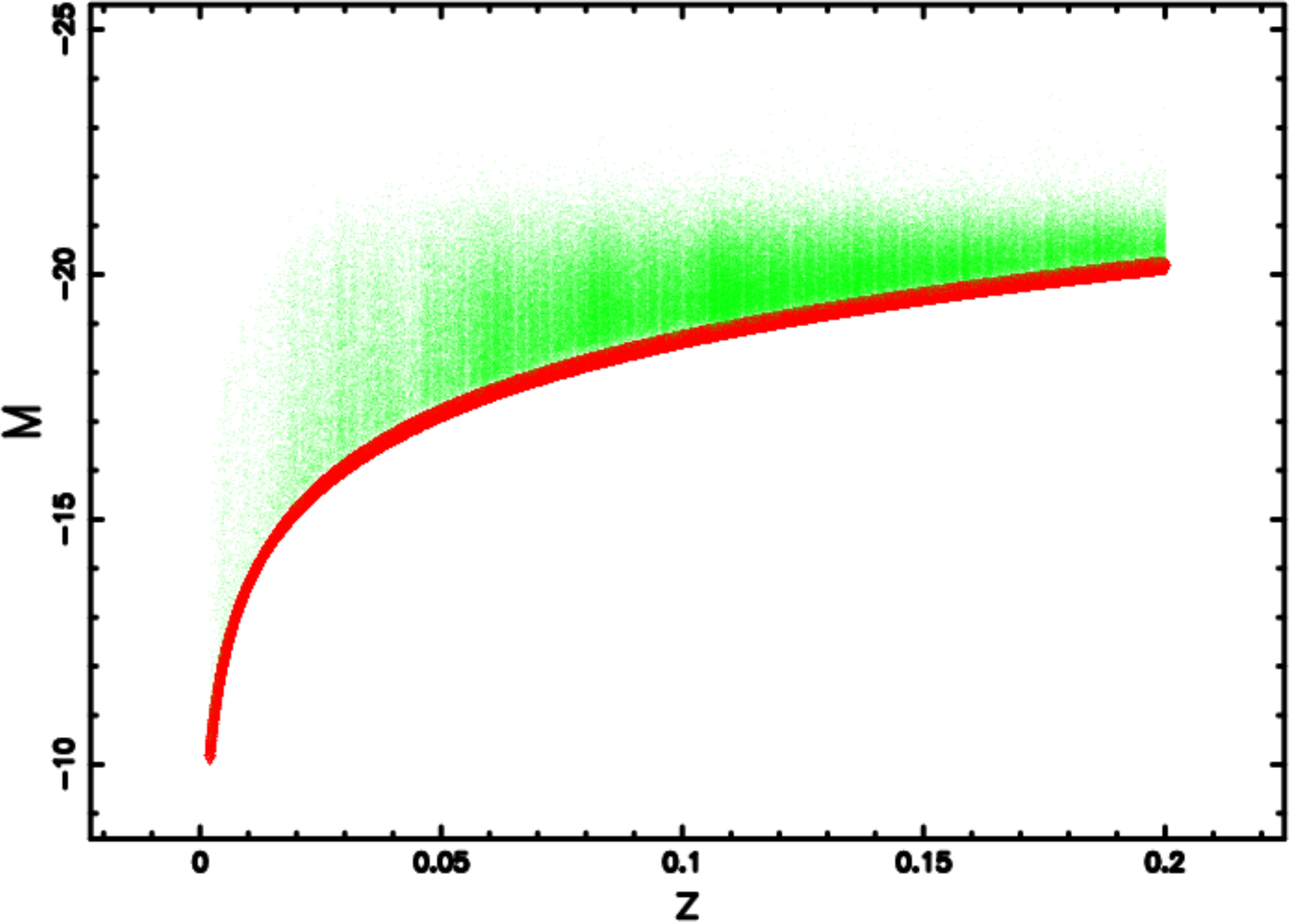}
\end{center}\caption{
The absolute magnitude $M$ of
202,923 galaxies belonging to the 2dFGRS
when $\mathcal{M_{\sun}}$ = 5.33 and
$H_{0}=66.04 \mathrm{\ km\ s}^{-1}\mathrm{\ Mpc}^{-1}$
(green points).
The upper theoretical curve as represented by
equation~(\ref{absolutel}) is reported as the
red thick line when $m_L$=19.61.
}
 \label{bias}%
 \end{figure}

This bias  was originally applied
to the stars,
see \cite{Malmquist_1920, Malmquist_1922},
 and was
then applied to the galaxies by \cite{Behr1951}.
We now introduce the concept of
limiting apparent magnitude and the corresponding
completeness in
absolute magnitude of the considered catalog
as a function of
redshift.
The observable absolute magnitude as a function of the
limiting apparent magnitude, $m_L$, is
\begin{equation}
M_L =
m_{{L}}-5\,{\it \log_{10}} \left( {\frac {{\it c}\,z}{H_{{0}}}}
 \right) -25
\quad .
\label{absolutel}
\end{equation}
The previous formula predicts, from a theoretical
point of view, an upper limit on the absolute
maximum magnitude that can be observed in a
catalog of galaxies characterized
by a given limiting
magnitude
and Fig. \ref{bias} reports such a curve
and the galaxies of the
2dFGRS.

The interval covered by the
LF of galaxies,
$\Delta M $,
is defined by
\begin{equation}
\Delta M = M_{max} - M_{min}
\quad ,
\end{equation}
where $M_{max}$ and $M_{min}$ are the
maximum and minimum
absolute
magnitude of the $LF$ for the considered catalog.
The real observable interval in absolute magnitude,
$\Delta M_L $,
 is
\begin{equation}
\Delta M_L = M_{L} - M_{min}
\quad .
\end{equation}
We can therefore introduce the range
of observable absolute maximum magnitude
expressed in percent,
$ \epsilon(z) $,
as
\begin{equation}
\epsilon_s(z) = \frac { \Delta M_L } {\Delta M } \times 100
\, \%
\quad .
\label{range}
\end{equation}
This is a number that represents
the completeness of the sample
and, given the fact that the limiting magnitude 
of the 2dFGRS is
$m_L$=19.61,
it is possible to conclude that the 2dFGRS is complete
for $z\leq0.0442$~.
In the case  of the 2MRS the limiting magnitude  is
$m_L$=11.19,
and therefore the  2MRS  is complete
for $z\leq0.00016$~.
This efficiency, expressed as a percentage,
can be considered to be a version  of the Malmquist bias.

\section{Photometric maximum}

\label{sec_maximum}
The parameters  of the Schechter $LF$ 
concerning the 2dFGRS
are reported in   Table~\ref{parameters}
and those  of the ${\mathcal M}-L$ 
$LF$  
are reported  in Table \ref{para_physical}.

\begin{table}
 \caption[]{The parameters of the Schechter function  for \\
      the 
     2dFGRS as in Madgwick et al. 2002. }
 \label{parameters}
 \[
 \begin{array}{lc}
 \hline
 \hline
 \noalign{\smallskip}
parameter            & 2dFGRS                                  \\ \noalign{\smallskip}
M^* - 5\log_{10}h ~ [mags]         &  ( -19.79 \pm 0.04)           \\ \noalign{\smallskip}
\alpha               &   -1.19  \pm 0.01                       \\ \noalign{\smallskip}
\Phi^* ~[h^3~Mpc^{-3}] &   ((1.59   \pm 0.1)10^{-2})      \\ \noalign{\smallskip}
 \hline
 \hline
 \end{array}
 \]
 \end {table}

\begin{figure}\begin{center}
\includegraphics[width=7cm]{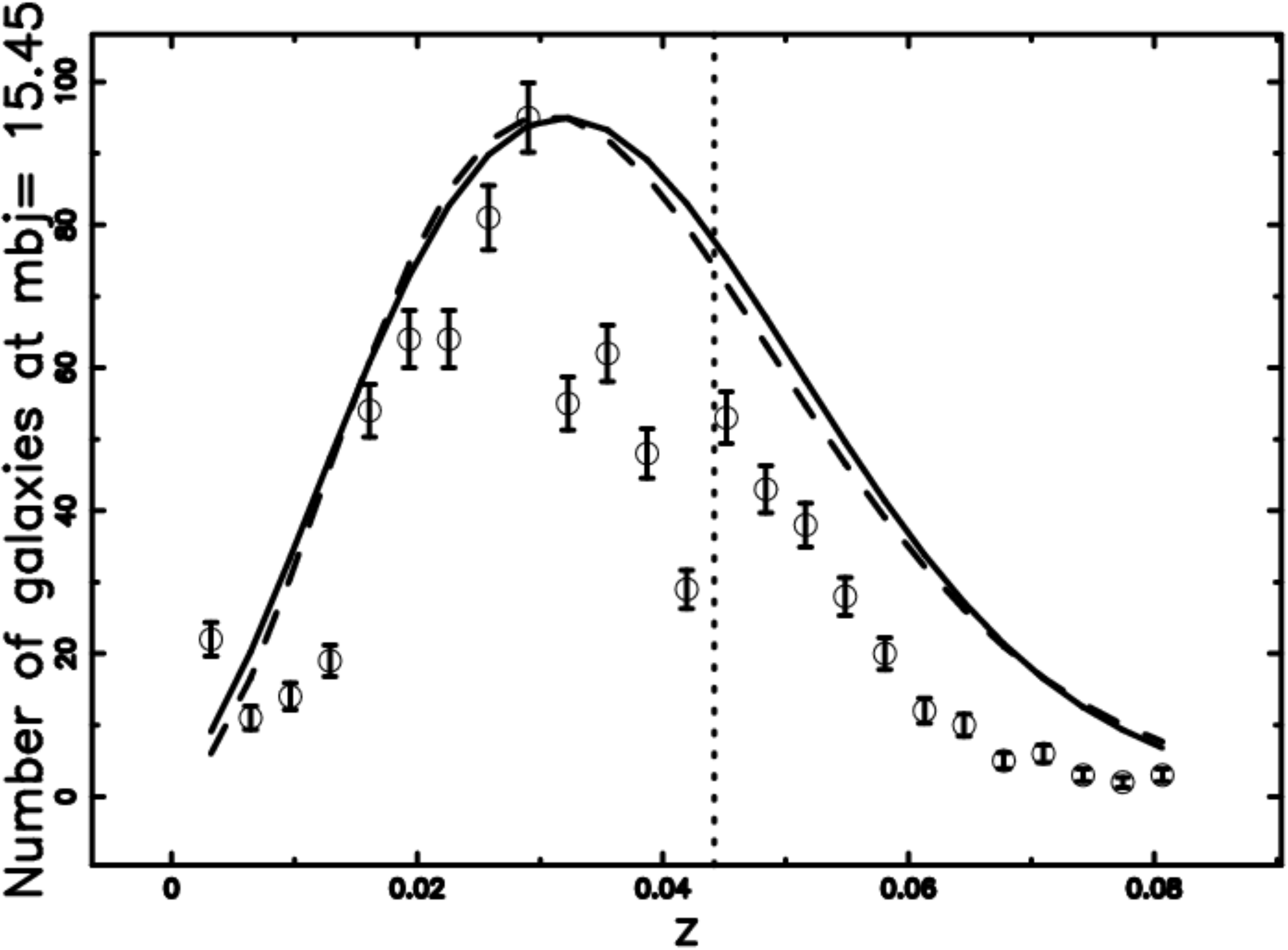}
\end{center}\caption{
The galaxies  of the 2dFGRS with 
$ 15.27  \leq  bJmag \leq 15.65 $  or 
$ 59253  \frac {L_{\sun}}{Mpc^2} \leq  
f \leq 83868  \frac {L_{\sun}}{Mpc^2}$
( with $bJmag$ representing  the 
relative magnitude  used in object selection),
are isolated 
in order to represent a chosen value of $m$ 
and then organized in frequencies versus
heliocentric  redshift,  
(empty circles);
the error bar is given by the square root of the frequency.
The maximum in the frequencies of observed galaxies is 
at  $z=0.03$.
The theoretical curve  generated by
the Schechter  $LF$  
(formula~(\ref{nfunctionz}) and parameters
as in column 2dFGRS of Table~\ref{parameters}) 
is drawn  (full line).
The theoretical curve  generated by
the ${\mathcal M}-L$   $LF$   (
formula~(\ref{nfunctionz_mia})
and  parameters as in column  
2dFGRS of Table~\ref{para_physical})
is drawn  (dashed line)
;
 $\chi^2$= 550  for the Schechter  function and $\chi^2$= 503
for the ${\mathcal M}-L$   function.
In this plot $\mathcal{M_{\sun}}$ = 5.33  and $h$=0.623.
The vertical  dotted line represents the boundary
between complete and incomplete samples.
}
          \label{maximum_flux}%
    \end{figure}

\begin{figure}\begin{center}
\includegraphics[width=7cm]{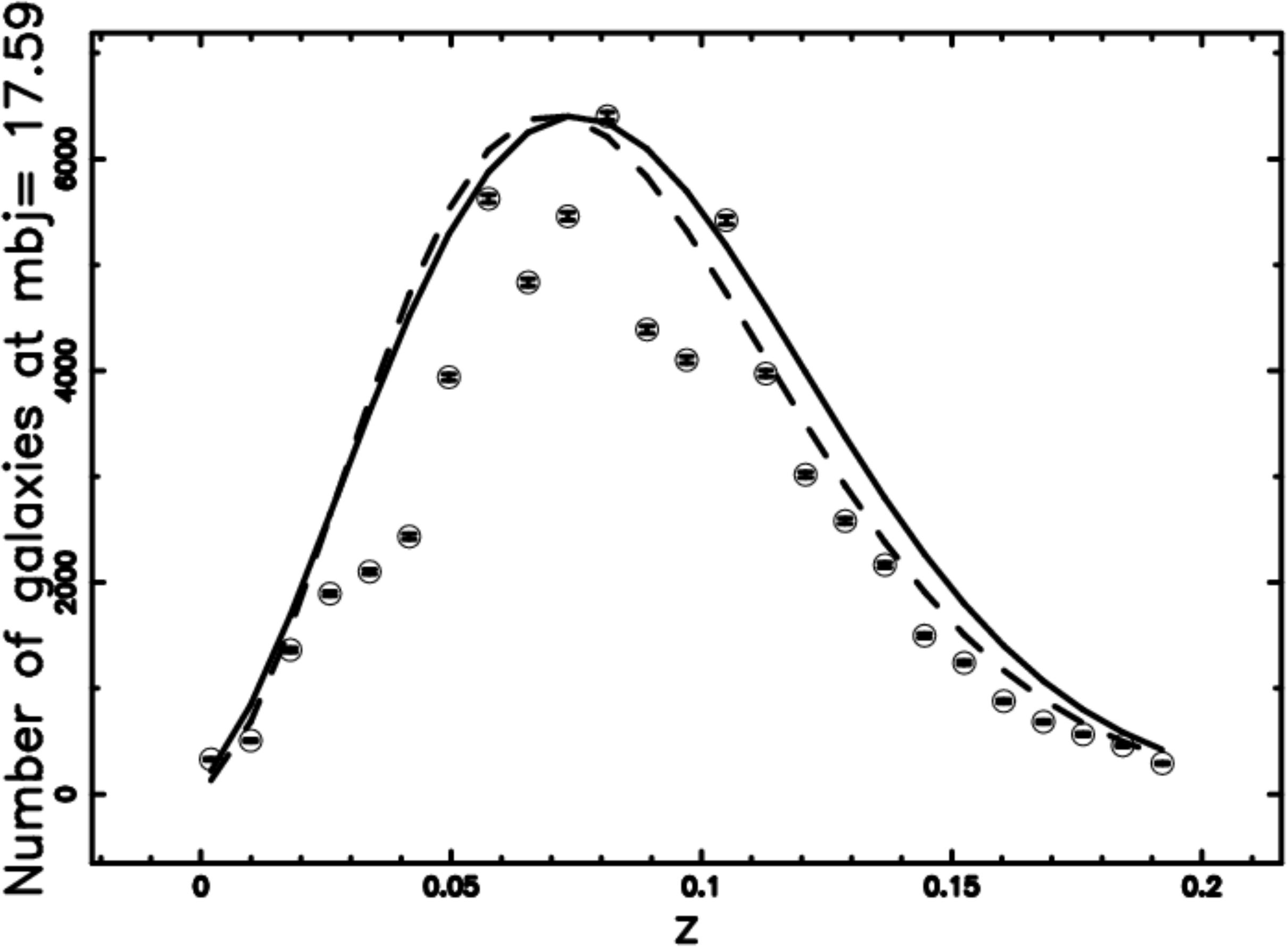}
\end{center}\caption{
The galaxies of the 2dFGRS database with 
$ 16.77 \leq bJmag \leq 18.40 $ or 
$ 4677 \frac {L_{\sun}}{Mpc^2} \leq 
f \leq 21087 \frac {L_{\sun}}{Mpc^2}$
(with $bJmag$ representing the 
relative magnitude used in object selection),
are isolated 
in order to represent a chosen value of $m$ 
and then organized as frequency versus
heliocentric redshift,
 (empty circles);
the error bar is given by the square root of the frequency.
The maximum in the frequencies of observed galaxies is 
at $z=0.085$ when $\mathcal{M_{\sun}}$ = 5.33 and
$h$=1 .
The theoretical curve generated by
the Schechter function of luminosity 
(formula~(\ref{nfunctionz}) and parameters
as in column 2dFGRS of Table~\ref{parameters}) 
is drawn (full line).
The theoretical curve generated by
generalized gamma LF,
 formula~(\ref{nfunctionz_gammagene}),
and parameters as in column 
2dFGRS of Table~\ref{para_physical})
is drawn (dashed line)
;
 $\chi^2$= 8078 for the Schechter function and $\chi^2$= 6654
for  generalized gamma LF.
}
 \label{maximumfluxgammagene}%
 \end{figure}

 \begin{table} 
 \caption[]{The parameters of the 
             ${\mathcal M}-L$  $LF$  \\
            based on the   2dFGRS data 
           ( triplets  generated by the author). } 
 \label{para_physical} 
 \[ 
 \begin{array}{lc} 
 \hline 
~     &   2dFGRS   \\ \noalign{\smallskip}  
 \hline 
 \noalign{\smallskip} 
c                   &    0.1                 \\ \noalign{\smallskip}
M^*   - 5\log_{10}h [mags]       &  -19  \pm 0.1       \\ \noalign{\smallskip}
\Psi^* [h^3~Mpc^{-3}] &  0.4  \pm 0.01      \\ \noalign{\smallskip}
a                   &  1.3   \pm 0.1       \\ \noalign{\smallskip} 
 \hline 
 \hline 
 \end{array} 
 \] 
 \end {table}
It is  interesting to point out  that 
other  values     for $h$  different from  1
shift all absolute magnitudes 
by $5\log_{10}h$ and change the 
number densities by the factor $h^3$.
Fig. \ref{maximum_flux}
reports the number of  observed  galaxies
of the 2dFGRS  catalog for  a  given    
apparent magnitude  and
two theoretical curves  
as represented by  formula~(\ref{nfunctionz})   
which is based on  the Schechter  $LF$
and  formula~(\ref{nfunctionz_mia})
which is based on the ${\mathcal M}-L$   $LF$.
A similar  plot  can be  done  
for the  generalized gamma  $LF$ ,
see  Fig. \ref{maximumfluxgammagene}.
The $\chi^2$ analysis allows to conclude that  
in the two cases  here  examined 
the  application of the ${\mathcal M}-L$   $LF$
and  the generalized gamma LF produce the same or better
results in respect to the use of the
Schechter LF.
More  details  can be found in  
\cite{Zaninetti2010a,Zaninetti2010f}.

\begin{figure}\begin{center}
\includegraphics[width=7cm]{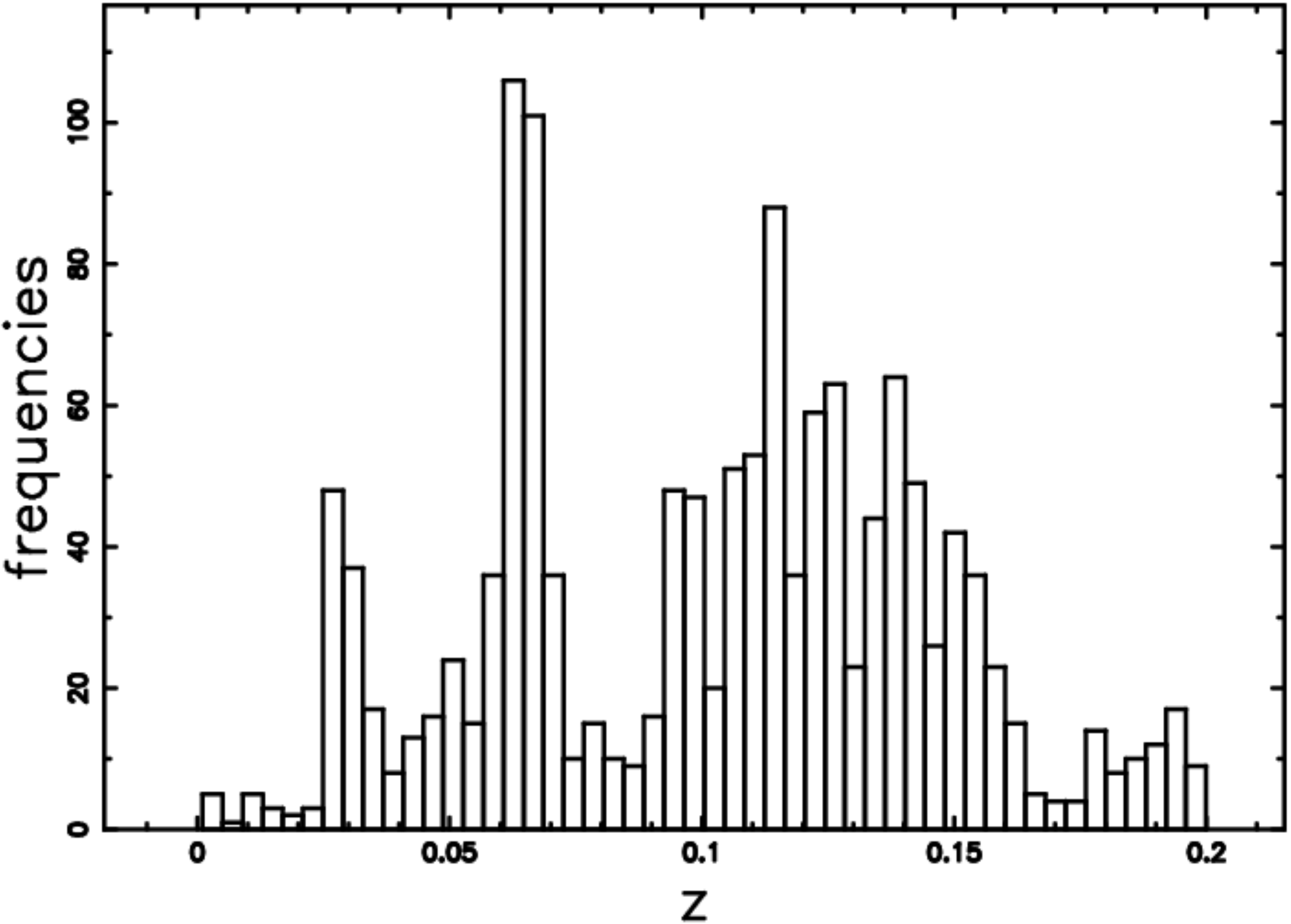} 
\end{center}\caption {
Histogram (step-diagram) of the 
number of galaxies of 2dFGRS as a function of the 
redshift in the  slice to the right of Fig.
\ref{2df_all}, the number of bins is 50.
The  circular sector has a central angle of
$1^\circ$. 
}
\label{isto2}
    \end{figure}

The non-homogeneous 
structure of the universe 
can be clarified by counting the number of galaxies
in one of the two slices of  2dFGRS
as a function of the redshift when a sector with
a central angle of $1^\circ$ is considered,
see Fig. \ref{isto2}.

\begin{figure}\begin{center}
\includegraphics[width=7cm]{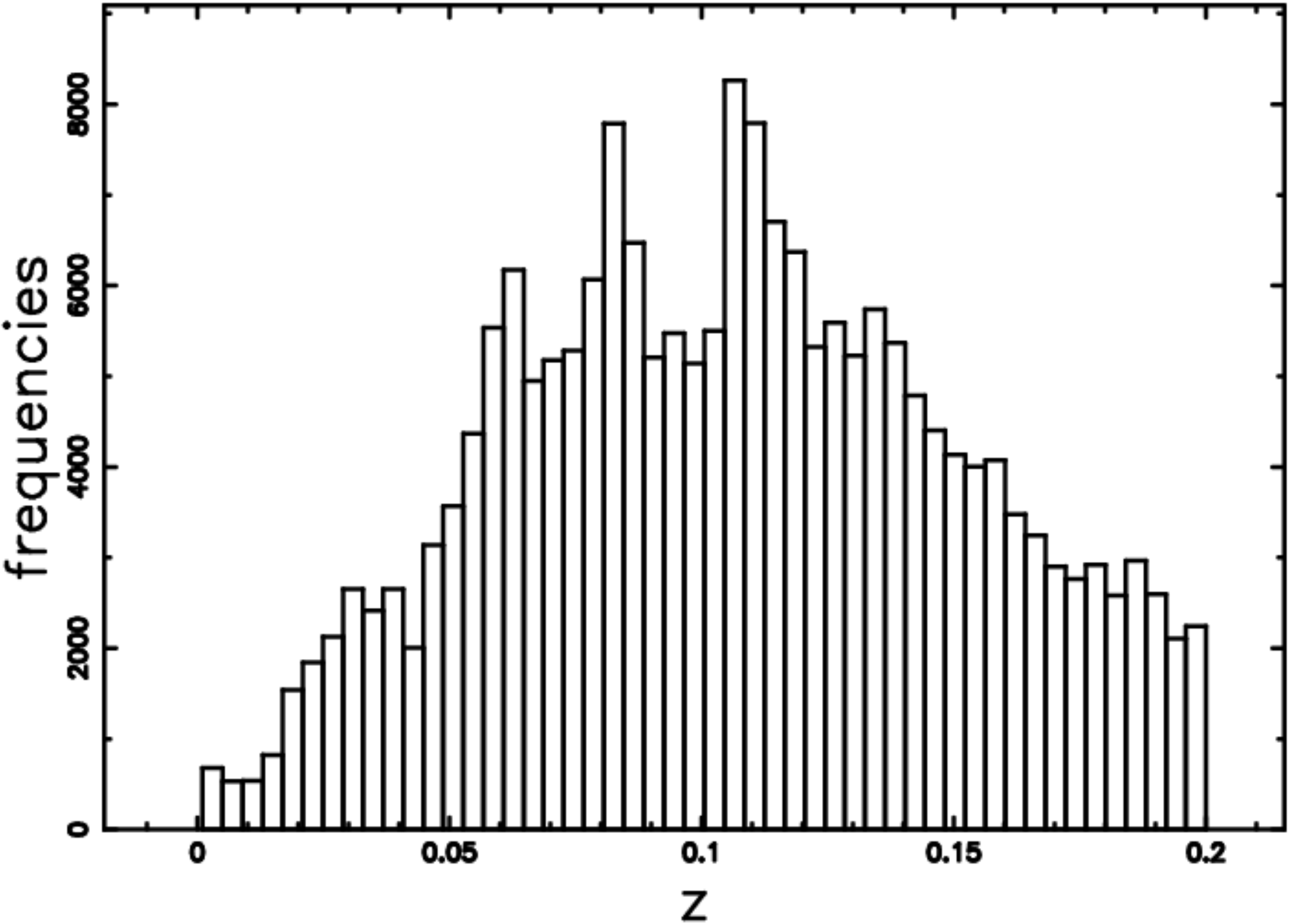} 
\end{center}\caption {
Histogram (step-diagram) of the 
number of galaxies  of  2dFGRS  as a function of the 
redshift when the two slices of Fig. \ref{2df_all} are added together, 
the number of bins is 50.
}
\label{isto1}
    \end{figure}

Conversely, when the two slices are considered together
the behavior of the number of galaxies 
as a function of the redshift is more continuous,
see Fig. \ref{isto1}. 

\section{The 3D Voronoi Diagrams}
\label{sec_voronoi} 
The faces of the Voronoi Polyhedra     
share the same property , i.e.
they are equally distant from 
two nuclei.  The intersection 
between a plane and the faces produces diagrams which are similar
to the edges'  displacement in  2D Voronoi diagrams.
From the point of view of the observations
it is very useful to study the intersection 
between a slice which crosses  the center of the box 
and the faces of  irregular polyhedrons where presumably
the galaxies reside.
The general definition of the  3D Voronoi Diagrams
is given in Section~\ref{general}.
The intersection between a slice of a given opening angle, 
for example $3^{\circ}$, 
and the faces of the Voronoi Polyhedra     
can be realized through an approximate  algorithm,
see  next  Section~\ref{faces}.

\subsection{General Definition}

\label{general}
The Voronoi diagram for a set of  seeds, $S$,
located at position $x_i$  
in $\mathcal{R}^3$ space  is the
partitioning of that space into regions such that 
all locations within any
one region are closer to the generating point 
than to any other.
In the following we will work on a three dimensional
lattice defined by 
$ pixels \times pixels \times pixels $ 
points,
$L_{kmn}$.
The Voronoi polyhedron $V_i$ around
a given center $i$ , 
is the set of lattice points $L_{kmn}$  closer
to $i$ than to any $j$: more formally,
\begin{equation}
  L_{kmn} \; \epsilon \;   V_i \leftrightarrow \mid x_{kmn} - x_i \mid 
\leq \mid x_{kmn} - x_j
\quad  ,  
\end{equation}
where $ x_{kmn}$ denotes the lattice point position. 
Thus,  the Polyhedra  are intersections
of half-spaces. Given a center $i$ and its neighbor $j$,  
the line
$ij$ is cut perpendicularly at its midpoint $ y_{ij}$ by
the plane $h_{ij}$. 
$H_{ij}$ is the half-space generated 
by the plane $h_{ij}$, which  consists of 
the subset of lattice points on 
the same side of $ h_{ij}$ as  $i$ ; therefore 
\begin{equation}
     V_i = \cap _j H_{ij }    ,    
\end{equation}
$V_i$ is bounded by faces , 
with each face $ f_{ij} $ belonging 
to a distinct plane $h_{ij}$. 
Each face will be characterized 
by its vertexes and edges.

\subsection{The adopted algorithm} 

\label{faces}
Our  method considers a  3D lattice with 
${\it pixels}^{3}$ points:
present in this lattice are $N_s$ seeds generated
according to a random process.
All the computations are usually performed on this mathematical
lattice; 
the conversion to the physical lattice
is obtained   by multiplying the unit
by $\delta=\frac{side}{pixels -1}$, where {\it side}
is the  length of the cube expressed in the physical
unit  adopted.
In order to minimize boundary  effects 
introduced by those
polyhedron which  cross the cubic boundary,
the cube in which the seeds are inserted 
is amplified
by  a factor {\it amplify}. 
Therefore the $N_s$  seeds are inserted in a volume
   $  pixels^3 \times
amplify$,  
which is bigger than  the box over which  the
scanning is performed; 
{\it amplify } is generally taken to be equal to 1.2.
This procedure inserts periodic boundary conditions to our
cube. 
A sensible and solid discussion of what such an 
extension of a cube should be 
can be  found  in 
\cite{Neyrinck2005b}.
The set $S$ of the seeds can be of
Poissonian   or non-Poissonian type.
Adopting  the point of view that the universe should be the same
from each point of view of the observer the 
Poissonian seeds can represent the best choice
in order to reproduce the large scale structures.

The Poissonian   seeds  are generated independently on the $X$, 
$Y$  and
$Z$ axis in 3D through a subroutine  which returns a
pseudo-random real number taken from a uniform distribution
between 0 and 1. For practical purposes,
 the subroutine
RAN2  was used, see \cite{press}.
Particular attention  should be paid to
the average observed diameter of voids, $\overline{DV^{obs}}$,
here chosen 
as
\begin{equation}
\overline{DV^{obs}} \approx  36.46  Mpc/h 
\quad ,
\label{dvobserved}
\end {equation} 
see Section \ref{secstat}.
The number of Poissonian     seeds is chosen in such 
a way that the averaged 
volume occupied by a Voronoi polyhedron is equal to 
the averaged observed volume of  the voids 
in the spatial distribution
of galaxies; 
more details can be found in
\cite{Zaninetti2006}.

\begin{figure}\begin{center}
\includegraphics[width=7cm]{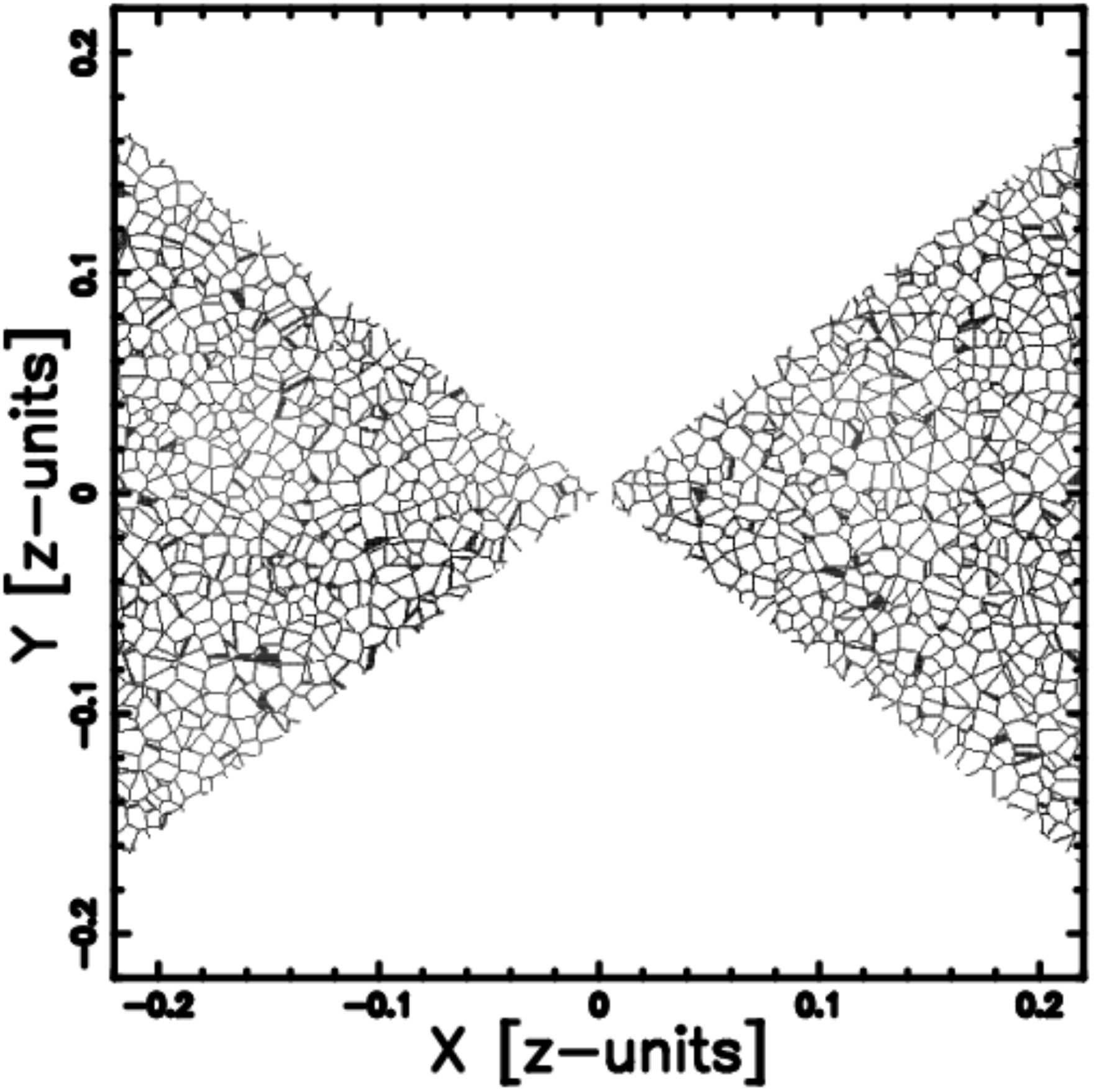}
\end{center}\caption{
Portion of the  Poissonian Voronoi--diagram $V_p(2,3)$ ;
cut on the  X-Y plane   when two strips of 
$75^{\circ}$ are considered.
The  parameters
are      $ pixels$= 600
       , $ N_s   $   = 137998
       , $ side  $   = 131908 $Km/sec$
  and    $ amplify$~= 1.2~.}
          \label{cut_middle}%
    \end{figure}

\begin{figure}\begin{center}
\includegraphics[width=7cm]{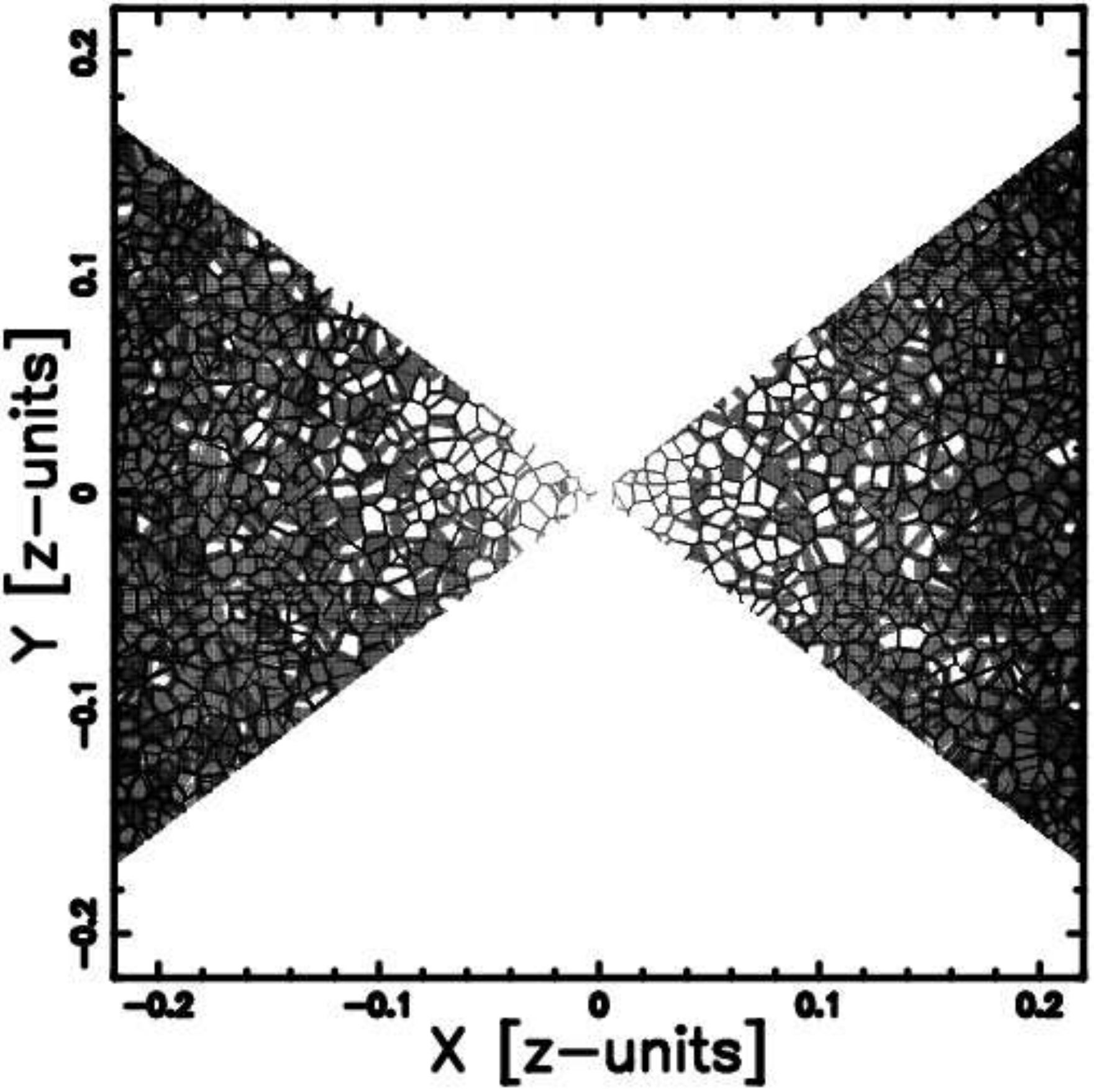}
\end{center}\caption{
The same as Fig. \ref{cut_middle}
but now   two slices  of 
$75^{\circ}$  long and  $3^{\circ}$
wide are considered.} 
          \label{voro_fetta_tutte}%
    \end{figure}

We now work on a 3D lattice  L$_{k,m,n}$ of $pixels^3$
 elements.
Given a section  of the cube
(characterized, for example, by $k=\frac{pixels}{2}$)
the  various $V_i$ (the volume belonging
to the seed i)
 may or may not cross the  pixels 
 belonging to the two dimensional lattice.
A typical  example of   a 2D cut  organized in   two  strips 
about $75^{\circ}$ long  
is  visible  in Fig. \ref{cut_middle} 
where 
the Cartesian coordinates $X$ and $Y$ with the origin 
of the axis at the center of the box  has been used.
The previous cut has an extension on the 
$Z$-axis equal to zero.
Conversely Fig. \ref{voro_fetta_tutte} reports 
two slices
of $75^{\circ}$  long and  $3^{\circ}$
wide. 
In this case the extension of the enclosed region 
belonging to the $Z$-axis increases with distance 
according to
\begin{equation}
\Delta Z = \sqrt {X^2 +y^2} \tan \frac {\alpha} {2} 
\quad ,
\end{equation}
where $\Delta Z $ is the thickness of the slice and
$\alpha$ is  the opening angle, 
in our case $3^\circ$.
In order to simulate the slices of observed galaxies 
 a subset is extracted ( randomly chosen)   
of the pixels belonging
to a slice  as represented, for example,
in  Fig. \ref{voro_fetta_tutte}. 
In this operation of extraction of the galaxies from the pixels
of the slice, the photometric rules as represented by 
formula~(\ref{nfunctionz}) must  be respected.   

\begin{figure}\begin{center}
\includegraphics[width=7cm]{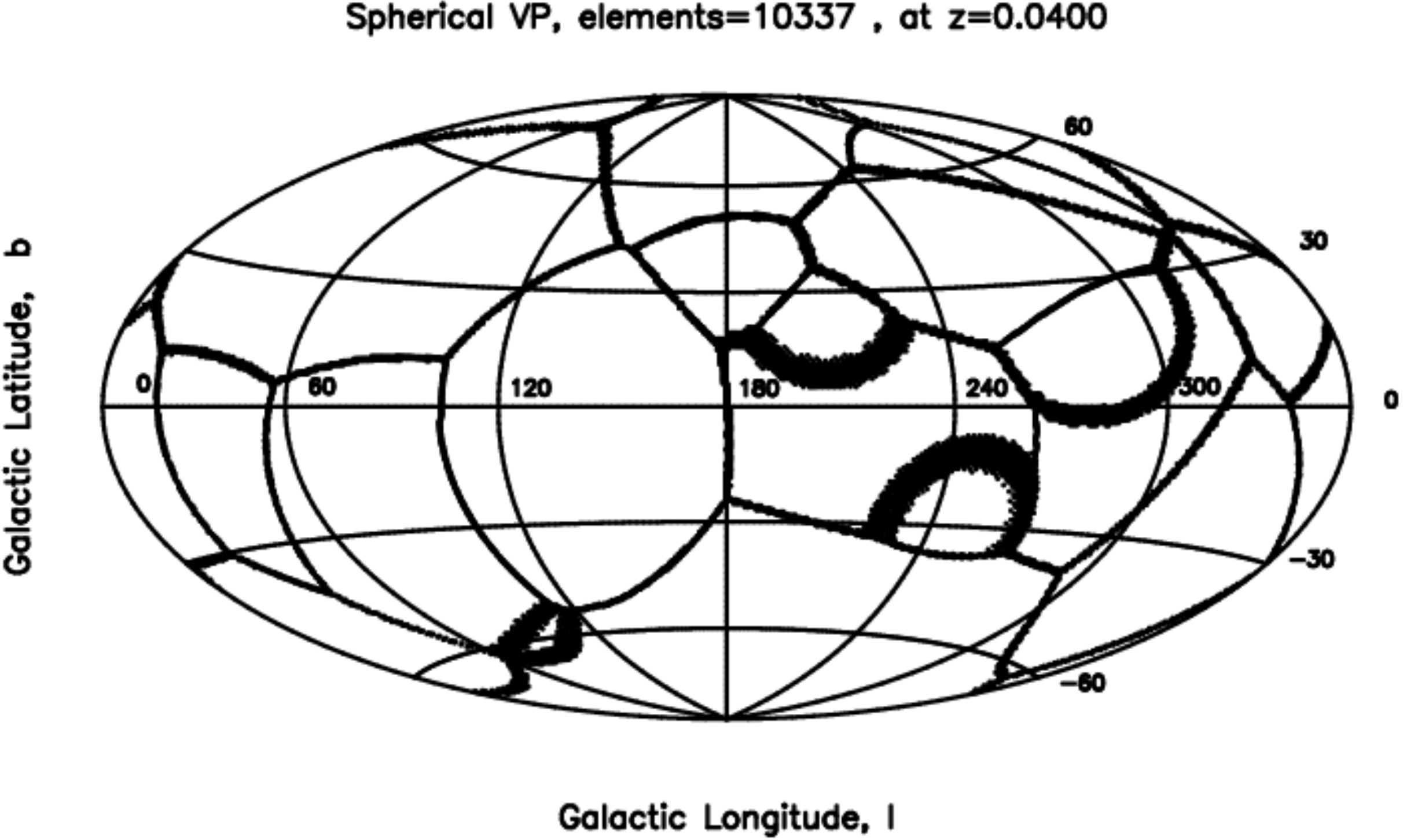} 
\end{center}\caption {
The Voronoi--diagram $V_s(2,3)$ 
in the Hammer-Aitoff  projection
at $z$ = 0.04.
The  parameters
are      $ pixels$= 400, 
         $ N_s   $   = 137998, 
         $ side  $   = 131908 $Km/sec$
and    $ amplify$= 1.2.}
          \label{aitof_sphere}%
    \end{figure}

\begin{figure}\begin{center}
\includegraphics[width=7cm]{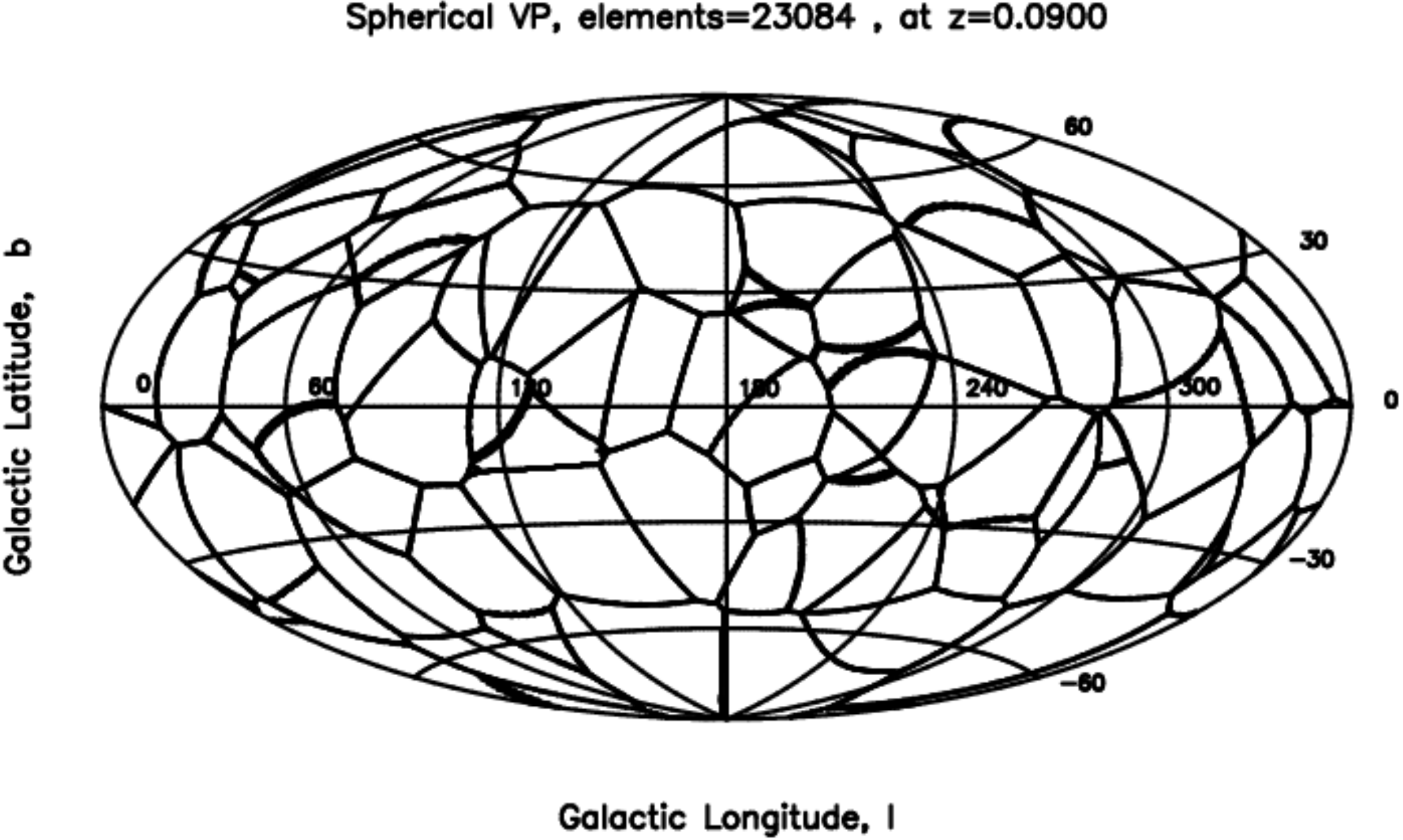} 
\end{center}\caption {
The Voronoi--diagram $V_s(2,3)$ 
in the Hammer-Aitoff  projection
at $z$ = 0.09;
other  parameters as in Fig. \ref{aitof_sphere}.
        }
          \label{aitofsphere2}%
    \end{figure}

The cross sectional area of the VP can also
be visualized through 
a spherical cut characterized by a constant value 
of the distance to the center of the box,
in this case expressed in $z$ units, 
see Fig. \ref{aitof_sphere}
and Fig. \ref{aitofsphere2};
 this intersection is called $V_s(2,3)$  where the 
index $s$ stands for sphere.

\section{The statistics of the Voronoi Diagrams}

\label{sec_statistics}

A probability density function
(PDF) is the first derivative
of a distribution function (DF)
$F(x)$
with respect to $x$.
In the case where the PDF is known
but the DF is unknown,
the following integral is evaluated
\begin{equation}
F(x) = \int_0^x f(x) dx
\quad .
\end{equation}
As a consequence the survival function (SF)
is
\begin{equation}
SF = 1 - F(x) \quad .
\end{equation}
We recall that the  PVT  is  a particular case of
the Voronoi tessellation in which the seeds  are generated
independently on the $X$, $Y$  and $Z$ axes in 3D.

\subsection{The Kiang function}

The gamma variate $H (x ;c )$  (\cite{kiang})
is
\begin{equation}
 H (x ;c ) = \frac {c} {\Gamma (c)} (cx )^{c-1} \exp(-cx)
\quad ,
\label{kiang}
\end{equation}
where $ 0 \leq x < \infty $, $ c~>0$,
and $\Gamma (c)$ is the gamma function with argument $c$.
The Kiang  PDF has a mean of
\begin{equation}
\mu = 1
\quad ,
\end{equation}
and variance
\begin{equation}
\sigma^2 = \frac{1}{c}
\quad .
\end{equation}
This PDF  can be  generalized by 
introducing  the 
dimension of the considered space,  $d(d=1,2,3)$,
and  $c=2d$  
\begin{equation}
 H (x ;d  ) = \frac {2d} {\Gamma (2d)} (2dx )^{2d-1} \exp(-2dx)
\quad .
\label{kiangd}
\end{equation}
In the case of a 1D PVT,
$c=2$  is an
exact analytical result
and conversely
$c$ is supposed to be 4 or 6
for  2D or  3D  PVTs,
respectively, \cite{kiang}.
The  DF of the Kiang function, DF$_K$,
is
\begin{equation}
DF_K =
1 -{\frac {\Gamma  \left( c,cx \right) }{\Gamma  \left( c \right) }}
\quad ,
\end{equation}
where the incomplete
Gamma  function, $ \Gamma (a,z) $,   is defined by
\begin{equation}
\Gamma (a,z) =
\int _{z}^{\infty }\!{{\rm e}^{-t}}{t}^{a-1}{dt}
\quad .
\end{equation}

The survival function $S_K$  is
\begin{equation}
S_K =
{\frac {\Gamma  \left( c,cx \right) }{\Gamma  \left( c \right) }}
\quad  .
\label{survival_kiang}
\end{equation}

\subsection{Generalized gamma}

The generalized gamma
PDF with three parameters $a,b,c$,
\cite{Hinde1980,Ferenc_2007,Tanemura2003}, is
\begin{equation}
f(x;b,c,d) =
 c \frac {b^{a/c}} {\Gamma (a/c) } x^{a-1} \exp{(-b
x^c)} \quad . \label{gammag}
\end{equation}

The generalized gamma
 has a mean of
\begin{equation}
\mu = \frac
{
{b}^{-\frac{1}{c} }\Gamma \left( \frac {1+a}{c} \right)
}
{
\Gamma \left( {\frac {a}{c}} \right)
}
\quad ,
\end{equation}
and a variance of
\begin{equation}
\sigma^2 = 
\frac
{
{b}^{- \frac{2}{c} } \left( +\Gamma \left( {\frac {2+a}{c}} \right)
\Gamma \left( {\frac {a}{c}} \right) - \left( \Gamma \left( {\frac {
1+a}{c}} \right) \right) ^{2} \right)
}
{
\left( \Gamma \left( {\frac {a}{c}} \right) \right) ^{2}
}
\quad .
\end{equation}


The  SF of the generalized gamma
is
\begin{equation}
S_{GG} =
\frac { \Gamma  \left( {\frac {a}{c}},b{x}^{c} \right) }
 {
  \Gamma
  \left( {\frac {a}{c}} \right)
}
\quad .
\label{survivalgg}
\end{equation}

\subsection{Ferenc--Neda function }

A new PDF
has been recently introduced, \cite{Ferenc_2007},
in order to model the normalized area/volume
in a 2D/3D PVT
\begin{equation}
FN(x;d) =  
 C \times x^{\frac {3d-1}{2} } \exp{(-(3d+1)x/2)}
\quad ,
\label{rumeni}
\end{equation}
where $C$ is a constant,
\begin{equation}
C =   
\frac
{
\sqrt {2}\sqrt {3\,d+1}
}
{
2\,{2}^{3/2\,d} \left( 3\,d+1 \right) ^{-3/2\,d}\Gamma \left( 3/2\,d+
1/2 \right)
}
\quad ,
\end{equation}
and $d(d=1,2,3)$ is the
dimension of the space under consideration.
We will call this
function the  Ferenc--Neda  PDF;
it has a mean of
\begin{equation}
\mu = 1
\quad ,
\end{equation}
and variance
\begin{equation}
\sigma^2 = \frac{2}{3d+1}
\quad .
\end{equation}
The  SF of the
Ferenc--Neda function when $d=3$ is
\begin{eqnarray}
S_{FN3} =
{{\rm e}^{-5\,x}}+5\,{{\rm e}^{-5\,x}}x
\nonumber  \\
+{\frac {25}{2}}\,{{\rm e}^{-5
\,x}}{x}^{2}
+{\frac {125}{6}}\,{{\rm e}^{-5\,x}}{x}^{3}
+{\frac {625}{
24}}\,{x}^{4}{{\rm e}^{-5\,x}}
\quad .
\label{survivalfn3}
\end{eqnarray}

\subsection{Kiang function of the radius }

\label{secradius}

We now analyze the distribution in effective radius $R$ of
the 3D PVT.
We assume that the volume of each cell, $v$,
is
\begin{equation}
v = \frac{4}{3} \pi (\frac{R}{\rho})^3 \quad ,
\end{equation}
where $\rho$ is a length  that connects the normalized radius to
the observed one.
In the following, we derive the PDF for radius
and related quantities relative to the Kiang function and
Ferenc--Neda function.
The PDF, $ H_R (R ;c )$,
of the radius corresponding
to the Kiang function as represented
by (\ref{kiang}) is
\begin{eqnarray}
 H_R (R ;c )   =  
\frac {
 4\,c \left( 4/3\,{\frac {c\pi \,{R}^{3}}{{\rho}^{3}}}
\right) ^{c-1}{e ^{-4/3\,{\frac {c\pi \,{R}^{3}}{{\rho}^{3}}}}}\pi
\,{R}^{2} } { \Gamma  \left( c \right) {\rho}^{3} }
 \quad , \label{kiangr}
\end{eqnarray}
where $ 0 \leq R < \infty $, $ c~>0$ and $ \rho ~>0$.
The Kiang PDF of the radius has a mean of
\begin{equation}
\mu = 1/2\,{\frac {\sqrt [3]{2}\sqrt [3]{3}\Gamma  \left( 1/3+c
\right) }{ \sqrt [3]{c}\sqrt [3]{\pi }\Gamma  \left( c \right) }}
\rho \quad ,
\end{equation}
and variance
\begin{eqnarray}
\sigma^2 =   \frac{1}{4} 
{\frac {{3}^{\frac{2}{3}}{2}^{\frac{2}{3}} ( \Gamma  ( 2/3+c )
\Gamma  ( c ) - ( \Gamma  ( 1/3+c )
 ) ^{2} ) }{{c}^{2/3}{\pi }^{2/3} ( \Gamma  ( c
 )  ) ^{2}}}
 \rho^2
\quad .
\end{eqnarray}

 The survival function of  the Kiang function in radius is
 \begin{equation}
 S_{KR} = {\frac {\Gamma  \left( c,4/3\,c\pi
 \,(\frac{R}{\rho})^{3}  \right) }{\Gamma \left( c \right) }} \quad
 . \label{survival_kiangr}
 \end{equation}

\subsection{The Ferenc--Neda function of the radius }

The PDF as a function of the radius,
obtained from (\ref{rumeni}) and inserting
$d=3$, is
\begin{equation}
FN_R(R;d) = \frac {400000\,{\pi }^{5}{R}^{14}{e^{-{\frac
{20}{3}}\,{\frac {\pi \,{R}^{3}} {{\rho}^{3}}}}} } {
243\,{\rho}^{15}}
  \quad .
\label{rumenir}
\end{equation}
The mean of the Ferenc--Neda function
is
\begin{equation}
\mu = 0.6  \rho
\quad ,
\end{equation}
and the variance is
\begin{equation}
\sigma^2 = 0.0085 \rho^2\quad .
\end{equation}


 The SF of the
 Ferenc--Neda function of the radius when $d=3$ is
 \begin{eqnarray}
 S_{FN3R} =
 {e^{-{\frac {20}{3}}\,{\frac {\pi
 \,{R}^{3}}{{\rho}^{3}}}}}+{\frac {20 }{3}}\,{e^{-{\frac
 {20}{3}}\,{\frac {\pi \,{R}^{3}}{{\rho}^{3}}}}}{R}^ {3}\pi
 {\rho}^{-3}
 +{\frac {200}{9}}\,{e^{-{\frac {20}{3}}\,{\frac { \pi
 \,{R}^{3}}{{\rho}^{3}}}}}{R}^{6}{\pi }^{2}{\rho}^{-6} 
 \nonumber\\
 +{\frac {4000 }{81}}\,{e^{-{\frac {20}{3}}\,{\frac {\pi
 \,{R}^{3}}{{\rho}^{3}}}}}{R} ^{9}{\pi }^{3}{\rho}^{-9}
 +{\frac
 {20000}{243}}\,{e^{-{\frac {20}{3}}\, {\frac {\pi
 \,{R}^{3}}{{\rho}^{3}}}}}{R}^{12}{\pi }^{4}{\rho}^{-12}
 \quad .
 \label{survivalfn3r}
 \end{eqnarray}

\subsection{Kiang distribution of $V_p(2,3)$ in radius }

Here, we first model the normalized area-distribution
$V_p(2,3)$  with Kiang  PDFs as represented
by  (\ref{kiang}),
see Table~\ref{tablev23}.
 \begin{table}
 \caption[]{
Values of $\chi^2$ for
the cell normalized area-distribution function of $V_p(2,3)$;
here $T_i$ are the  theoretical frequencies and
$O_i$ are the  sample frequencies.
Here we have
8517 Poissonian seeds and 40 intervals in the histogram.
}
 \label{tablev23}
 \[
 \begin{array}{llll}
 \hline
PDF ~  &  parameters & \chi^2  \\ \noalign{\smallskip}
 \hline
 \noalign{\smallskip}
 H (x ;c ) ~(Eq. (\ref{kiang}))
    &  c=2.07 & 114.41   \\
 p (x ; b ) ~(Eq. (\ref{exponential}))
    &  d=1 & 85.38   \\
\noalign{\smallskip}
 \hline
 \hline
 \end{array}
 \]
 \end {table}

The PDF, $ H_{R23}  (R ;c )$,
as a function of the radius corresponding
to the Kiang function as represented
by (\ref{kiang}) for
$V_p(2,3)$
 is
\begin{equation}
 H_{R23}  (R ;c )   =
\frac
 {
2\,c \left( {\frac {c\pi \,{R}^{2}}{{\rho}^{2}}} \right) ^{c-1}{
{\rm e}^{-{\frac {c\pi \,{R}^{2}}{{\rho}^{2}}}}}\pi \,R
 }
 {
\Gamma  \left( c \right) {\rho}^{2}
 }
 \quad ,
\label{kiang23}
\end{equation}
where $ 0 \leq R < \infty $, $ c~>0$
and $ \rho ~>0$.
The Kiang  PDF of the radius
for
$V_p(2,3)$
has a mean of
\begin{equation}
\mu =
{\frac {\rho\,\Gamma  \left( c+1/2 \right) }{\sqrt {c}\sqrt {\pi }
\Gamma  \left( c \right) }}
 \quad ,
\end{equation}
and variance
\begin{equation}
\sigma^2 =
\frac
{
{\rho}^{2} \left( c \left( \Gamma  \left( c \right)  \right) ^{2}-
 \left( \Gamma  \left( c+1/2 \right)  \right) ^{2} \right)
}
{
c\pi \, \left( \Gamma  \left( c \right)  \right) ^{2}
}
\quad .
\end{equation}

The survival function
of the Kiang function
, $S_{KR23}$ , in  radius
for
$V_p(2,3)$
 is
\begin{equation}
S_{KR23} =
\frac
{
\Gamma  \left( c,2\,{\frac {c\pi \,{R}^{2}}{{\rho}^{2}}} \right)
}
{
\Gamma  \left( c \right)
}
\quad .
\label{survival_kiangr23}
\end{equation}

\begin{figure}\begin{center}
\includegraphics[width=7cm]{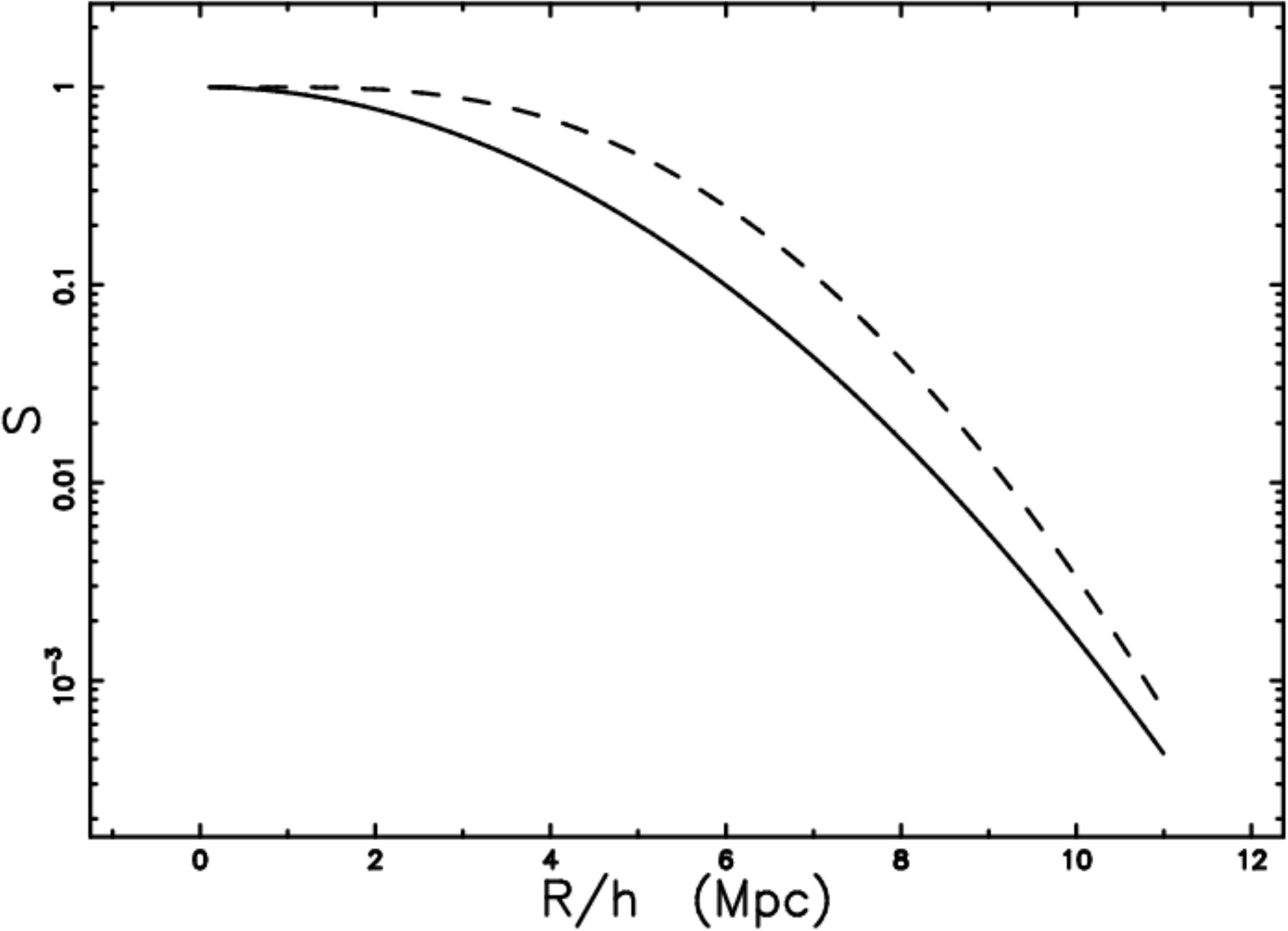}
\end{center}\caption
{
The survival function, $S_{KR23} $,
of the radius of the Kiang 
function for $V_p(2,3)$
as represented by
(\ref{survival_kiangr23})
when $\rho=13$~Mpc, $c=2.25$
and $\chi^2=67.1$ with
100 points
 (dashed line).
The survival function, $S_{ER23} $,
of the radius of the exponential distribution
for $V_p(2,3)$
as represented by
(\ref{survival_expr23})
when $\rho=7$~Mpc
and $\chi^2=9.27$ with
100 points
(full line).
}
 \label{comparison_cut2}%
 \end{figure}

A comparison of the survival function
$S_{KR23}$
of the radius
and  the exponential
distribution is
reported in Fig. \ref{comparison_cut2}.

\subsection{Exponential distribution of $V_p(2,3)$ in radius }

Another PDF that can be considered in order to
model
the normalized area distribution of $V_p(2,3)$
is the exponential distribution,
\begin{equation}
p(x) = \frac{1} {b} \exp {-\frac{x} {b}}
\label{exponential}
\quad ,
\end{equation}
which has an average value
\begin{equation}
\overline {x} = b
\quad .
\end{equation}
In the case of the normalized areas $b=1$,
Table \ref{tablev23} reports the $\chi^2$ values
of the two distributions adopted here.
The PDF, $ p_{R23}  (R ;c )$,
as a function of the radius corresponding
to the exponential distribution as represented
by  (\ref{exponential}) for
$V_p(2,3)$
 is
\begin{equation}
 p_{R23}  (R ;c )   =
\frac
 {
2\,{{\rm e}^{-{\frac {\pi \,{R}^{2}}{{\rho}^{2}}}}}\pi \,R
 }
 {
{\rho}^{2}
 }
 \quad ,
\label{exponential23}
\end{equation}
where $ 0 \leq R < \infty $, $ \rho>0$.
The exponential PDF of the radius
for
$V_p(2,3)$
has a mean of
\begin{equation}
\mu =
\frac
{
{\pi }^{3/2}
}
{
2\,{\rho}^{2} \left( {\frac {\pi }{{\rho}^{2}}} \right) ^{3/2}
}
 \quad ,
\label{rmedioexp}
\end{equation}
and variance
\begin{equation}
\sigma^2 =
\frac
{
{\rho}^{2} \left( 4 - \pi  \right)
}
{
4\,\pi
}
\quad .
\end{equation}

The survival function 
of the exponential distribution 
, $S_{ER23}$,  in  radius
for
$V_p(2,3)$
 is
\begin{equation}
S_{ER23} =
{{\rm e}^{-{\frac {\pi \,{R}^{2}}{{\rho}^{2}}}}}
\quad .
\label{survival_expr23}
\end{equation}
Fig. \ref{comparison_cut2}
reports a
 comparison between the survival
function of Kiang distribution 
and the exponential  distribution
for $V_p(2,3)$.
More details can be found  in \cite{Zaninetti2010g}.

\section{Stereology}

\label{secstereology}
We first briefly review   how a
PDF   $f(x)$ changes to
$g(y)$  when a new variable $y (x)$  is introduced. 
We limit  ourselves
to the case in which $y(x)$ is a one-to-one transformation.
The rule for
transforming  a PDF  is
\begin{equation}
g(y) =  \frac  {f(x) } {\vert\frac {dy } {dx} \vert}.
\label{trans} 
\end{equation}

Analytical results  have  shown 
that sections through D-dimensional Voronoi 
tessellations are
not themselves D-1 Voronoi tessellations,
see \cite{Moller1989,Moller1994,Weygaert1996}.
According to \cite{Blower2002}, the
probability of a plane intersecting a given 
sphere is proportional
to the sphere's radius, $R$. 
Cross-sections of radius $r$ may be
obtained from any sphere with a radius greater than or equal to
$r$. We may now write a general expression for 
the probability of
obtaining a cross-section of radius $r$ from the whole
distribution (which is denoted $F(R)$):
\begin{equation}
f(r) = \int_r^{\infty} F(R)  R \frac{1}{R} \frac{r} {\sqrt{R^2
-r^2}} dR,
\label{fundamental}
\end{equation}
which is formula (A7) in \cite{Blower2002}. 
That is to say, $f(r)$
is the probability of finding a bubble of radius $R$,
multiplied
by the probability of intersecting this bubble, 
multiplied by the
probability of obtaining a slice of radius $r$ from this bubble,
integrated over the range of $R\ge r$. 
A first  example is  given
by  the so called monodisperse bubble size distribution (BSD)
which are bubbles of constant radius $R$ and therefore
\begin{equation}
F(R) =\frac{1}{R},
\end{equation}
which is defined in  the interval $[0, R]$
and
\begin{equation}
f(r)= {\frac {r}{\sqrt {{R}^{2}-{r}^{2}}R}},
\end{equation}
which is defined in  the interval $[0, R]$,
see  Eq.~(A4) in \cite{Blower2002}.
The average value  of the radius of the  2D-slices is
\begin{equation}
\overline{r}  =
1/4\,R\pi,
\end{equation}
the variance  is
\begin{equation}
\sigma^2  =
2/3\,{R}^{2}-1/16\,{R}^{2}{\pi }^{2},
\end{equation}
and finally,
\begin{equation}
Skewness = -1.151, \quad
Kurtosis = 0.493.
\end{equation}

\subsection{PVT stereology}

In order to  find our $F(R)$, we now analyze 
the distribution in
effective radius $R$ of the 3D PVT. We assume that 
the volume of
each cell, $v$, is
\begin{equation}
v = \frac{4}{3} \pi R^3.
\end{equation}
In the following, we derive the PDF for
the radius and related
quantities relative to the
 Ferenc--Neda function.
The PDF as a function of the radius 
according to the rule of change of variables (\ref{trans}),
is obtained from
(\ref{rumeni}) on inserting $d=3$
and  was already defined , see equation \ref{rumenir}.
The average radius  is
\begin{equation}
\overline{R} = 0.6065,
\end{equation}
and the variance is
\begin{equation}
\sigma^2(R) = 0.00853.
\end{equation}
The  introduction of the scale factor, $b$,
with the new variable
$R=R^{\prime}/b$
transforms Eq.~(\ref{rumenir})  into
\begin{equation}
F(R^{\prime},b) = 
\frac
{
400 000\,{\pi }^{5}{R^{\prime}}^{14}{{\rm e}^{-{\frac {20}{3}}\,{\frac {\pi \,{R^{\prime}
}^{3}}{{b}^{3}}}}}
}
{ 
243\,{b}^{15}
}.
\label{RUMENIRB}
\end{equation}

We now have $F(R)$  as given by Eq.~(\ref{rumenir}) and 
the fundamental integral (\ref{fundamental}), as derived in \cite{Zaninetti2011b},  is 
\begin{eqnarray}
f(r)=&2/3\,{\it K }\,\sqrt [6]{3}\sqrt [3]{10}\sqrt [3]{\pi }r
G^{4, 1}_{3, 5}\left({\frac {100}{9}}\,{\pi }^{2}{r}^{6}\,
\Big\vert\,^{5/6, 1/6, 1/2}_{7/3, 2/3, 1/3, 0, {\frac
{17}{6}}}\right)  \\ 
     & \quad 0 \leq r  \leq 1,
\nonumber
\label{FRMEIJER}
\end{eqnarray}
where ${K}$ is a constant,
\begin{equation}
{K} = 1.6485,
\end {equation}
and  the Meijer $G$-function  is defined
as  in  \cite{Meijer1936,Meijer1941,NIST2010}.
Details  on the real or complex parameters of the 
Meijer $G$-function are  given in the Appendix
of  \cite{Zaninetti2012e}.
Table~\ref{table_parameters} shows the average value, variance, mode, skewness, and kurtosis of the already derived  $f(r)$.
\begin{table}
 \caption[]{
The parameters   of  \lowercase{f(r)}, Eq.~(\ref{FRMEIJER}), relative to
the  $PVT$ case. }
 \label{table_parameters}
 \[
 \begin{array}{ll}
 \hline
Parameter & value   \\ \noalign{\smallskip}
 \hline
 \noalign{\smallskip}
Mean   &  0.4874   \\
\noalign{\smallskip}
\hline
Variance     & 0.02475  \\
\noalign{\smallskip}
\hline
Mode        & 0.553 \\
 \hline
Skewness        & -.5229  \\
 \hline
Kurtosis        &  -.1115  \\
 \hline
 \end{array}
 \]
 \end {table}
Asymptotic series are
\begin{eqnarray}
f(r) \sim
 2.7855\,  r    \\
when \quad \quad r
\rightarrow  0,
\nonumber
\end{eqnarray}
and
\begin{eqnarray}
f(r) \sim - 0.006\, \left( r-1 \right) + 0.136 \, \left( r-1
\right) ^{2}
\\
when \quad r  \rightarrow  1.
\nonumber
\end{eqnarray}
The  distribution function (DF) is
\begin{eqnarray}
DF(r) =   \nonumber  \\
 {\frac {1}{90}}\,{\it K}\,{3}^{5/6}{10}^{2/3} G^{4, 2}_{4,
6}\left({\frac {100}{9}}\,{\pi }^{2}{r}^{6}\, \Big\vert\,^{1, 7/6,
1/2, 5/6}_{8/3, 1, 2/3, 1/3, {\frac {19}{6}}, 0}\right) {\frac
{1}{\sqrt [3]{\pi }}}\\
 \quad 0 \leq r  \leq 1. 
\nonumber  
\end{eqnarray}
The already  defined PDF is  defined in the interval
$0 \leq r  \leq 1$.
In order to make a comparison with a normalized sample
which has a unitarian mean or an 
astronomical sample which  has the mean expressed in 
Mpc, a transformation of scale
should  be introduced.
The  change of variable is $r=x/b$ 
and the resulting PDF is 
\begin{eqnarray}
f(x,b)=  \nonumber  \\
\frac{2}{3}\,{\it K}\,\sqrt [6]{3}\sqrt [3]{10}\sqrt [3]{\pi }x
G^{4, 1}_{3, 5}\left({\frac {100}{9}}\,{\frac {{\pi
}^{2}{x}^{6}}{{b}^{6}}}\, \Big\vert\,^{5/6, 1/6, 1/2}_{7/3, 2/3, 1/3, 0,
{\frac {17}{6}}}\right) (\frac{1}{b})^{2}
\\
 \quad 0 \leq r  \leq b.
\nonumber
\label{FRMEIJERB}
\end{eqnarray}
As an example, Table \ref{table_parameters_twob} 
shows
the statistical parameters for two  different 
values of $b$. Skewness and kurtosis 
do not change with a transformation of scale.
\begin{table}
 \caption[]{
Parameters   of  \lowercase{f(x,b)}, 
Eq.~(\ref{FRMEIJERB}), relative to
the  $PVT$ case. }
 \label{table_parameters_twob}
 \[
 \begin{array}{lll}
 \hline
Parameter ~& b=2.051 & b=34         \\ \noalign{\smallskip}
 \hline
 \noalign{\smallskip}
Mean   &  1.         & 16.57  Mpc   \\
\noalign{\smallskip}
\hline
Variance     & 0.104 & 28.62  Mpc^2 \\
\noalign{\smallskip}
\hline
Mode        & 1.134  &  18.80 Mpc   \\
 \hline
 \end{array}
 \]
 \end {table}

We  briefly  recall  that  a  PDF $ f(x) $ is the first 
derivative of a distribution function (DF) 
$F(x) $ with respect to $x$.
When the DF is unknown  but the PDF known, we have
\begin{equation}
F(x) = \int_0^x f(x) dx.
\end{equation} 
The  survival function (SF)  $S(x)$  is
\begin{equation}
S(x) = 1 -F(x),
\label{defsurvival}
\end{equation}
and  represents the  probability that the variate 
takes a value  greater than $x$.
The  SF  with the
scaling parameter $b$  is
\begin{eqnarray}
SF(x,b)=  \nonumber  \\
1- 0.01831\,{3}^{5/6}{10}^{2/3}
G^{4, 2}_{4, 6}\left({\frac {100}{9}}\,{\frac {{x}^{6}{\pi
}^{2}}{{b}^{6}}}\, \Big\vert\,^{1, 7/6, 1/2, 5/6}_{8/3, 1, 2/3, 1/3, {\frac
{19}{6}}, 0}\right)
{\frac {1}{\sqrt [3]{\pi }}}
\\
 \quad 0 \leq r  \leq b.
\nonumber
\label{sfb}
\end{eqnarray}

\begin{figure*}
\begin{center}
\includegraphics[width=7cm]{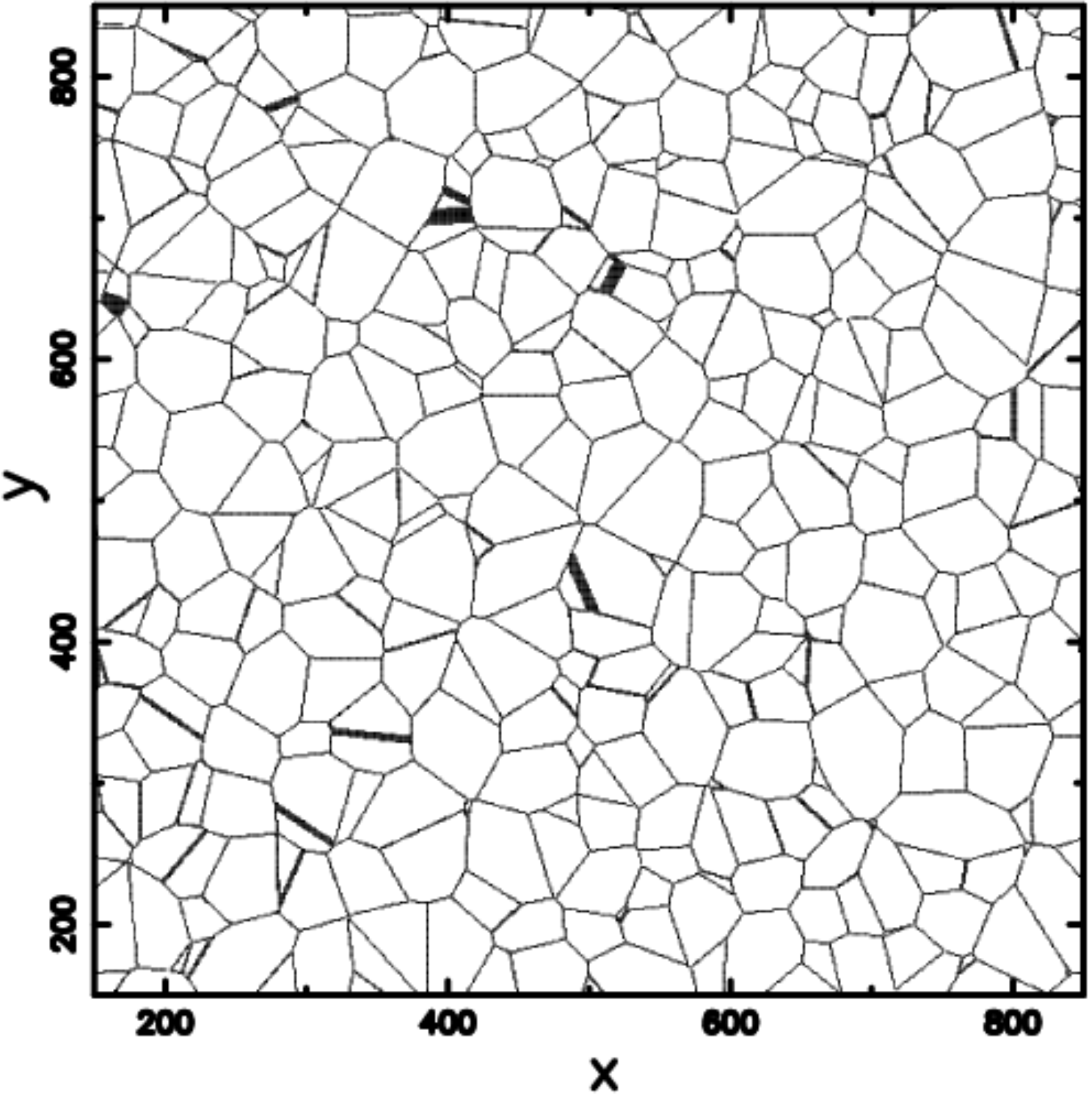}
\end {center}
\caption{
PVT diagram $V_p(2,3)$
when 789  2D cells generated by  15 000 3D seeds are considered.
}
\label{cut_middle_cut}%
    \end{figure*}

\begin{figure*}
\begin{center}
\includegraphics[width=7cm]{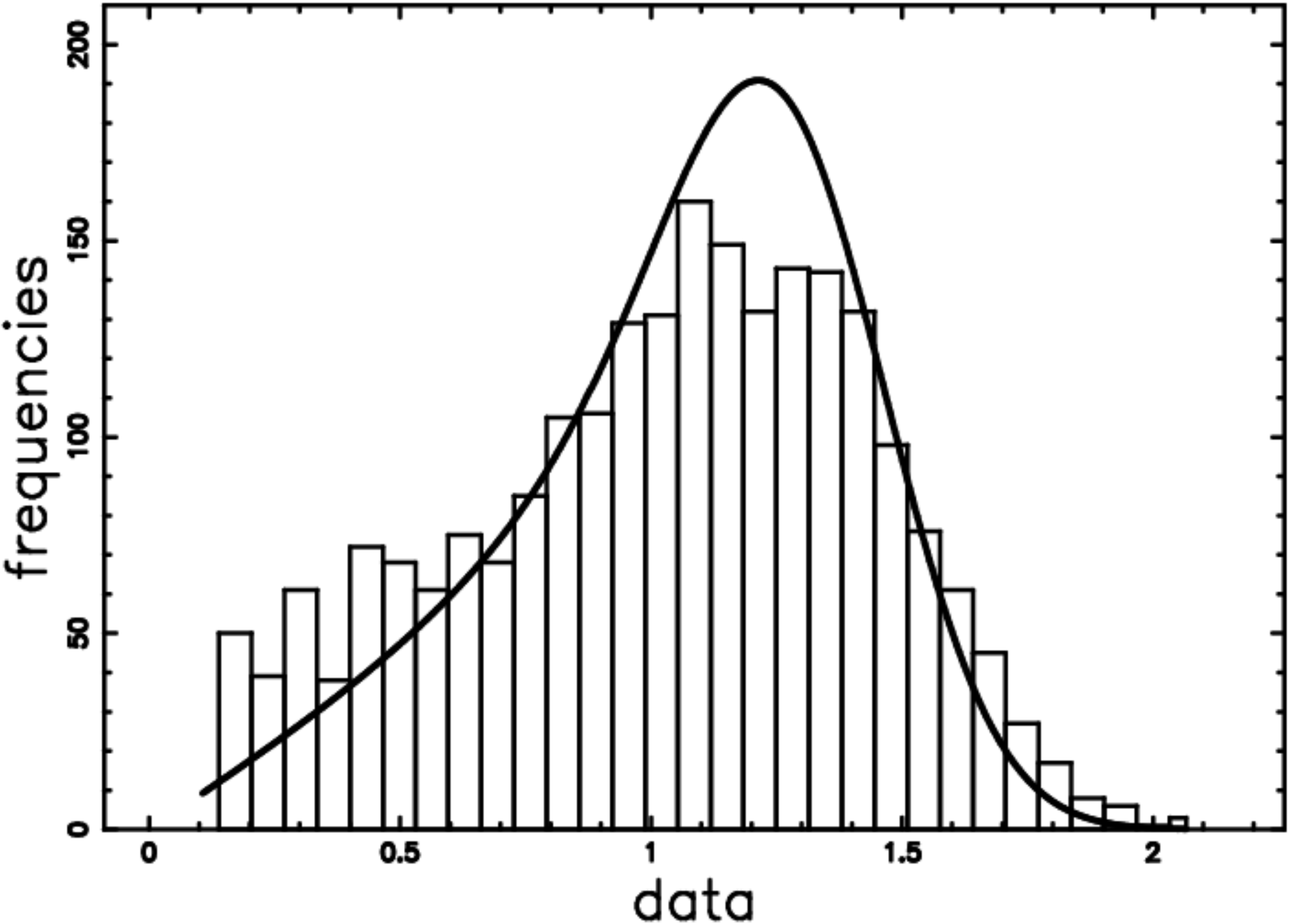}
\end {center}
\caption { Histogram (step-diagram) of PVT  $V_{\lowercase{p}}(2,3)$ when 789 2D
cells, generated by  15 000 3D seeds, are considered. The
superposition of the $f(x,b)$, Eq.~(\ref{FRMEIJERB}), is
displayed. }
\label{frequencies}%
    \end{figure*}

A first  application can be a comparison
between the  real distribution of radii  of
$V_p(2,3)$, see  Fig. \ref{cut_middle_cut},
and the already obtained
rescaled PDF $f(x,b)$.
The  fit  with the rescaled $f(x,b)$ is  shown in
Fig. ~\ref{frequencies} and
Table \ref{tablechisquare} shows  the $\chi^2$
of three different fitting functions.

 \begin{table}
 \caption[]
{
The values of \lowercase {$\chi^2$}  for
the cell normalized area-distribution
of $V_{\lowercase{p}}(2,3)$.
The number of 2D cells is  789,
the 3D seeds are   15 000 and
the number of bins in  the histogram is 30.
}
 \label{tablechisquare}
 \[
 \begin{array}{lll}
 \hline
PDF ~& parameters  & \chi^2  \\ \noalign{\smallskip}
 \hline
 \noalign{\smallskip}
 H (x ;c ) (Eq.~(\ref{kiang}))   &  c=5.8  & 250.8   \\
\noalign{\smallskip}
\hline
f(x;d) (Eq.~(\ref{rumeni}))       &  d=3.53 & 250.8  \\
\noalign{\smallskip}
\hline
f(x,b) (Eq.~(\ref{FRMEIJERB})) & b= 2.0514   &  127 \\
 \hline
 \end{array}
 \]
 \end {table}

The PDF  $f_A$ of the  areas of 
$V_p(2,3)$  can be obtained from 
$f(r)$  by means of the transformation, see \cite{Zaninetti2011b}, 
\begin{equation}
\label{farea}
f_A(A)=f(r) \left ( \left (\frac{A}{\pi}\right )^{1/2}  
\right )\frac{\pi^{-1/2}}{2} {A}^{-1/2},
\end{equation}
\noindent that is, 
\begin{equation}
\label{FAREAG}
f_A(A)=
 0.549\,\sqrt [6]{3}\sqrt [3]{10}
G^{4, 1}_{3, 5}\left({\frac {100}{9}}\,{\frac {{A}^{3}}{\pi }}\,
\Big\vert\,^{5/6, 1/6, 1/2}_{7/3, 2/3, 1/3, 0, {\frac {17}{6}}}\right)
{\pi }^{-2/3}. 
\end{equation}

The already derived  $f_A(A)$ has
average value,  variance, mode, skewness
and kurtosis
as shown in  Table~\ref{table_parameters_area}.
 \begin{table}
 \caption[]{
Parameters   of  {\lowercase{$f}_A(A)$}, Eq.~(\ref{FAREAG}),  relative to
the  $PVT$ case. }
 \label{table_parameters_area}
 \[
 \begin{array}{ll}
 \hline
Parameter ~& value   \\ \noalign{\smallskip}
 \hline
 \noalign{\smallskip}
Mean      &  0.824  \\
\noalign{\smallskip}
\hline
Variance  & 0.204  \\
\noalign{\smallskip}
\hline
Mode      & 0.858 \\
 \hline
Skewness        &0.278    \\
 \hline
Kurtosis        & -0.337  \\
 \hline
 \end{array}
 \]
 \end {table}

\begin{figure*}
\begin{center}
\includegraphics[width=7cm]{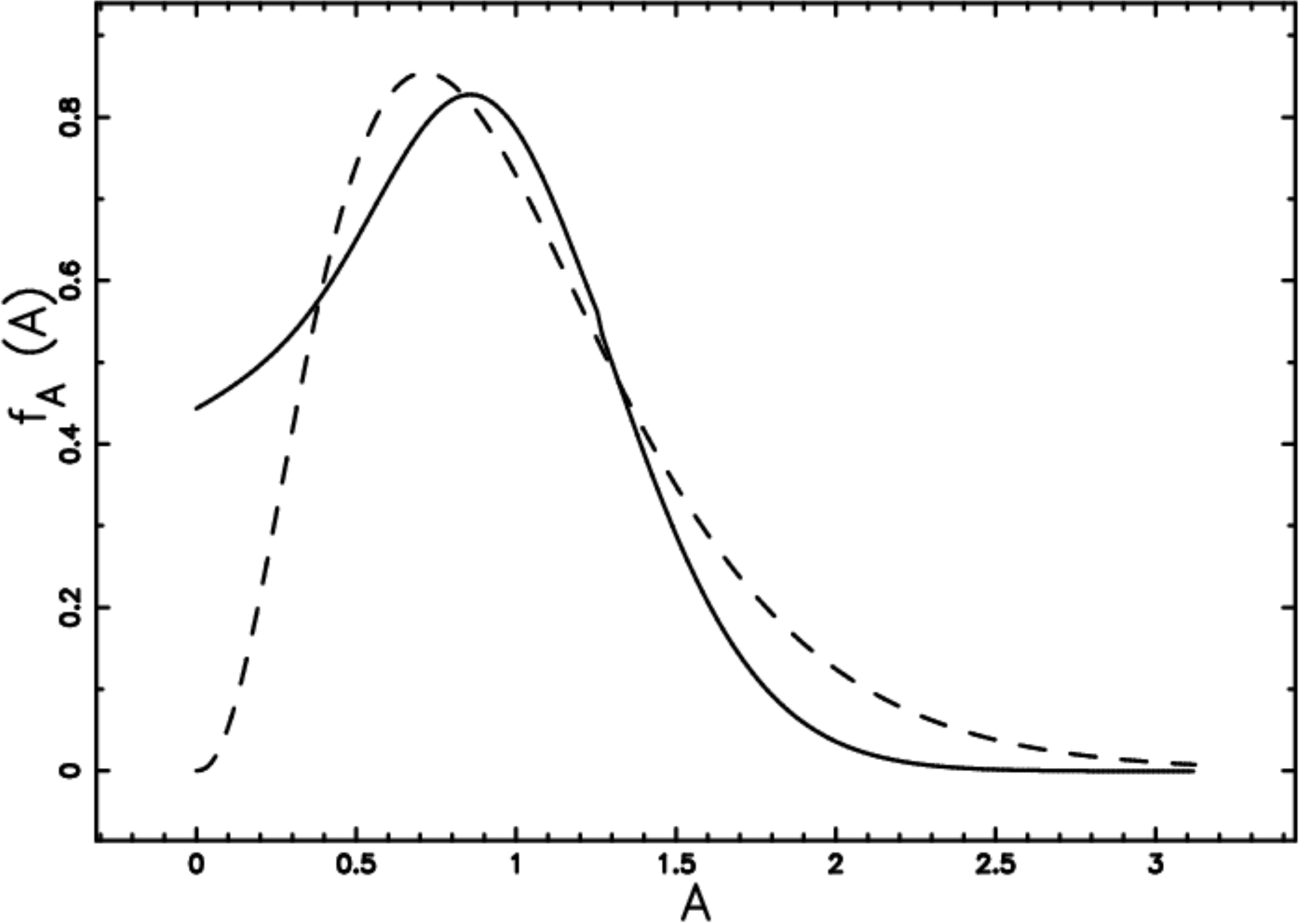}
\end {center}
\caption{
The PDF $f_A$, Eq.~(\ref {FAREAG}), as a function of $A$ (full line) and 
FN(x;d), Eq.~(\ref{rumeni}),  when d=2 
(dotted line).
}
\label{cut_due}
    \end{figure*}

Since, for $r$ close to $0$, $f(r) \sim r$ from Eq.~(\ref{FAREAG}) 
it follows that $f_A(0) \neq 0$, in particular
 $f_A(0)=0.443$   and 
Fig. ~\ref{cut_due} shows the graph of $f_A$.  

The previous figure shows that sections 
through 3-dimensional Voronoi 
tessellations are
not themselves 2-dimensional  Voronoi tessellations because
$f_A(0)$ has a finite value  rather 
than 0 as does the 2D area distribution;
this fact can be  considered a numerical demonstration 
in agreement with   
\cite{Weygaert1996}.
The distribution function $F_A$ is given by  
\begin{equation}
\label{dfa}
F_A  =
0.018\,{3}^{5/6}{10}^{2/3}
G^{4, 2}_{4, 6}\left({\frac {100}{9}}\,{\frac {{A}^{3}}{\pi }}\,
\Big\vert\,^{1, 7/6, 1/2, 5/6}_{8/3, 1, 2/3, 1/3, {\frac {19}{6}}, 0}\right)
{\frac {1}{\sqrt [3]{\pi }}}.
\end{equation}
Consider a three-dimensional Poisson Voronoi diagram  and suppose 
it intersects a randomly oriented plane $\gamma$:
the resulting cross sections are  polygons.

\begin{figure*}
\begin{center}
\includegraphics[width=7cm]{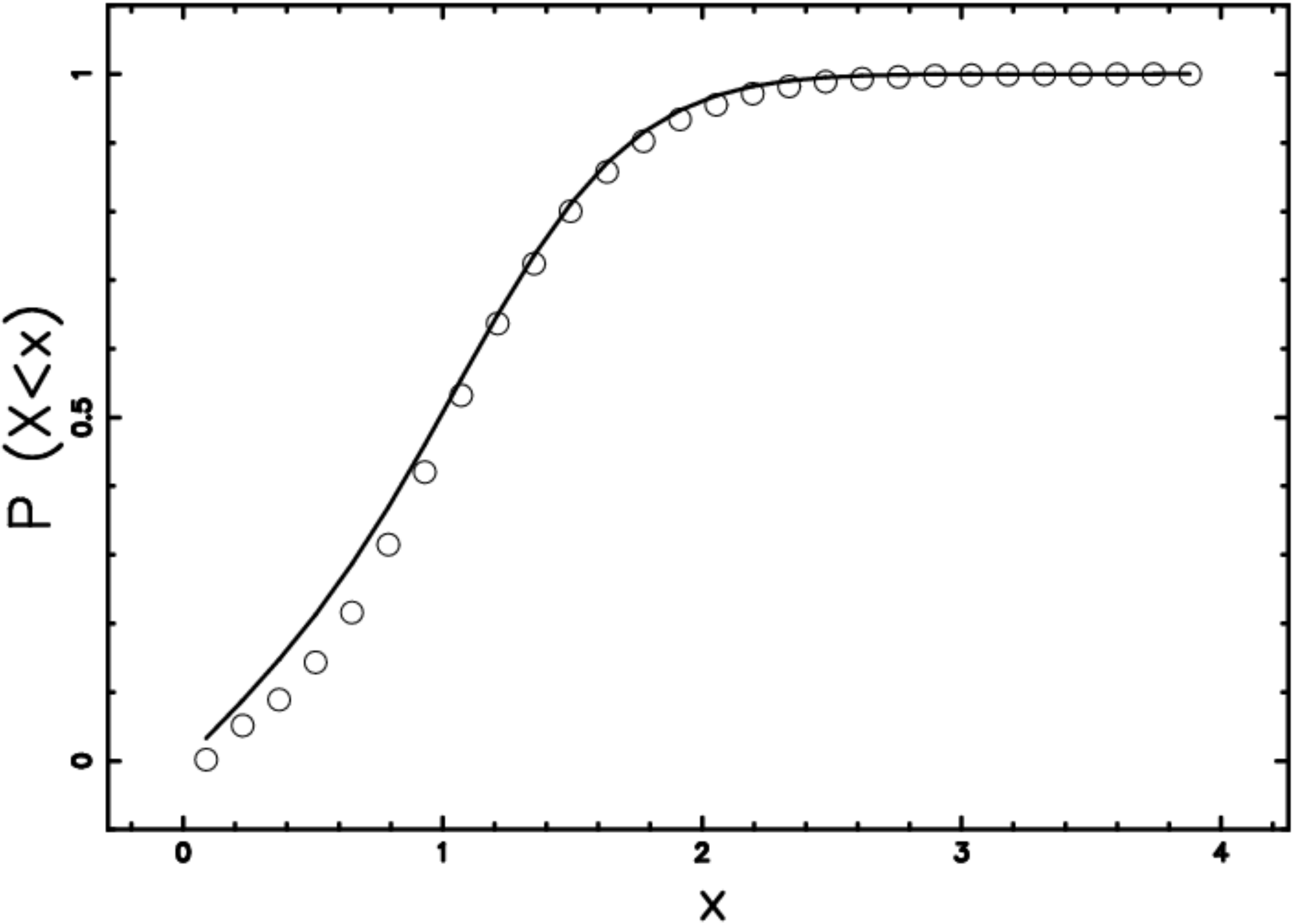}
\end {center}
\caption
{
Comparison between data (empty circles) 
and theoretical curve
(continuous line) 
of  the distribution of areas of the
planar cross  sections.
}
\label{area_xyz}
    \end{figure*}

A  comparison between $F_A$ and 
the area of the irregular polygons 
is shown in Fig. ~\ref{area_xyz}. 
In this case the number of seeds is $300 000$ 
and 
we  processed $100168$  irregular polygons
obtained by adding together  results of cuts by  
$41$ triples of  mutually perpendicular  planes.    
The maximum distance  between the two curves 
is $d_{max}=0.039$.  

As concerns the linear dimension, in our approximation 
the two-dimensional cells were considered 
circles and thus, for consistency, the radius $r$ 
of an irregular  polygon was defined as 
\begin{equation}
r=\left (\frac{A}{\pi} \right )^{1/2},
\label{empradius}
\end{equation}
that is, $r$  is the radius of a circle with the same area, $A$, as the polygon.
The assumption of  sphericity can be considered 
an axiom of the theory here
presented, but for a more realistic 
situation  the stereological 
results will be far more complex.
More details can be found  in  \cite{Zaninetti2012e}

\subsection{NPVT stereology}

\begin{figure}
\begin{center}
\includegraphics[width=7cm]{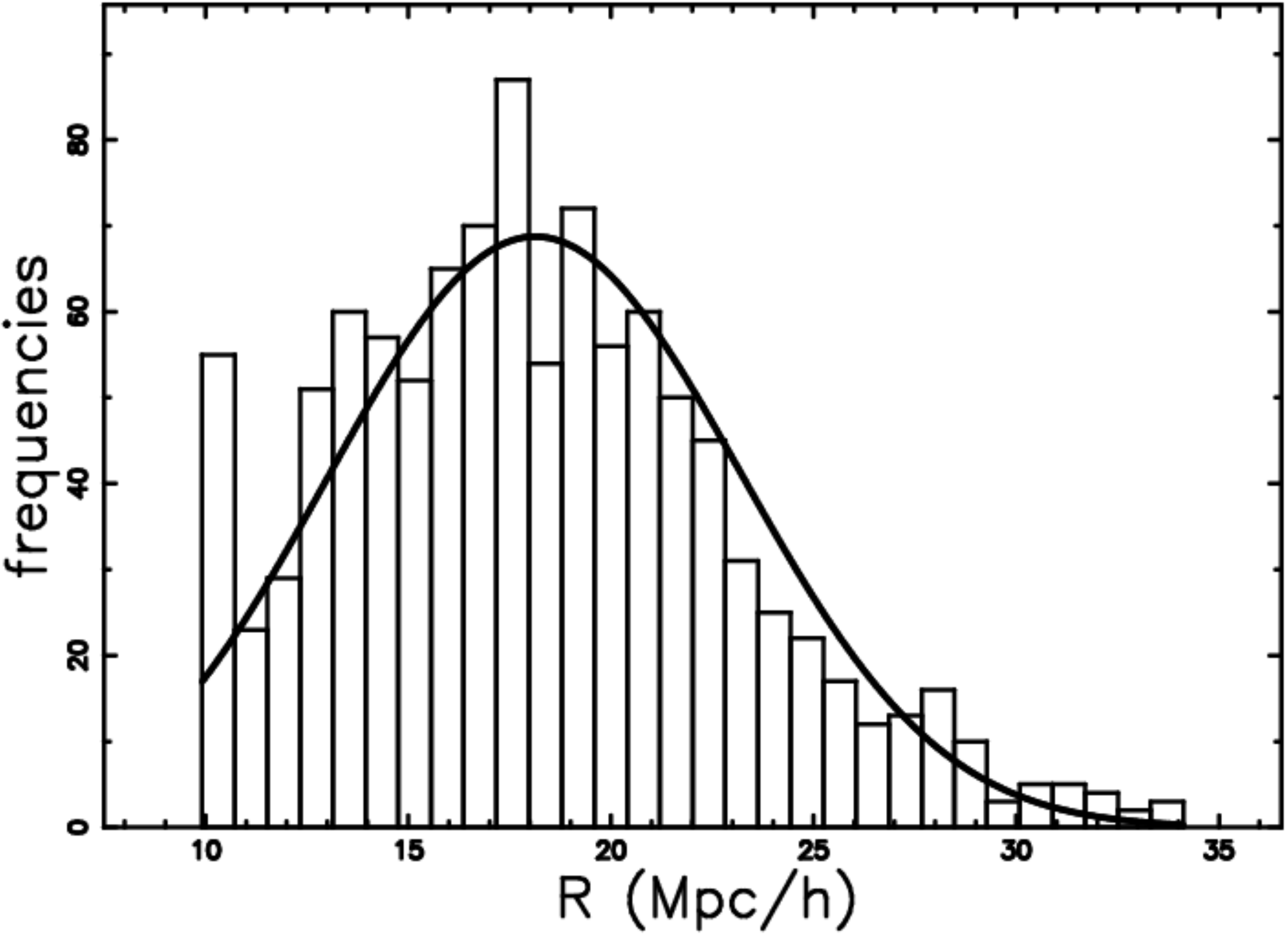}
\end {center}
\caption
{
Histogram (step-diagram)  of
the
effective radius in the  SDSS DR7
with a superposition of the
PDF  in radius of the NPVT spheres, $F_K(R,b,c)$,
 as represented by Eq.~(\ref{KIANGVARC}).
The number of bins is 30, $b$=31.33 Mpc,
and  $c=1.768$. 
}
\label{stat_kiangvarc_sdss}
    \end{figure}

An   example  of  NPVT is
represented  by a distribution  in volume
which  follows a Kiang function as given
by Eq.~(\ref{kiang}).
The case  of PVT volumes indicates $c=5$,
or  $c=6$, 
the so called Kiang conjecture:
we will  take  $c$ as a variable.
The resulting  distribution in radius  
once  the scaling parameter $b$ is introduced 
is  
\begin{equation}
F_K(R,b,c) = 
\frac
{
4\,c \left( 4/3\,{\frac {c\pi \,{R}^{3}}{{b}^{3}}} \right) ^{c-1}{
{\rm e}^{-4/3\,{\frac {c\pi \,{R}^{3}}{{b}^{3}}}}}\pi \,{R}^{2}
}
{ 
\Gamma  \left( c \right) {b}^{3}
}.
\label{KIANGVARC}
\end{equation}
The average radius  is
\begin{equation}
\overline{R} = 
\frac
{
\sqrt [3]{2}\sqrt [3]{3}b\Gamma  \left( 1/3+c \right)
}
{
2\,\sqrt [3]{c}\sqrt [3]{\pi }\Gamma  \left( c \right)
},
\end{equation}
and the variance is
\begin{equation}
\sigma^2(R) =
\frac   
{
-{3}^{2/3}{2}^{2/3}{b}^{2} \left( -\Gamma  \left( 2/3+c \right) 
\Gamma  \left( c \right) + \left( \Gamma  \left( 1/3+c \right) 
 \right) ^{2} \right) 
}
{
4\,{c}^{2/3}{\pi }^{2/3} \left( \Gamma  \left( c \right)  \right) ^{2}
}.
\end{equation}
The skewness  is
\begin{equation}
\gamma =
\frac
{
 \left( \Gamma  \left( c \right)  \right) ^{3}c-3\,\Gamma  \left( c
 \right) \Gamma  \left( 1/3+c \right) \Gamma  \left( 2/3+c \right) +2
\, \left( \Gamma  \left( 1/3+c \right)  \right) ^{3}
}
{
 \left( \Gamma  \left( 2/3+c \right) \Gamma  \left( c \right) -
 \left( \Gamma  \left( 1/3+c \right)  \right) ^{2} \right) ^{3/2}
},
\end{equation}
and the  
kurtosis is  given by a complicated analytical expression.
Fig. ~\ref{stat_kiangvarc_sdss} shows  a superposition 
of   the effective
radii of the voids in  SDSS DR7 with a 
superposition  of the curve of
the theoretical PDF in the radius, $F_K(R,b,c) $, 
as  represented by
Eq.~(\ref{KIANGVARC}).
Table~\ref{statkiangvarc} shows
the theoretical  statistical  parameters.

\begin{table}
 \caption[]
{
The statistical  parameters
of the theoretical radius of the  NPVT spheres 
as represented by Eq.~(\ref{KIANGVARC})
when ${\lowercase{b}}$=31.33 M\lowercase{pc}
and  ${\lowercase{c}}=1.768$.
}
 \label{statkiangvarc}
 \[
 \begin{array}{lc}
 \hline
 \hline
 \noalign{\smallskip}
parameter                  &   value                          \\ \noalign{\smallskip}
mean                       &  18.23h^{-1}~ Mpc   \\ \noalign{\smallskip}
variance                   &   23.31h^{-2}~ Mpc^2 \\ \noalign{\smallskip}
standard~ deviation        &   4.82h^{-1} ~ Mpc   \\ \noalign{\smallskip}
skewness                   &   0.072         \\ \noalign{\smallskip}
kurtosis                   &  -0.162         \\ \noalign{\smallskip}
 \hline
 \end{array}
 \]
 \end {table}
The result of the integration of the fundamental Eq.~(\ref{fundamental}) inserting $c$=2 gives 
the following PDF for the radius of
the cuts
\begin{eqnarray}
f(r)_{NPVTK} = 
 3.4148\,\sqrt [6]{3}\sqrt [3]{\pi }r{2}^{2/3}
G^{4, 0}_{2, 4}\left({\frac {16}{9}}\,{\pi }^{2}{r}^{6}\, \Big\vert\,^{1/6,
1/2}_{4/3, 2/3, 1/3, 0}\right)
 \\
 \quad 0 \leq r  \leq 1. \label{FRMEIJERNPVTK}
\nonumber
\end{eqnarray}
The statistics of NPVT cuts with $c$=2  are shown in
Table~\ref{table_parametersnpvtk}.
 \begin{table}
 \caption[]{
NPVT parameters   of 
\lowercase{$f(r)}_{NPVTK}$, Eq.~(\ref{FRMEIJERNPVTK}). }
 \label{table_parametersnpvtk}
 \[
 \begin{array}{ll}
 \hline
Parameter ~& value   \\ \noalign{\smallskip}
 \hline
 \noalign{\smallskip}
Mean   &  0.488   \\
\noalign{\smallskip} \hline
Variance     & 0.0323 \\
\noalign{\smallskip} \hline
Mode        & 0.517 \\
 \hline
Skewness        & -.114  \\
 \hline
Kurtosis        &  2.614  \\
 \hline
 \end{array}
 \]
 \end {table}

\begin{figure}
\begin{center}
\includegraphics[width=7cm]{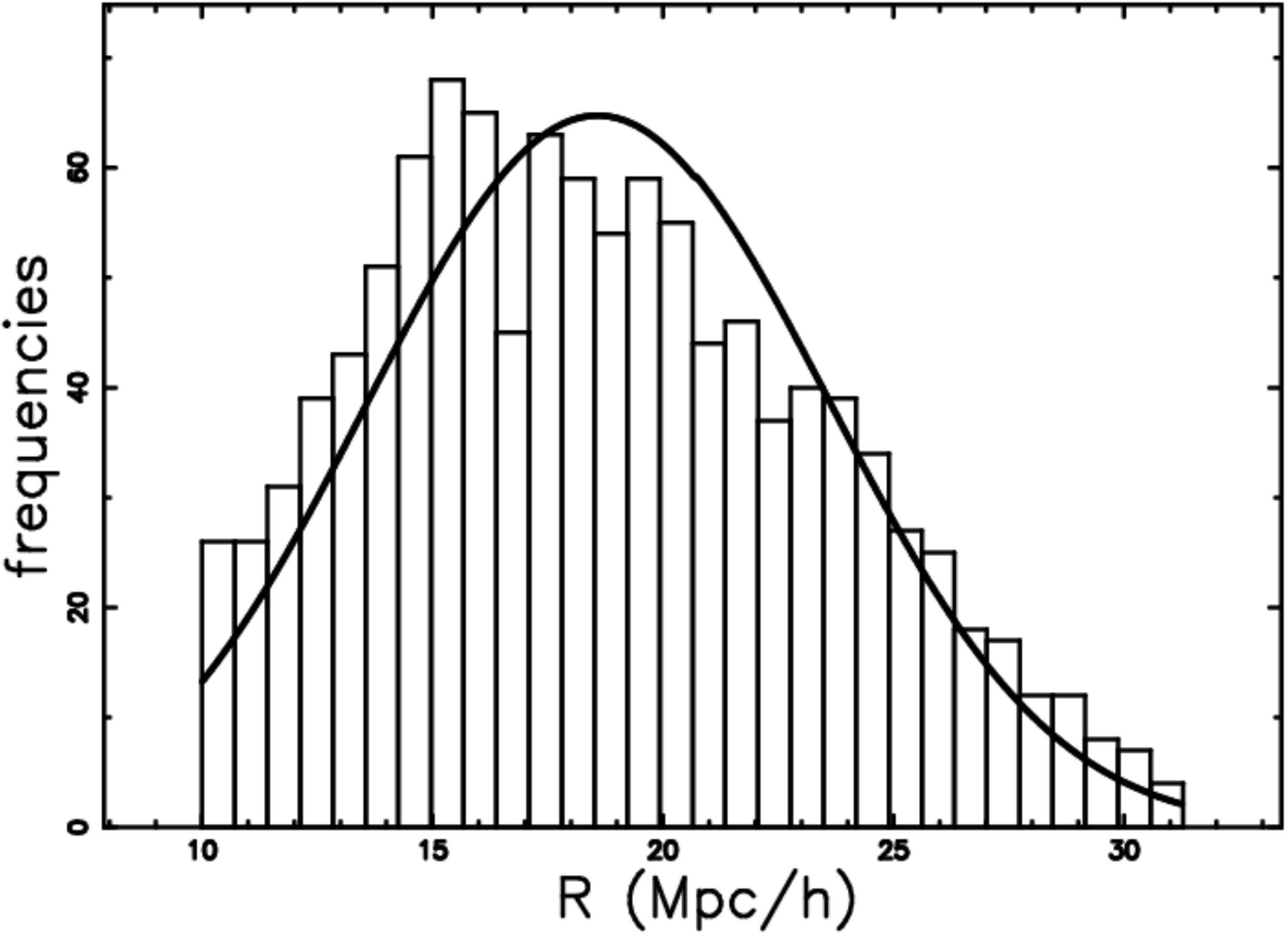}
\end {center}
\caption
{
Histogram (step-diagram)  of
the
simulated  effective radius of  SDSS DR7
with a superposition of the
PDF  in radius of the PVT spheres
as represented by Eq.~(\ref{KIANGVARC}).
The artificial sample has a minimum value   
of  10/h  Mpc,
the number of bins is 30, $b$= 31.5/h Mpc, and $c=1.3$.
}
\label{stat_cvar_gene}
    \end{figure}

On introducing the scaling parameter $b$, the  
PDF which describes
the radius of the cut becomes
\begin{eqnarray}
f(x,b)_{NPVTK} = 
3.4148\,\sqrt [6]{3}\sqrt [3]{\pi }x{2}^{2/3}
G^{4, 0}_{2, 4}\left({\frac {16}{9}}\,{\frac {{\pi }^{2}{x}^{6}}{{b}^{6}}}\,
\Big\vert\,^{1/6, 1/2}_{4/3, 2/3, 1/3, 0}\right)
{b}^{-2}
 \\
 \quad 0 \leq r  \leq b. \label{frmeijerbnpvtk}
\nonumber
\end{eqnarray}
The  SF  of the {\it second} 
NPVT case, $SF_{NPVTK}$, with the
scaling parameter $b$, is
\begin{eqnarray}
SF(x,b)_{NPVTK}=   
1- 0.2845\,{3}^{5/6}\sqrt [3]{2}
G^{4, 1}_{3, 5}\left({\frac {16}{9}}\,{\frac {{\pi }^{2}{x}^{6}}{{b}^{6}}}\,
\Big\vert\,^{1, 1/2, 5/6}_{5/3, 1, 2/3, 1/3, 0}\right)
{\frac {1}{\sqrt [3]{\pi }}}
\\
 \quad 0 \leq r  \leq b.
 \nonumber
\label{sfbnpvtk}
\end{eqnarray}
A careful  exploration of the distribution  
in effective radius
 of
SDSS DR7  reveals that the detected  voids have  radius
$\geq $ 10/h Mpc.
This observational fact demands the  
generation of random numbers
in  the distribution in radii of the 3D cells as
given by Eq.~(\ref{KIANGVARC})   with a minimal value  of
10/h  Mpc.
The  artificial sample is generated
through a numerical computation
of the inverse function~\cite{Brandt1998} and  displayed 
in Fig. \ref{stat_cvar_gene};  the sample's  statistics   
are shown
in Table~\ref{statsimulatedcvar}.

\begin{table}
 \caption[]
{
The statistical  parameters
of the artificially generated   radius 
with a lower bound
of   10 /\lowercase{h}  M\lowercase{pc}, \lowercase{$c=1.3$} 
and \lowercase{$b$} =31.5/\lowercase{h}  M\lowercase{pc}.
}
 \label{statsimulatedcvar}
 \[
 \begin{array}{lc}
 \hline
 \hline
 \noalign{\smallskip}
parameter                  &   value                          \\ \noalign{\smallskip}
mean                       & 18.69  h^{-1}~ Mpc   \\ \noalign{\smallskip}
variance                   & 22.74  h^{-2}~ Mpc^2 \\ \noalign{\smallskip}
standard~ deviation        & 4.76 h^{-1} ~ Mpc   \\ \noalign{\smallskip}
skewness                   & 0.33           \\ \noalign{\smallskip}
kurtosis                   & -0.623               \\ \noalign{\smallskip}
maximum ~value             & 31.27  h^{-1}~ Mpc   \\ \noalign{\smallskip}
minimum ~value             & 10    h^{-1}~   Mpc   \\ \noalign{\smallskip} \hline
 \hline
 \end{array}
 \]
 \end {table}
More details can be found  in  \cite{Zaninetti2012e}.

\section{The cellular structure of the Universe }
\label{sec_cellular}

\begin{figure}\begin{center}
\includegraphics[width=7cm]{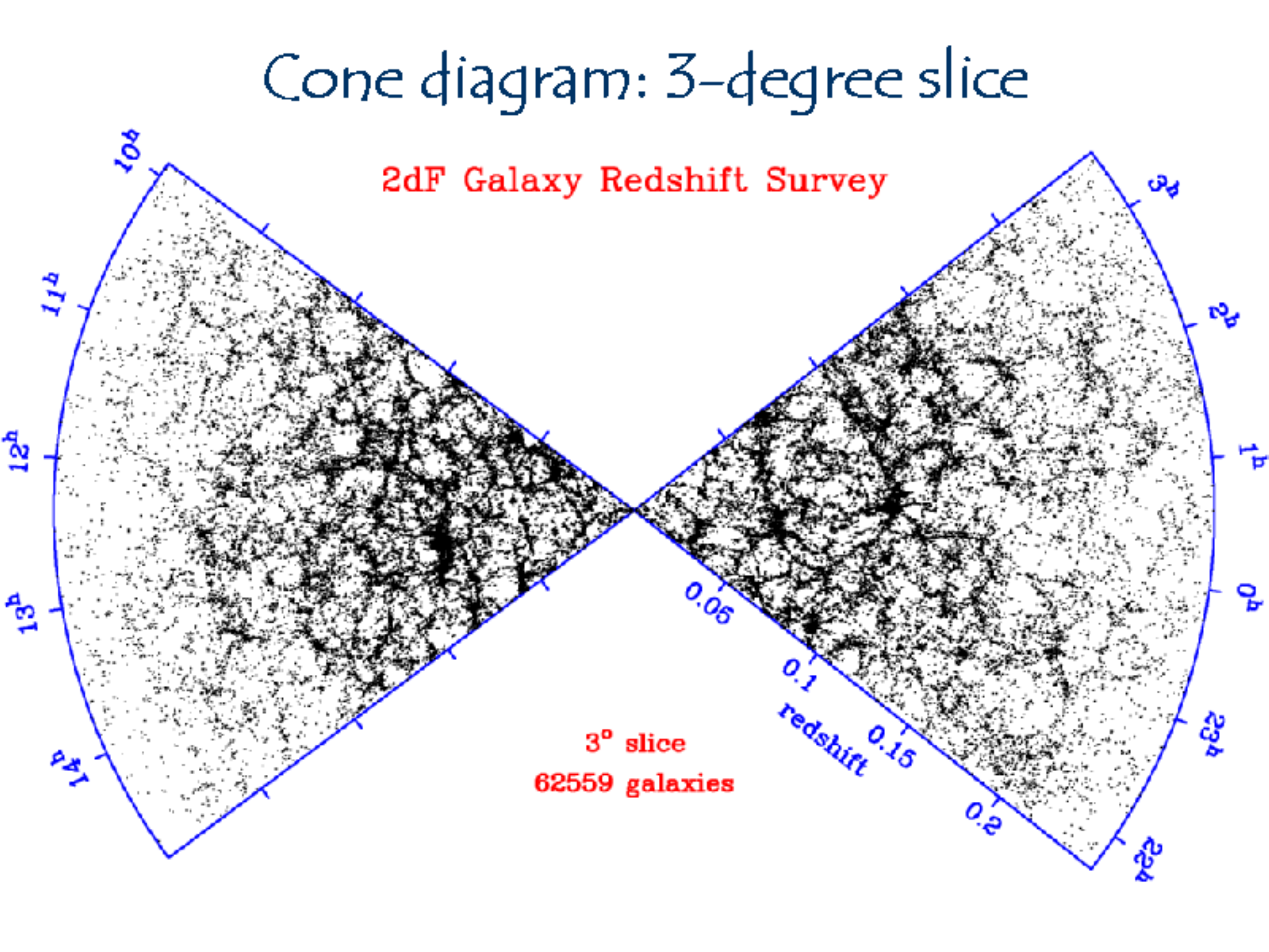}
\end{center}\caption{Slice  of   $75^{\circ} \times 3^{\circ}$  
in the 2dFGRS.
This plot contains  62559  galaxies and belongs to the   
 2dFGRS Image Gallery
available at  the  web site: http://msowww.anu.edu.au/2dFGRS/.
}
          \label{2df_cone}%
    \end{figure}

From a simplified point of view the galaxies belonging 
to a given catalog are characterized by the  astronomical
coordinates, 
the redshift and the apparent magnitude.
Starting from the  second CFA2 redshift   Survey,
the catalogs were organized in slices of a given 
opening angle, $3^{\circ}$ or $6^{\circ}$,
and a given  angular
extension, for example $130^{\circ}$.
When plotted in polar coordinates of $c_lz$ 
the spatial distribution of galaxies
is not random but distributed on filaments.
Particular attention should be paid to the fact that 
the astronomical slices are not a plane which intersects 
a Voronoi Network. 
In order to quantify this effect we introduce a confusion
distance, $DV_c$, as the distance  after which 
the half altitude of the slices equalizes the 
observed average diameter  $\overline{DV^{obs}}$
\begin{equation}
DV_c \tan (\alpha) = \frac{1}{2} \overline{DV^{obs}}
\quad  ,
\end{equation}
where $\alpha$ is the opening angle  of the slice 
and  $\overline{DV^{obs}}$ the averaged diameter of
voids.
In the case of  2dFGRS   $\alpha=3^{\circ}$ 
and therefore 
$DV_c=2.57~10^{4} \frac {Km}{sec} $
when   $\overline{DV^{obs}}=2700\frac {Km}{sec} $.
For values of $c_lz$ greater than $DV_c$ 
the voids in the distribution
of galaxies are dominated by the confusion.
For values of $c_lz$ lower  than $DV_c$
the filaments of galaxies  can be considered  
the intersection
between a plane and the faces of the Voronoi Polyhedrons.
A measure of the portion of the sky covered by 
a catalog of galaxies is the area covered 
by a  unitarian sphere  which  is 
$4\pi$ steradians or $\frac{129600}{\pi}$ square degrees.
In  the case of 2dFGRS  the covered area 
of two slices  of $75^{\circ}$  long and  $3^{\circ}$
wide,
as in Fig. \ref{2df_cone},
is  $\frac{1414}{\pi}$ square degrees or 0.13 $sr$.
In the case of RC3 the covered area 
it is $4\pi$ steradians with the 
exclusion of the {\it Zone of Avoidance}, see 
Fig. \ref{rc3_all}.

In the following we will simulate 
the 2dFGRS, a catalog that occupies a small area of the sky,
the  RC3  a catalog that occupies all the sky
and  the CFA2 catalog.

In the case of 3C3 we  demonstrate  
how it is possible to 
simulate the {\it Zone of Avoidance}
in the theoretical simulation.
The paragraph ends with a discussion 
on the Eridanus super-void also known as "Cold Spot".

\subsection{The 2dFGRS }

\label {cat2dFGRS}

\begin{figure}
\begin{center}
\includegraphics[width=7cm]{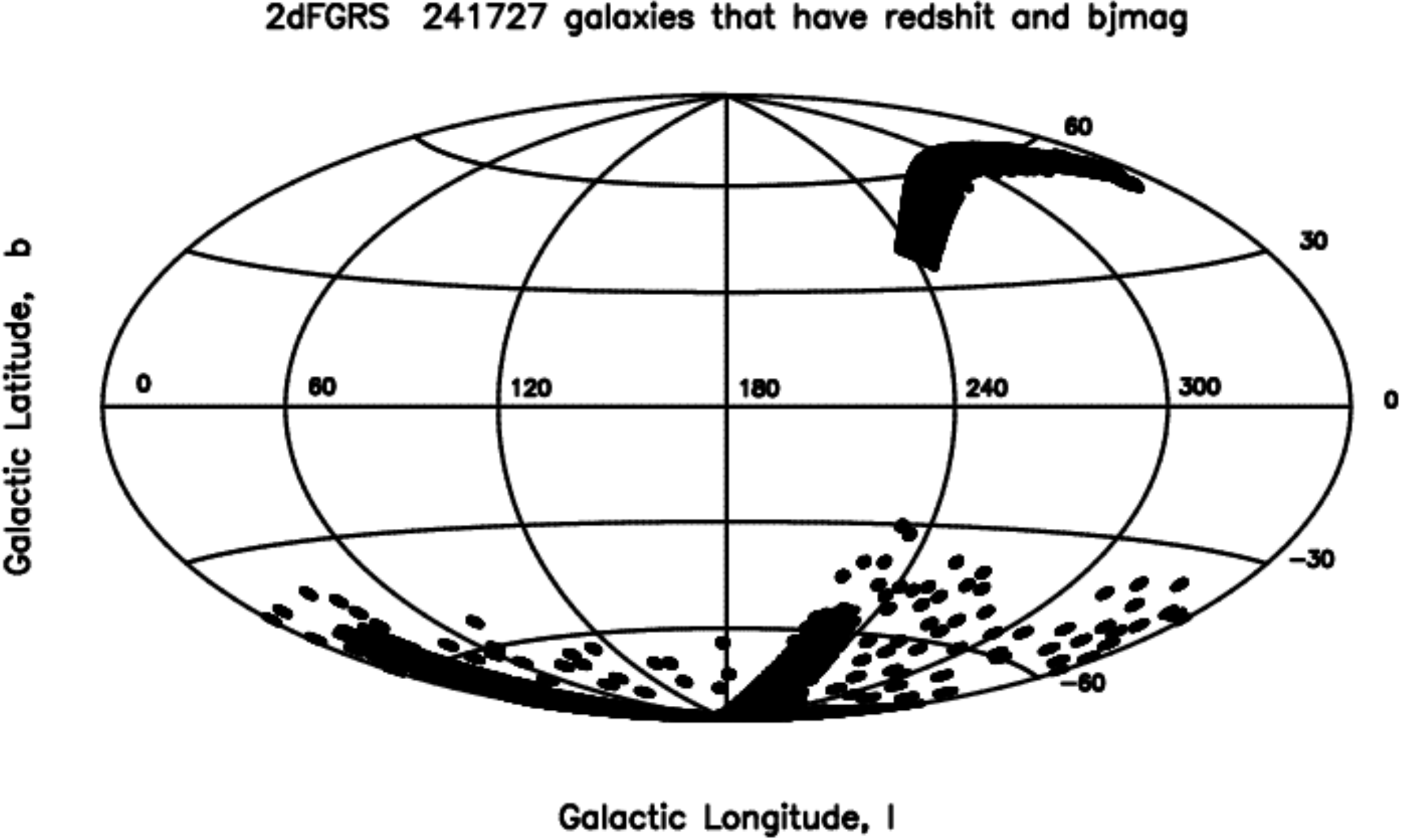}
\end{center}\caption{Hammer-Aitoff  projection  in galactic coordinates 
  of 230540 galaxies  
 in the 2dfGRS   which have bJmag  and redshift $<0.3$.
}
          \label{2dftuttegal}%
    \end{figure}

\begin{figure}
\begin{center}
\includegraphics[width=7cm]{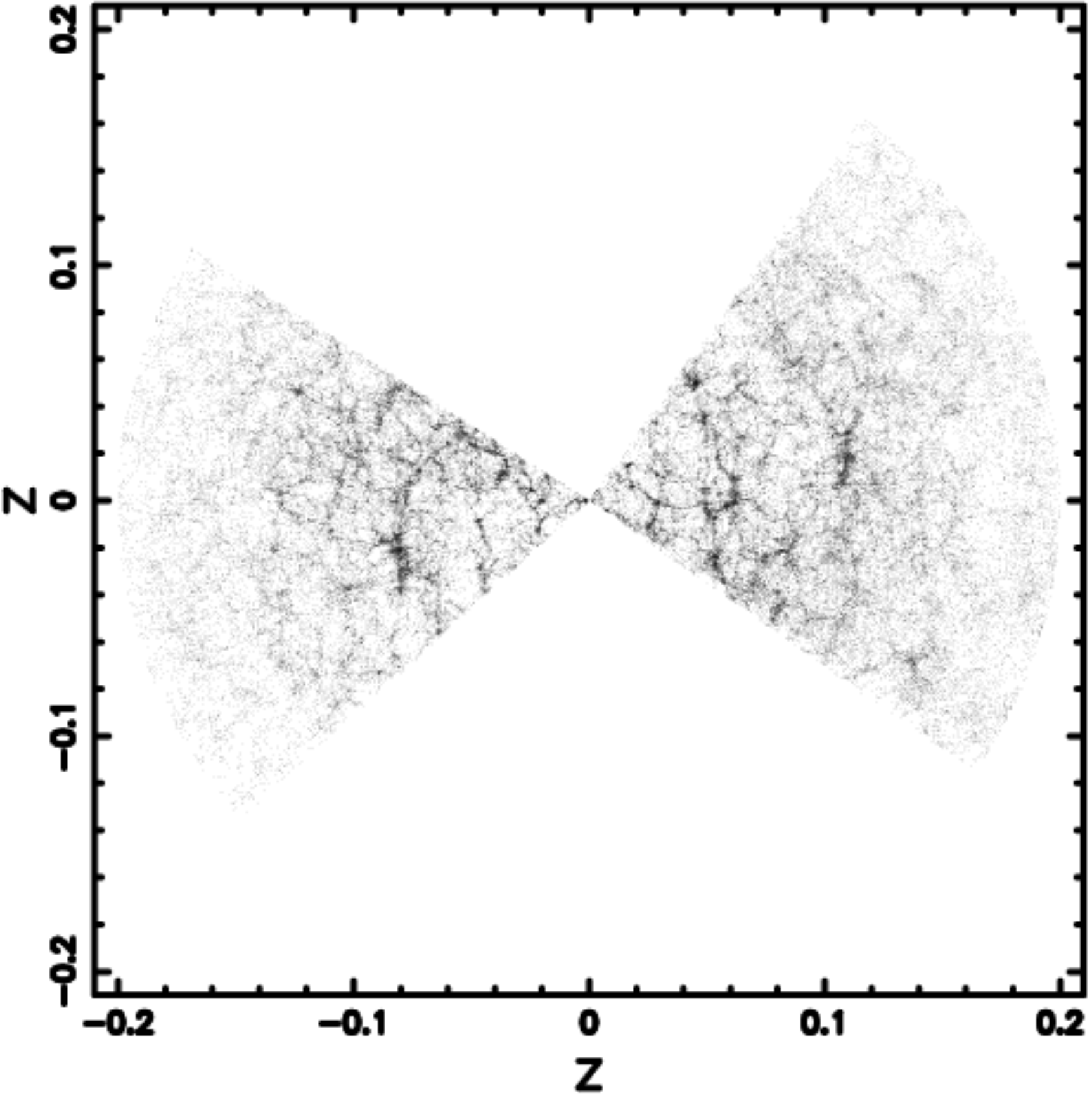}
\end{center}\caption{Cone-diagram  of all the  galaxies  
in  the 2dFGRS.
This plot contains  203249  galaxies.  
}
          \label{2df_all}%
    \end{figure}

Fig. \ref{2dftuttegal} shows the galaxies 
of  the 2dFGRS with $z<0.3$ in galactic coordinates
and the  two strips in the  2dFGRS are 
shown in Fig. \ref{2df_all}.

Fig. \ref{2df_cone}  conversely reports  
the 2dfGRS catalog   when a slice of
$75^{\circ} \times 3^{\circ}$
is taken into account.
This slice represents 
the object to simulate.

The previous observational slice can  be simulated
by adopting the Voronoi  network reported 
in Fig. \ref{voro_fetta_tutte}.

\begin{figure}\begin{center}
\includegraphics[width=7cm]{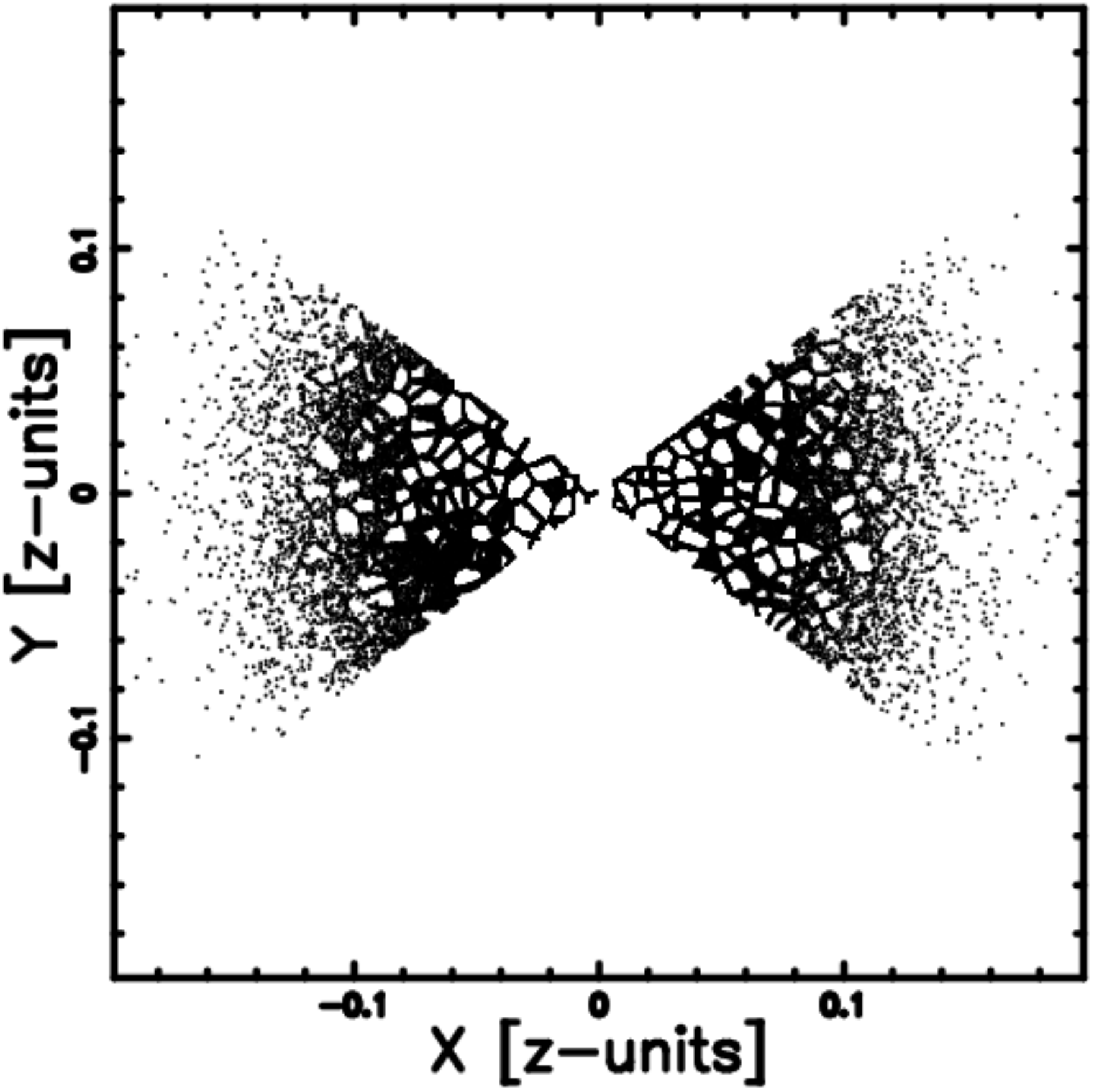}
\end{center}\caption{
Polar plot
of the  pixels  belonging to a
slice   $75^{\circ}$~long  and $3^{\circ} $ wide.
This plot contains  62563
galaxies,
the maximum in the frequencies of theoretical
galaxies is at  $z=0.043$.
In this plot $\mathcal{M_{\sun}}$ = 5.33  and $h$=0.623.
}
          \label{voro_2df_cones}%
    \end{figure}

The  distribution of the galaxies as  
given by the Voronoi Diagrams
is reported in Fig. \ref{voro_2df_cones} 
where  all the galaxies  are considered.
In this case the  galaxies are extracted according 
to the integral of the Schechter  function  in flux  
(formula~(\ref{integrale_schechter})  with parameters
as  in  Table~\ref{parameters}). 
Table~\ref{zvalori} reports the basic data of the
astronomical and  simulated data of the 
$75^{\circ} \times 3^{\circ}$ slice.

\begin{table}
\caption { 
Real and  Simulated  data 
of the slice   $75^{\circ}$~long  and $3^{\circ}$ wide .
     }
 \label{zvalori}
 \[
 \begin{array}{ccc}
 \hline
 \hline
 \noalign{\smallskip}
~~   &   2dFGRS   & simulation~         \\
 \noalign{\smallskip}
 \hline
 \noalign{\smallskip}
elements    &  62559  &  62563  \\
z_{min}     &  0.001  &  0.011  \\
z_{pos-max} &  0.029  &  0.042  \\
z_{ave}     &  0.051  &  0.058  \\
z_{max}     &  0.2    &  0.2    \\
\noalign{\smallskip}
\noalign{\smallskip}
 \hline
 \hline
 \end{array}
 \]
 \end {table}

\begin{figure}\begin{center}
\includegraphics[width=7cm]{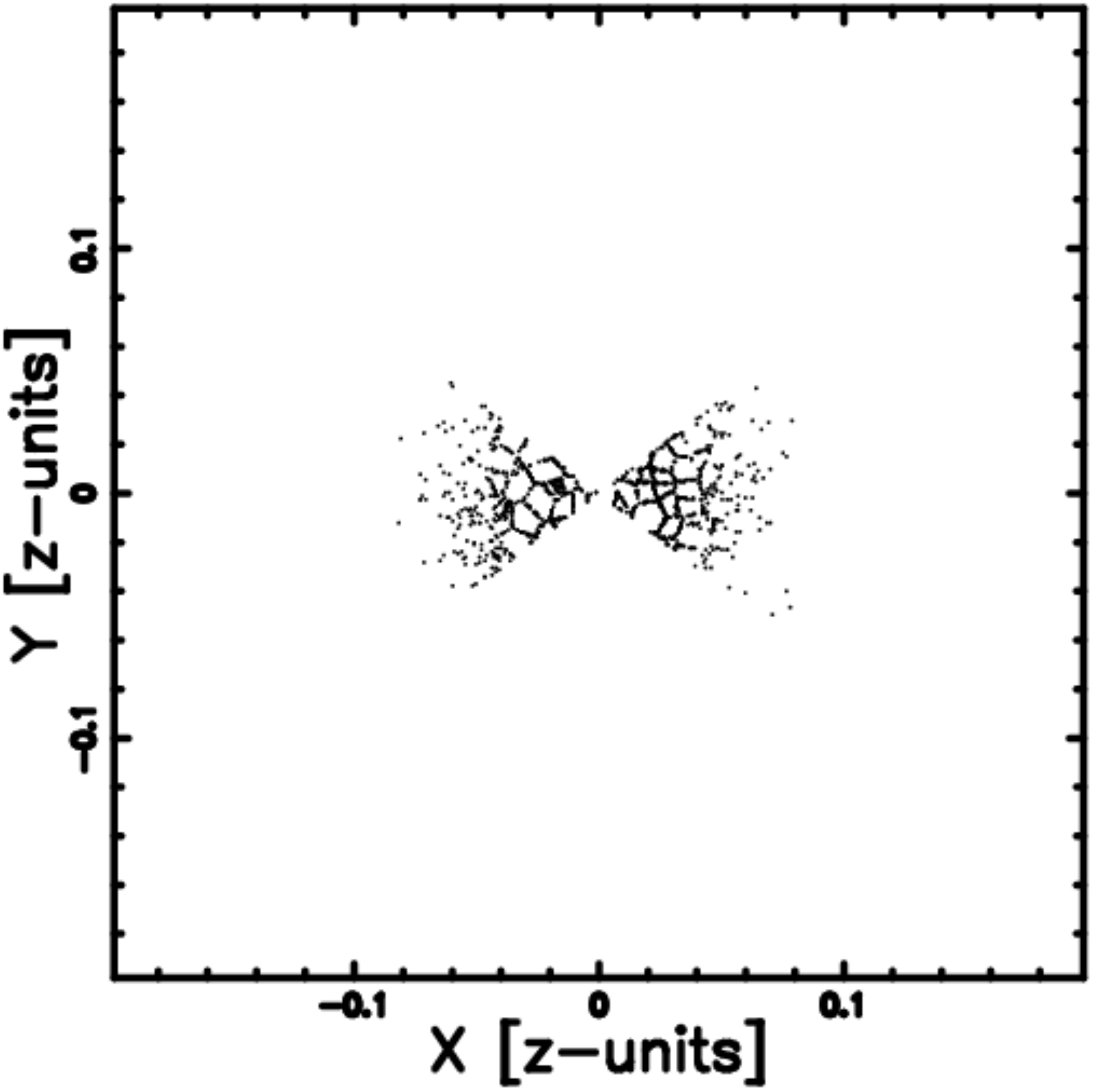}
\end{center}\caption{
Polar plot
of the  pixels  belonging to a
slice   $75^{\circ}$~long  and $3^{\circ}$
wide.
Galaxies  with magnitude 
$ 15.02  \leq  bJmag \leq 15.31 $
or 
$ 46767  \frac {L_{\sun}}{Mpc^2} \leq  f 
  61063 \frac {L_{\sun}}{Mpc^2}$.
The maximum in the frequencies of theoretical
galaxies is at  $z=0.029$, 
$N_{SC}$=2186  and    $N_C$=62559.
In this plot $\mathcal{M_{\sun}}$ = 5.33  and $h$=0.623.
}
          \label{voro2dfconessel}%
    \end{figure}

When conversely a given interval in flux  (magnitudes) 
characterized by  $f_{min}$ and  $f_{max}$
is considered the number of galaxies, $N_{SC}$,
of a $3^{\circ}$ slice
can be found 
with the following formula 
\begin{equation}
N_{SC}  = N_C
\frac  
{ 
\int_{f_{min}} ^{f_{max}}
4 \pi  \bigl ( \frac {c_l}{H_0} \bigr )^5    z^4 \Phi (\frac{z^2}{z_{crit}^2})
df} 
{ 
\int_{f_{min,C}} ^{f_{max,C}}
4 \pi  \bigl ( \frac {c_l}{H_0} \bigr )^5    z^4 \Phi (\frac{z^2}{z_{crit}^2})
df} 
\quad ,
\end{equation}
where   $f_{min,C}$ 
and     $f_{max,C}$  represent
the minimum and maximum flux of the considered catalog
and  $N_C$ all the galaxies of the considered catalog;
a typical  example is reported in 
Fig. \ref{voro2dfconessel}.

\subsection{The Third Reference Catalog of Bright Galaxies}

\begin{figure}\begin{center}
\includegraphics[width=7cm]{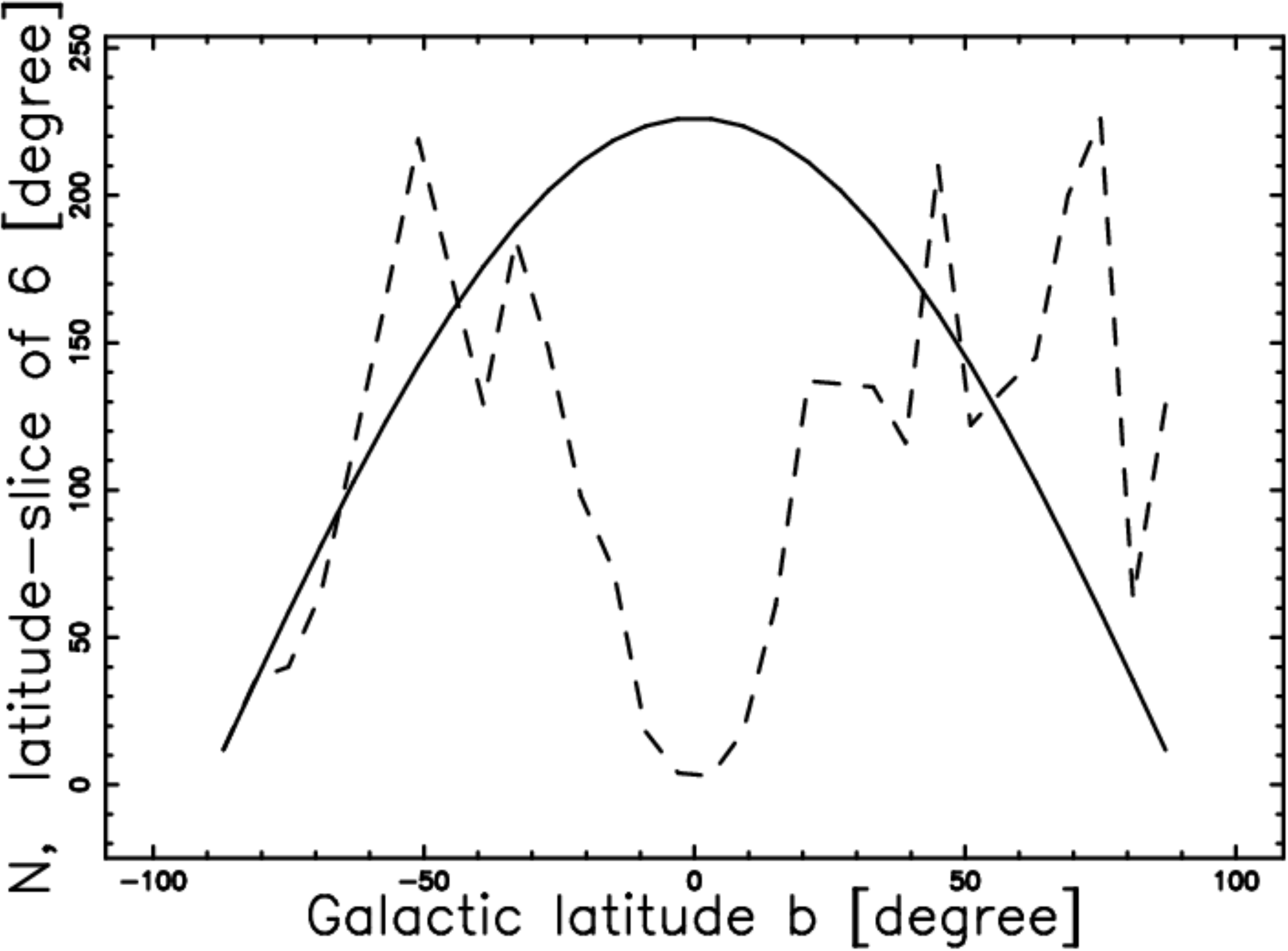}
\end{center}\caption{
The galaxies  
in the  RC3  which have BT and redshift
are organized in  frequencies versus  galactic latitude  $b$
(dashed line).
The theoretical fit  represents  
$N_i$ ( full line).
}
          \label{teta}%
    \end{figure}

We now test the concept  of an   isotropic universe.
This can be by  done by  plotting  the  number of galaxies  
comprised in
a slice of $360^{\circ}$ in galactic longitude  versus  
a variable number $\Delta b$ in galactic latitude, 
for example $6^{\circ}$.
The number of galaxies in the  RC3 versus galactic latitude is 
plotted  in Fig. \ref{teta}.

The solid angle $d\Omega$ in spherical  coordinates 
(r,$\theta , \phi $) is
\begin{equation}
d \Omega = \sin (\theta) d\theta d \phi \quad .
\end {equation}
In a slice of     $360^{\circ} \times \Delta b $  
the  amount of solid angle,   $\Delta \Omega$,   is 
\begin{eqnarray}
\Delta \Omega
 =  2\pi \bigl (
 (\cos (90^{\circ}) - \cos ( b + \Delta b  ) )     
-(\cos (90^{\circ}) - \cos ( b   ) )
         \bigr )   
\,\mathrm {steradians} 
\label{solid}
\quad .
\end {eqnarray}
The  approximate number  of galaxies  in each  slice 
can be found through the following  approximation.
Firstly, we find the largest value  of the  frequencies 
of galaxies, $F_i$,
versus  $b$, $max(F_i)$
 where the index $i$ denotes  a class in
latitude.
We therefore find  the largest value  of  $\Delta \Omega_i$,
$max(\Delta \Omega_i)$.
The introduction of  the multiplicative  factor $M$
\begin{equation}
M  =  \frac {max(F_i)}{ max(\Delta \Omega_i)}
\quad  ,
\end{equation}
obtains  the following theoretical evaluation of the 
number of galaxies $N_i$ as a  function of the latitude,
\begin{equation}
N_i =  M \times \Delta \Omega_i
\quad .
\label{nteorico}
\end{equation}
This number, $N_i$,  as a function of $b$ is plotted 
in Fig. \ref{teta}.

\begin{figure}\begin{center}
\includegraphics[width=7cm]{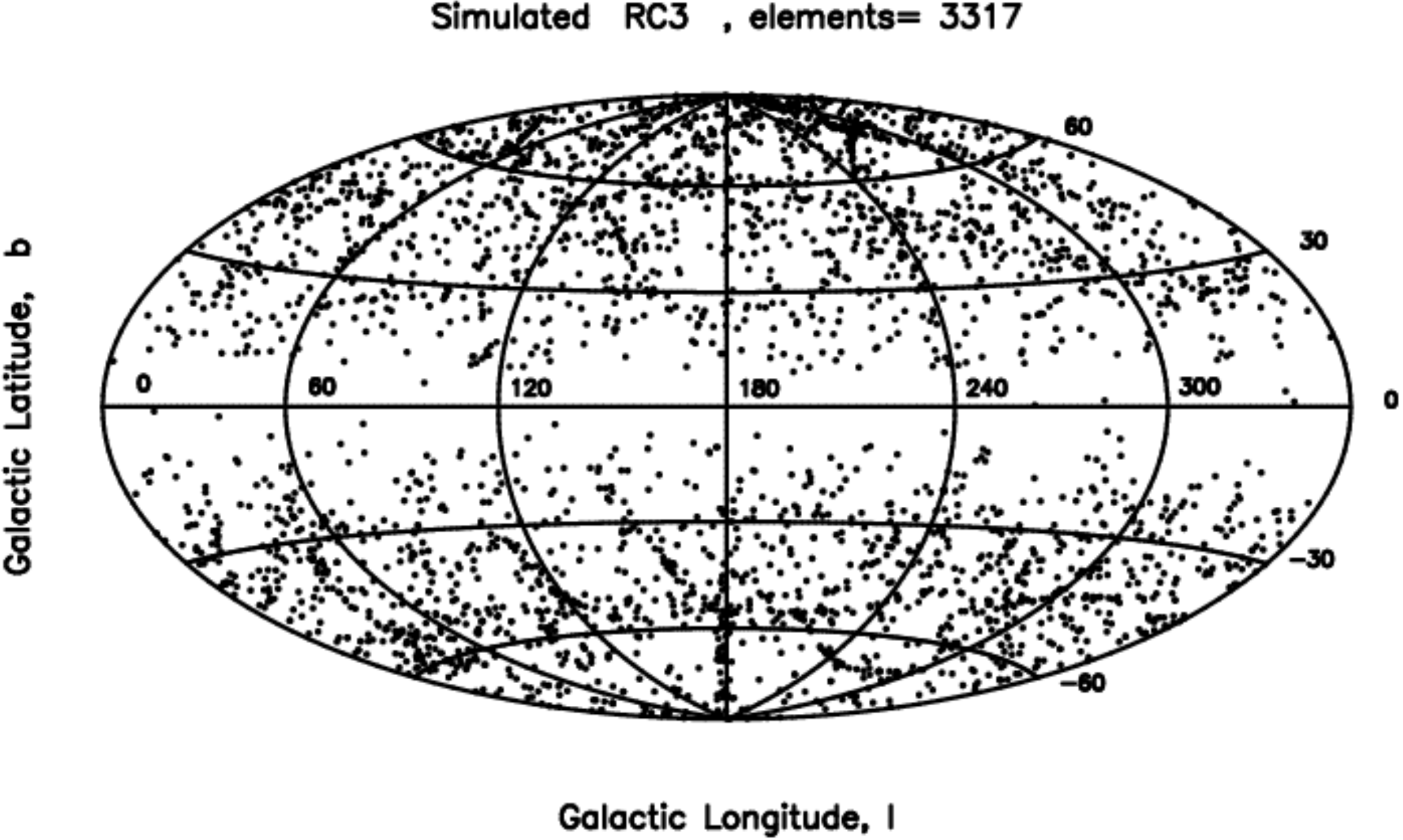}
\end{center}\caption{Hammer-Aitoff  projection
  of 3317   pixels  belonging to a
  face of an irregular  Voronoi Polyhedron.
  The  {\it Zone of Avoidance}  at the galactic plane  
  follows Fig. \ref{rc3_all}.
  This plot simulates the RC3 galaxies 
  which have BT and redshift.
  The  galaxies are extracted according 
  to the integral of the Schechter  function  in flux  
  (formula~(\ref{integrale_schechter})  with parameters
  as  in  Table~\ref{parameters}) 
}
          \label{mix_rc3}%
    \end{figure}

\begin{figure}\begin{center}
\includegraphics[width=7cm]{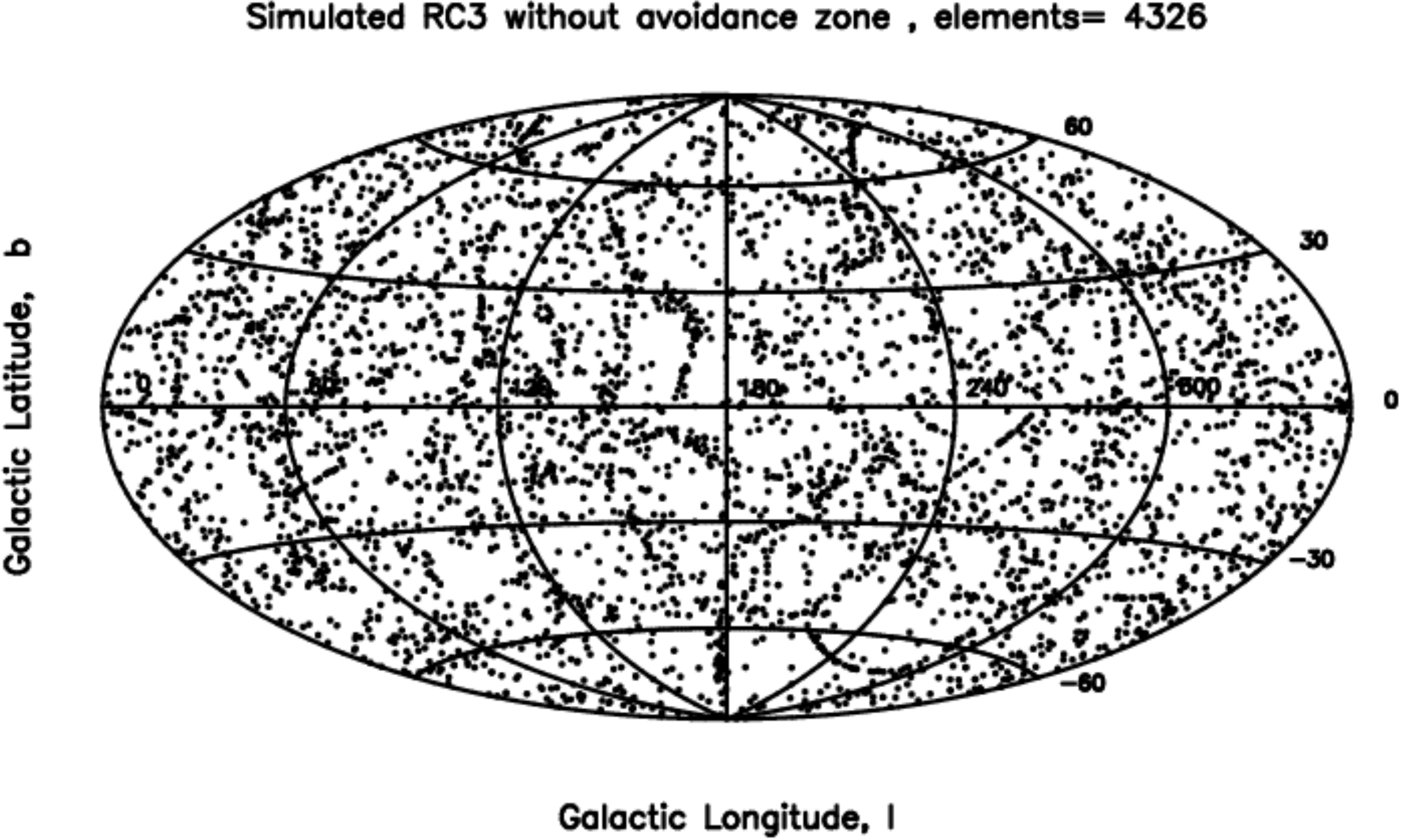}
\end{center}\caption{Hammer-Aitoff  projection
  of  4326  pixels  belonging to a
  face of an irregular  Voronoi Polyhedron.
  This plot simulates the RC3 galaxies 
  which have BT and redshift but 
  the  {\it Zone of Avoidance}  at the galactic plane  
  is absent.
  The  galaxies are extracted according 
  to the integral of the Schechter  function  in flux  
  (formula~(\ref{integrale_schechter})  with parameters
  as  in  Table~\ref{parameters}) 
}
          \label{noavoid_rc3}%
    \end{figure}

The simulation of this overall sky survey  can be done
in the following way:
\begin {itemize}
\item The pixels  belonging to the faces of  irregular
      polyhedron are  selected according 
      to the distribution in
      $z$ of the galaxies in the RC3 catalog  
      which have redshift and BT.
\item A second operation selects  the  pixels   according 
      to the distribution in latitude in the RC3 catalog,
      see  Fig. \ref{mix_rc3}.
\item In order to simulate  a theoretical  distribution of objects
      which represent the RC3 catalog without the {\it Zone of Avoidance} 
      we made a  series of $6^{\circ}$ slices in latitude 
      in the RC3 catalog, selecting   $N_i$  pixels 
      in  each slice, see  Fig. \ref{noavoid_rc3}.
      In order to ensure that  the range in 
       $z$ is correctly
      described  Table~\ref{rc3catalog} reports 
      $z_{min}$, $z_{pos-max} $, $z_{ave} $ and 
      $z_{max}$   which  represent 
      the minimum $z$, the  position in $z$ of the maximum
      in the number of galaxies,
      and  
      the maximum $z$  in the RC3 catalog or  the simulated sample.
\end {itemize}
\begin{table}
\caption { Real and  Simulated data without the {\it Zone of Avoidance} 
   in the RC3 catalog.   }
 \label{rc3catalog}
 \[
 \begin{array}{ccc}
 \hline
 \hline
 \noalign{\smallskip}
~~   &   RC3   &   no~  {\it ZOA}        \\
 \noalign{\smallskip}
 \hline
 \noalign{\smallskip}
elements    &  3316                 &  4326                    \\
z_{min}     &  5.7  \times 10^{-7}  & 8.9  \times 10^{-3}      \\
z_{pos-max} &  5.6  \times 10^{-3}  & 8.9  \times 10^{-2}     \\
z_{ave}     &  1.52 \times 10^{-2}  & 7.96  \times 10^{-2}     \\
z_{max}     &  9.4  \times 10^{-2}  & 0.14                    \\
\noalign{\smallskip}
\noalign{\smallskip}
 \hline
 \hline
 \end{array}
 \]
 \end {table}

\subsection{The CFA2 catalog}

\begin{figure}\begin{center}
\includegraphics[width=7cm]{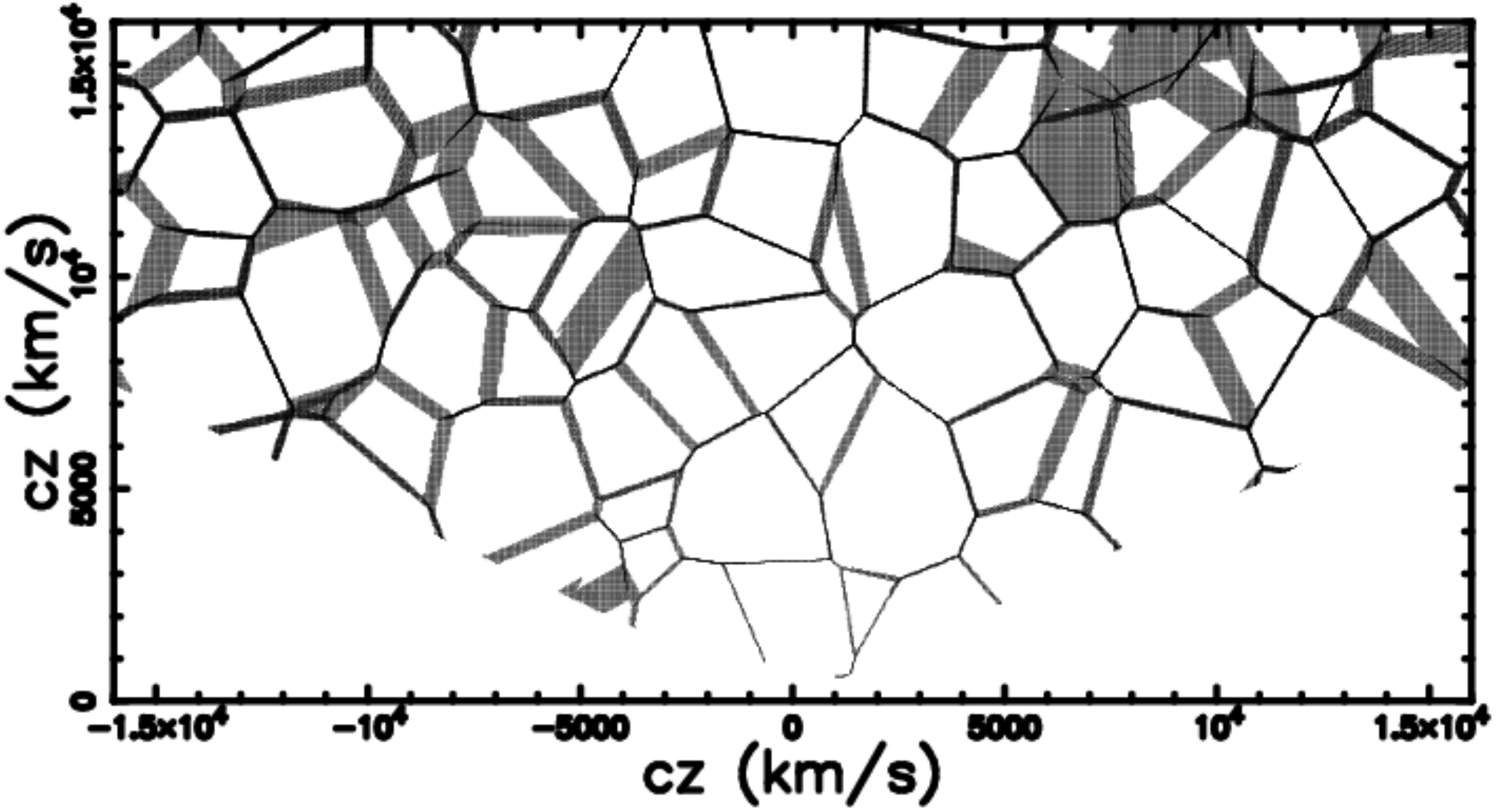}
\end{center}\caption {
Polar plot
of the  little cubes belonging to a   
slice   $130^{\circ}$~long  and $6^{\circ}$
wide.
}
          \label{cfaslices}%
    \end{figure}

The results of the simulation can be represented by  a
 slice
similar to that observed (a strip of $6^{\circ}$ wide and about
$130^{\circ}$ long) , see Fig. \ref{cfaslices}.

\begin{figure}\begin{center}
\includegraphics[width=7cm]{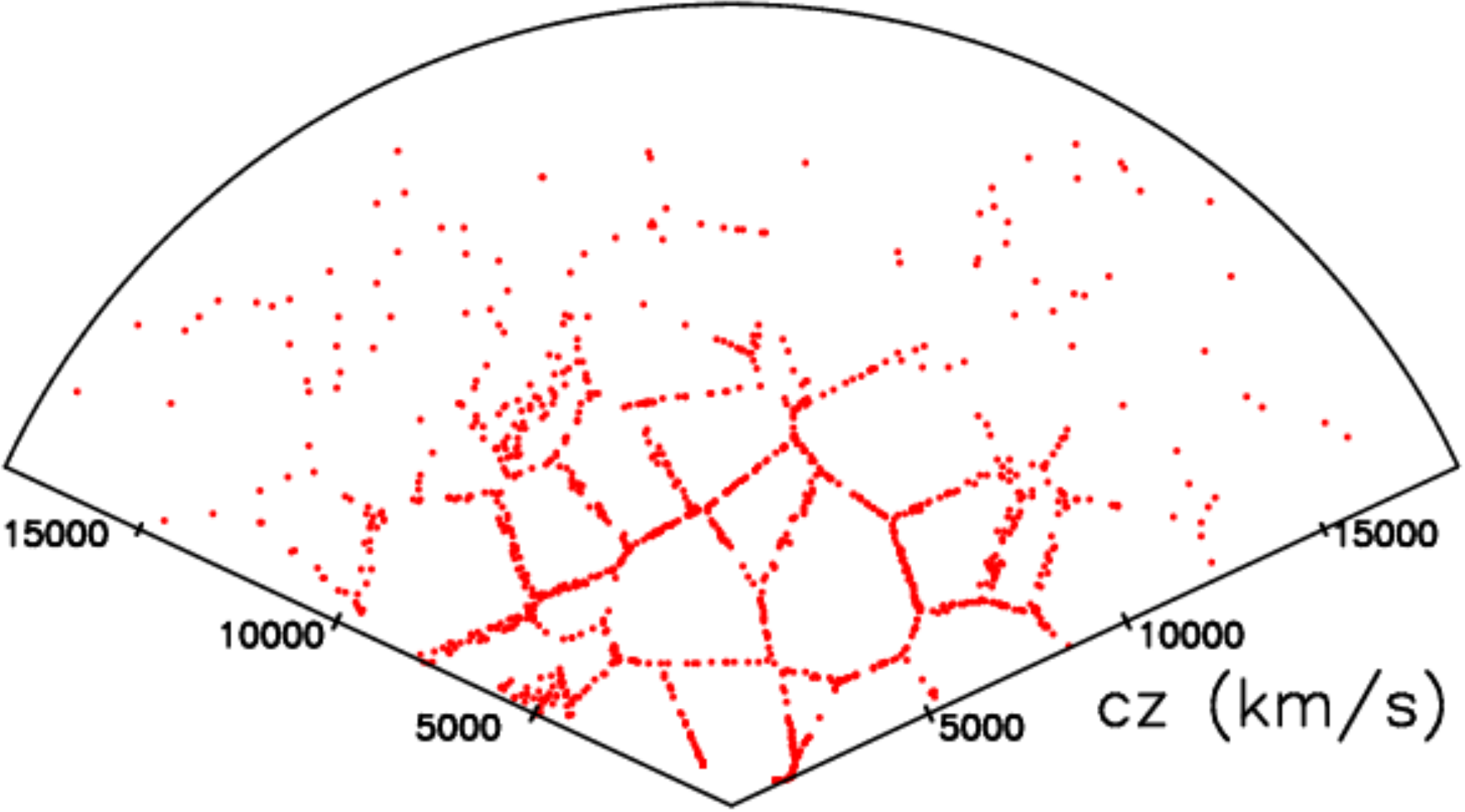}
\end{center}\caption 
{
Polar plot of the little cubes (red points) 
when  the "scaling" algorithm
is applied.
Parameters  as in Fig. \ref{cfaslices}~.
This plot simulates the CFA2  slice. 
}
          \label{simu_mia}%
    \end{figure}

\begin{figure}\begin{center}
\includegraphics[width=7cm]{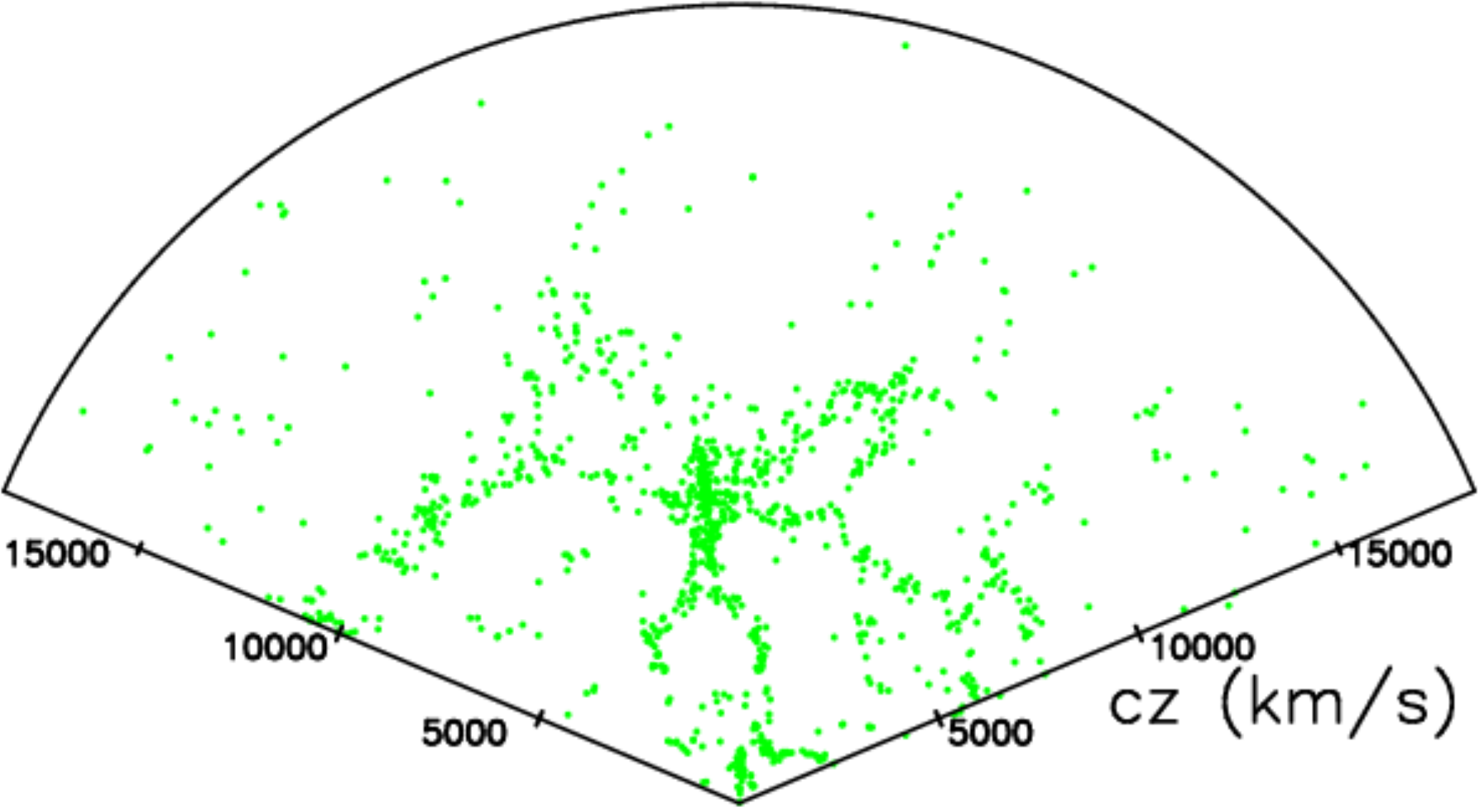}
\end{center}\caption { Polar plot of   the   real galaxies (green  points)
belonging to the second CFA2 redshift catalog.}
          \label{simu_cfa}%
    \end{figure}

A typical polar plot  once the "scaling" 
algorithm is  implemented , see  \cite{Zaninetti2006} , 
is  reported in Fig. \ref{simu_mia}; 
it should be compared 
with the observations  , see Fig. \ref{simu_cfa}. 

\begin{figure}\begin{center}
\includegraphics[width=7cm]{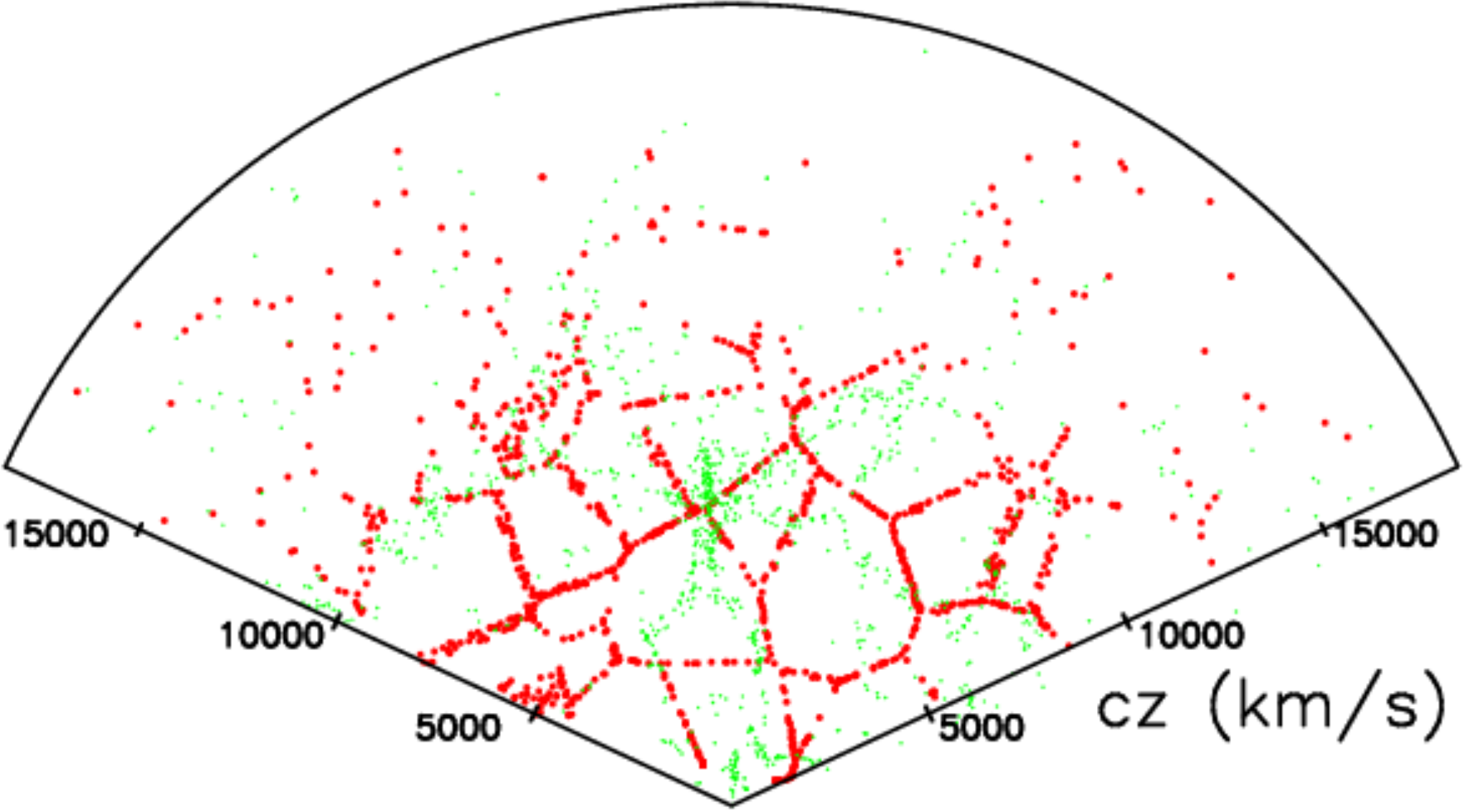}
\end{center}\caption 
{ 
Polar plot of   the   real galaxies (green  points)
belonging to the second CFA2 redshift catalog
and  the little cubes (red points).
}
          \label{true_simu_color}%
    \end{figure}

Fig. \ref{true_simu_color}  reports  both  the 
CFA2 slice as  well the simulated slice.

\subsection{The Eridanus Supervoid}

\label{eridanus}

A void can be defined as the 
empty space between filaments 
in a slice  and 
the typical  diameter has a range of   $[11-50]~Mpc/h$,
see 
\cite{Einasto1994} 
and 
\cite{Einasto1995}.
The probability,   for example,  of  having  a volume
3 times bigger than the average is $3.2~10^{-3}$ 
for PDF~(\ref{rumeni}) when $d=3$ and
$2.1~10^{-3}$ 
for PDF~(\ref{kiangd}) when $d=2.75$.  
Particularly  large voids are called super-voids and
have a range of  $[110-163]~Mpc/h$.

Special attention should be paid to  the Eridanus super-void
of 300~$Mpc $  in diameter.
This super-void was  detected by the 
Wilkinson Microwave Anisotropy Probe (WMAP), see 
\cite{Vielva2004,Cruz2005,Vielva2011} 
and was  named {\it Cold Spot}.
The  WMAP measures the temperature fluctuations
of the cosmic microwave background (CMB).
Later on the astronomers  confirmed the largest   void
due to the  fact that the density of radio sources 
at 1.4 GHz  is anomalously  low
in the direction of the {\it Cold Spot}, see 
\cite{Rudnick2007} 
and 
\cite{McEwen2008}.
The standard statistics of the 
Voronoi  normalized volume distribution in 3D covers the range
$[0.1-10]$.
In the case of a   Eridanus super-void the normalized volume 
is $\approx~\frac{300}{27}=1.37~10^3$ and the connected
probability of  having  such a super-void is $1.47~10^{-18}$ 
when the Ferenc \& Neda  function
with $d=3$, formula ~(\ref{rumeni}),
is used and $\approx 0 $ when the Kiang function
with $d=2.75$, formula (\ref{kiangd}) is
used. 

Due to this low probability of  having  such a 
large normalized volume  we 
 mapped a possible
spatial distribution  of the  SDSS-FIRST 
(the Faint Images of the Radio Sky at Twenty cm survey)
sources
with complex radio morphology 
 from
the theoretical distribution of galaxies belonging to the  RC3.
The fraction of galaxies belonging 
to the 2dFGRS   detected as  SDSS-FIRST sources 
with complex radio morphology 
is less than $10\%$  according to  Section 3.8 
in \cite{Ivezic2002}.
We therefore introduced a  probability, $p_{rs}$,
that a galaxy is a radio source.
The  number of  SDSS-FIRST sources $N_{rs}$  in the RC3 which 
are  SDSS-FIRST sources 
with complex radio morphology is 
\begin{equation}
N_{rs} = p_{rs} * N_g
\quad ,
\end{equation}
where $N_g$ is the number of galaxies 
in the theoretical RC3.

\begin{figure}\begin{center}
\includegraphics[width=7cm]{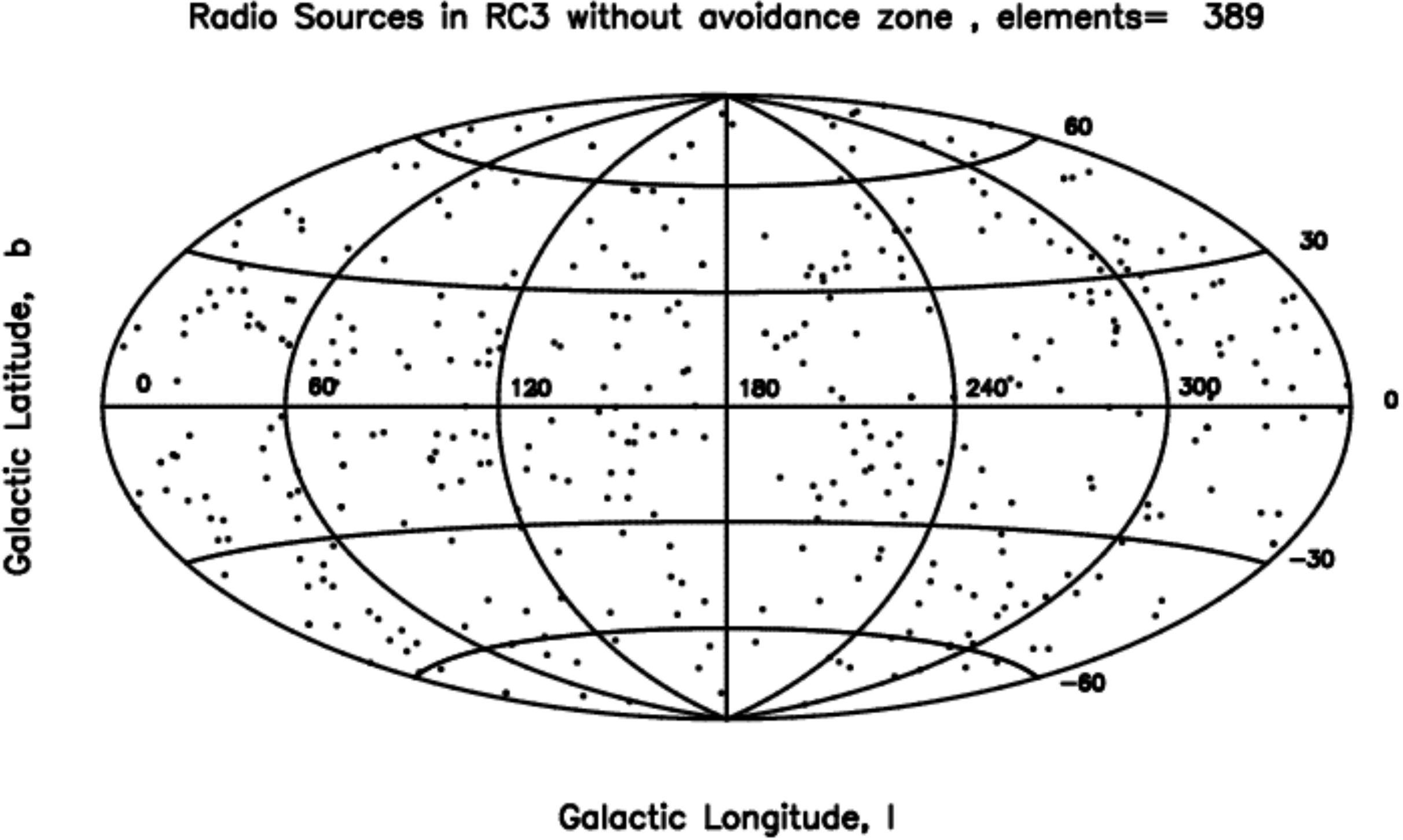}
\end{center}\caption{
  Hammer-Aitoff  projection
  of the  
  SDSS-FIRST sources
  with complex radio morphology 
  belonging to  the
  RC3, $p_{rs}=0.09$.
  Other parameters as in Fig. \ref{noavoid_rc3}.
}
          \label{rc3_radio}%
    \end{figure}
From a visual inspection of Fig. \ref{noavoid_rc3} 
and  Fig. \ref{rc3_radio} it is possible
to conclude that the voids increase in size when 
 radiogalaxies which  are a subset of the galaxies
are considered.

\section{The  correlation function for galaxies}
\label{sec_corr}

Galaxies have the tendency to be grouped in clusters 
and a typical measure is the computation of the
two-points  correlation
function for galaxies, see \cite{Peebles1993,Peebles1980}.
The correlation function can be computed in
two ways: a local analysis  in the range
$[0-16] Mpc/h$ and an extended analysis in the range 
$[0-200] Mpc/h$ .

\subsection { The local analysis}

A first  way  to describe the degree  of clustering 
of galaxies is  the two point  correlation  function
$\xi _{GG} (r)$, usually presented in the form
\begin{equation}
 \xi _{GG} = ({r \over r_G})^{-\gamma_{GG }}  
\quad ,
\end{equation}
where  $\gamma_{GG }$=1.8  and  $r_G = 5.77h^{-1} Mpc$   
(the correlation length) when the range 
$0.1 h^{-1} Mpc < r < 16 h^{-1} Mpc$ is considered,
see 
\cite{Zehavi_2004} 
where 118149 galaxies were
analyzed.

In order to compute the correlation function,  
two volumes were compared: one containing 
the little cubes belonging to a face, the other containing 
a random distribution of points. 
From an analysis of the distances of pairs,
the minimum and maximum   were computed
and  $ n_{DD}(r)$   was obtained, 
where $n_{DD}(r)$ is the number of pairs 
of galaxies with separation
within the interval $[r-dr/2, r+dr/2]$.
A similar procedure was applied to the random elements
in the same volume with the same number of elements
and  $n_{RR}(r)$ is the number of
pairs of the Poissonian Process.
According to formula~(16.4.6) in~\cite{coles} 
the correlation function is:  
\begin{equation}
 \xi _{GG} (r) = \frac { n_{DD}(r) } {n_{RR}(r)} -1 \quad . 
\end {equation}
To check whether  $\xi_{GG}$  obeys  a power law or
not we used a simple linear regression test with the formula:
\begin{equation}
 Log \,  \xi _{GG} = a + b\; Log \; r    \quad ,
\end {equation}
which allows us  to compute  $r_G = 10 ^{-a/b}$ and  
$\gamma_{GG }$=-b.

\begin{figure}\begin{center}
\includegraphics[width=7cm]{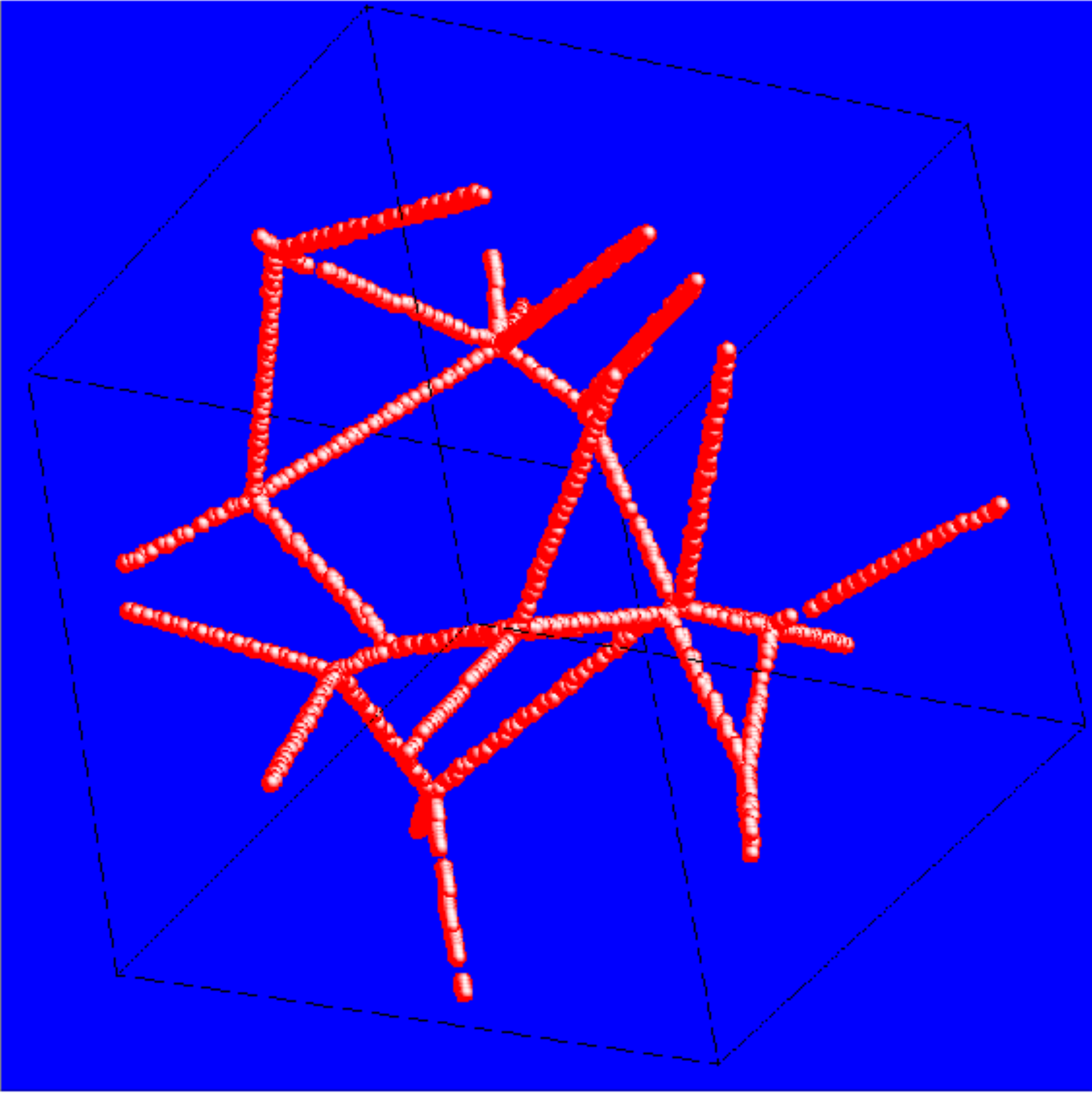}
\end{center}\caption{
3D visualization of the 
edges of the Poissonian Voronoi--diagram.
The  parameters
are      $ pixels$= 60, $ N_s   $   = 12, 
         $ side  $   = 96.24 $Mpc$, 
         $h=0.623$  and    $ amplify$= 1.2.}
          \label{spigoli3d_sb}%
    \end{figure}

\begin{figure}\begin{center}
\includegraphics[width=7cm]{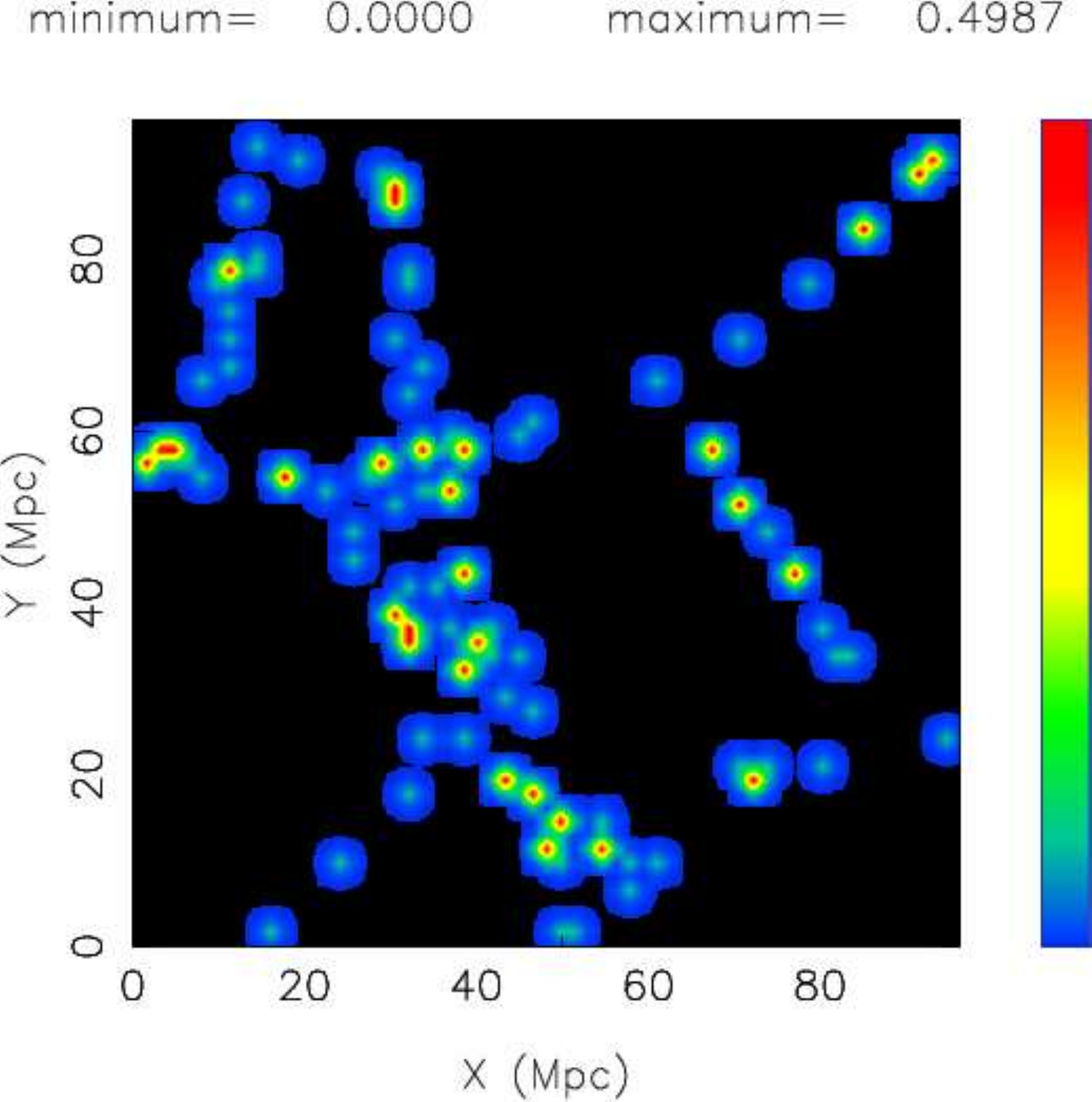}
\end{center}\caption{
Cut  in the middle of the 3D grid $P(i,j,k)$
which represents a theoretical 2D map 
of the probability of having  
a galaxy. 
The  Voronoi parameters are the same as in  
Fig. \ref{spigoli3d_sb} and $\sigma=0.8 Mpc$.
The X and Y units are in Mpc.
        }
    \label{probability2d}
    \end{figure}

We now outline the method that allow us 
to compute the correlation
function  using the concept of  thick faces, 
see \cite{zaninetti95}.
A practical implementation is to consider a decreasing probability
of  having  a galaxy  in the direction  perpendicular to the face.
As an example we assume  a  probability, 
$p(x)$, of  having  a galaxy 
outside the face distributed as a Normal (Gaussian) 
distribution
\begin{equation}
p(x) = 
\frac {1} {\sigma (2 \pi)^{1/2}}  \exp {- {\frac {x^2}{2\sigma^2}}} 
\quad  ,
\label{gaussian}
\end{equation}
where $x$ is the distance in $Mpc$ from the face and $\sigma$ 
the standard deviation in $Mpc$.
Once the complex 3D behavior of the faces of the Voronoi
Polyhedron is set up  we can memorize 
such a probability on a 3D grid $P(i,j,k)$ 
which  can be found in the following way  
\begin{itemize}
\item In each lattice point $(i,j,k)$ we search for  
      the nearest  element
      belonging to a Voronoi face. The probability of having  
      a galaxy
      is therefore computed according to formula~(\ref{gaussian}).
\item    A number of galaxies, $N_G=n_* \times side^3$
         is then inserted in the box; 
         here $n_*$ represents the   density of  galaxies 
\end{itemize}
Fig. \ref{spigoli3d_sb} visualizes  
 the edges belonging  to the Voronoi 
diagrams and Fig. \ref{probability2d} represents 
a cut in the middle of the probability, $P(i,j,k)$, 
of having  a galaxy to a given 
distance from a face.

\begin{figure}\begin{center}
\includegraphics[width=7cm]{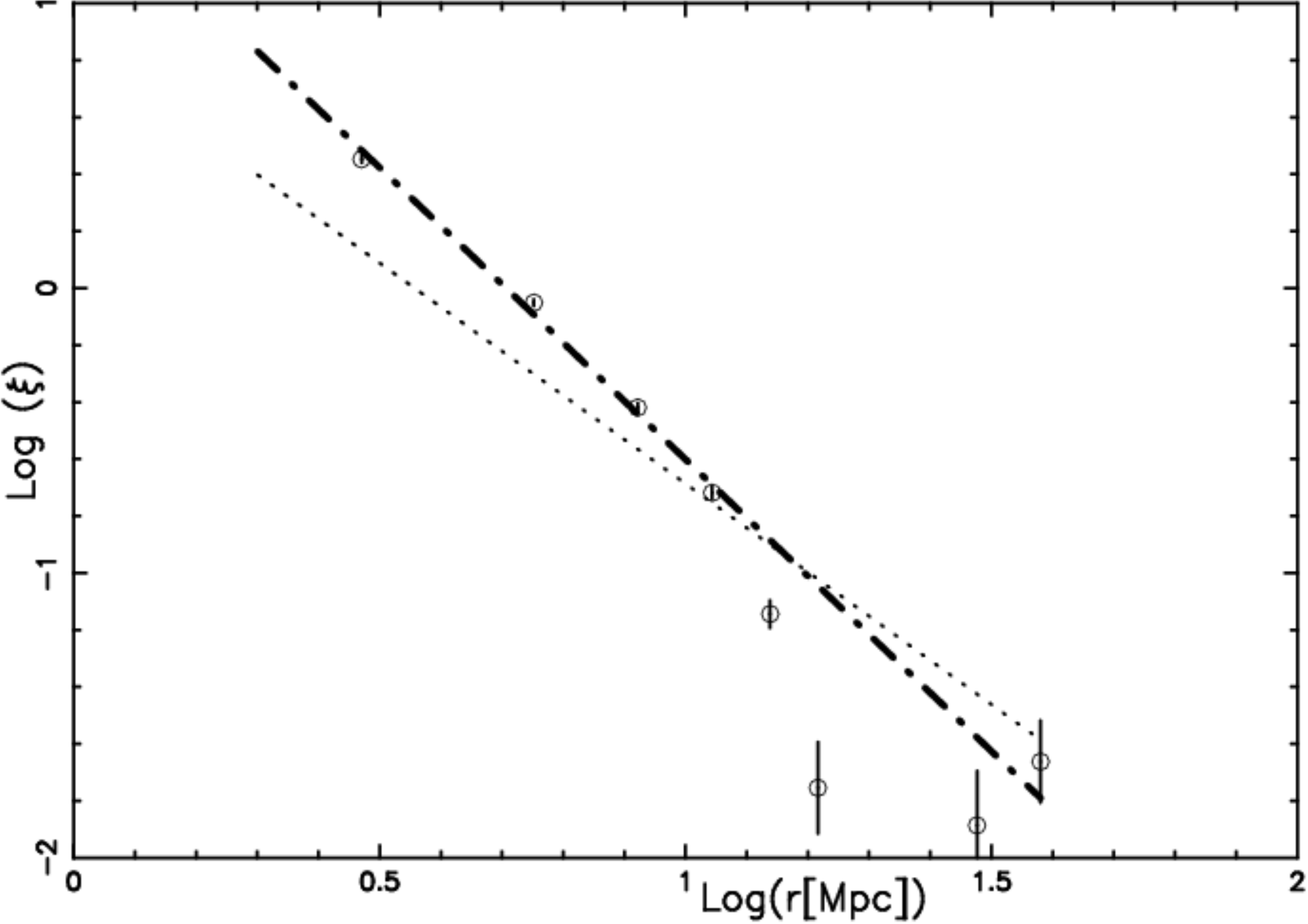}
\end{center}\caption 
{
The logarithm  of the correlation function is  visualized 
through   points with their uncertainty (vertical bar),
the asymptotic behavior  of the correlation 
function  $\xi_{GG}$  is reported as  dash-dot-dash line;
in our simulation $\gamma_{GG }$=2.04  and  $r_G$ = 5.08 Mpc.
The standard value 
of the correlation function 
is reported  as a  dotted line; 
from the point of view of the observations 
in average $\gamma_{GG }$=1.8  and  $r_G$ = 5 Mpc.
Parameters  of the simulation  as in Fig. \ref{spigoli3d_sb}.
 }
          \label{correlation}%
    \end{figure}

A typical result of the simulation is reported  in 
Fig. \ref{correlation}  where the center of the 
smaller box in which  the correlation function
is computed   is the  point belonging 
to a face nearest to 
the center of the big box.

From an  analysis of Fig. \ref{correlation}
we can deduce that the correlation 
function $\xi_{GG}$ of the simulation has a behavior  
similar to the standard one.
Perhaps  the value  $r_G$ is a simple measure 
of the face's thickness, $\Delta R_F$.
From this point of view on adopting  a standard value  of the
expanding shell thickness, $\Delta R$ = $\frac{R}{12}$ 
and assuming that the thickness of the shell is made 
by the superposition of two expanding shells
the following is obtained
\begin{equation}
\Delta R_F  \approx  \frac{R}{6} 
\approx \frac{\overline{D^{obs}}}{h\,12} = 3.62~Mpc
\quad ,
\end{equation}
where $h=0.623$ has  been used.
 The 
correlation dimension $D_2$, see~\cite{Jones2005}, 
is connected with the exponent $ \gamma$  
through the relation:
\begin{equation}
 D_2  = 3 - \gamma      
\quad .
\end{equation}
Here there is the case 
in which the  mass M(r) increases
as   $r^{1.2}$,  in the middle  of a one dimensional
structure ( $M(r) \propto r$) and a two dimensional
sheet ( $M(r) \propto r^2$), see \cite{coles}.
In this paragraph the dependence of the correlation
function  on  the magnitude is   not  considered.

\subsection {The extended analysis}

\begin{figure}\begin{center}
\includegraphics[width=7cm]{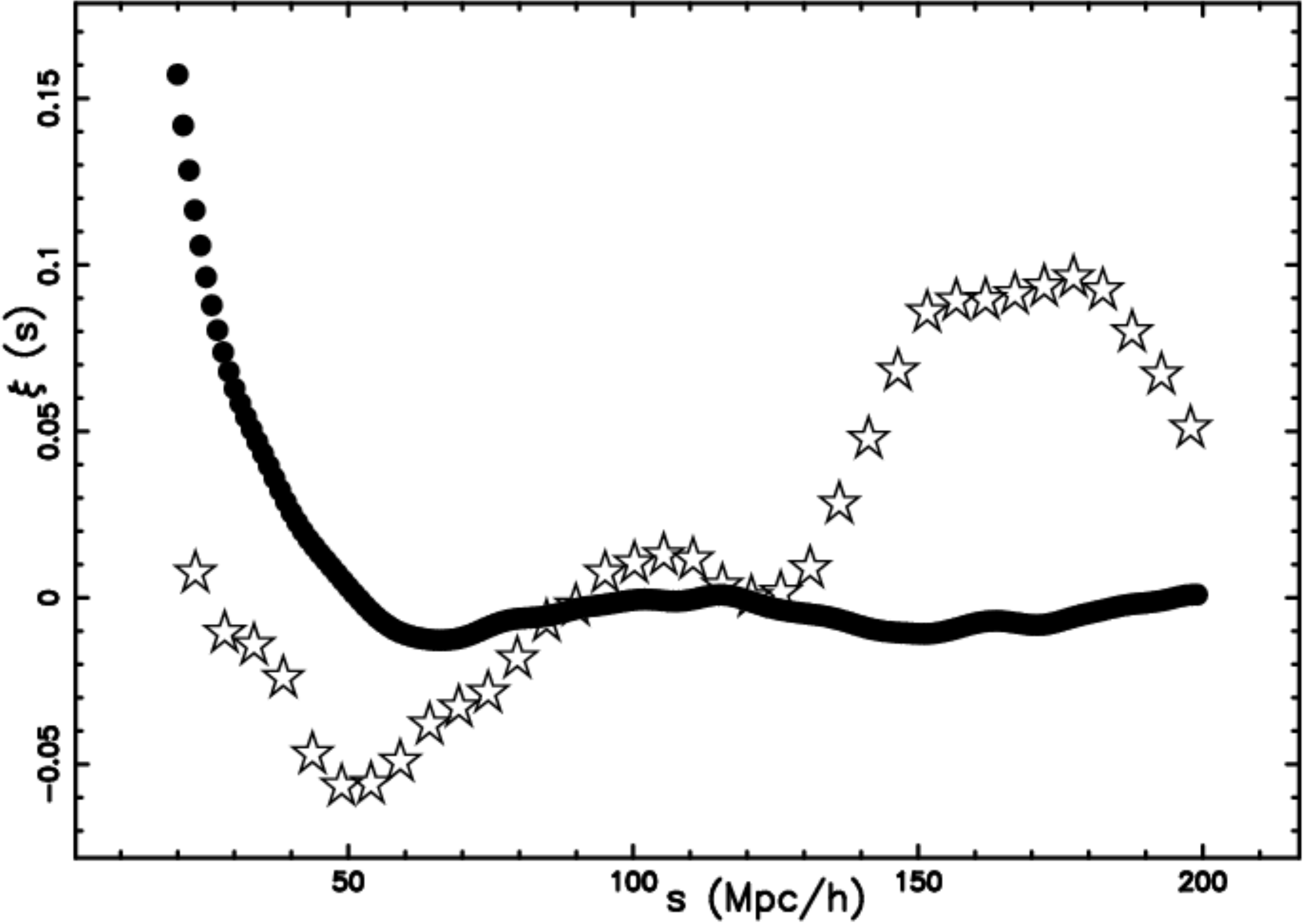}
\end{center}\caption{
Redshift-space correlation function 
for  the 2dFGRS sample limited 
at $z=0.12$ as given by  our code          ( empty stars )
 and the results  of \cite{Martinez2009}  ( full points )
for 2dFVL.  
The covered range is $[40-200] Mpc/h$ .
        }
    \label{2df_our_martinez}
    \end{figure}

A second definition of the correlation function 
takes account of the Landy-Szalay border correction,
see \cite{Szalay1993},
\begin{equation}
 \xi _{LS} (s) = 1
+ \frac { n_{DD}(s) } {n_{RR}(s)} 
- 2  \frac { n_{DR}(s) } {n_{RR}(s)} 
 \quad . 
\end {equation}
where $n_{DD}(s)$,  
$n_{DD}(s)$  and  $n_{DR}(s)$
are the number of  galaxy-galaxy ,random-random 
and galaxy-random pairs having distance $s$,
see \cite{Martinez2009}.
A random catalog of galaxies in polar coordinates
can built by generating 
a first random number $\propto~z^2$ in the z-space
and a second random angle in the interval $\bigl [0,75 \bigr ]$. 
A test of our code for the correlation 
function versus  a more sophisticated code 
is reported in Fig. \ref{2df_our_martinez}
for the 2dFVL volume limited (VL) sample,
where the data available at the   
Web site http://www.uv.es/martinez/ 
have been processed.

\begin{figure}
\begin{center}
\includegraphics[width=7cm]{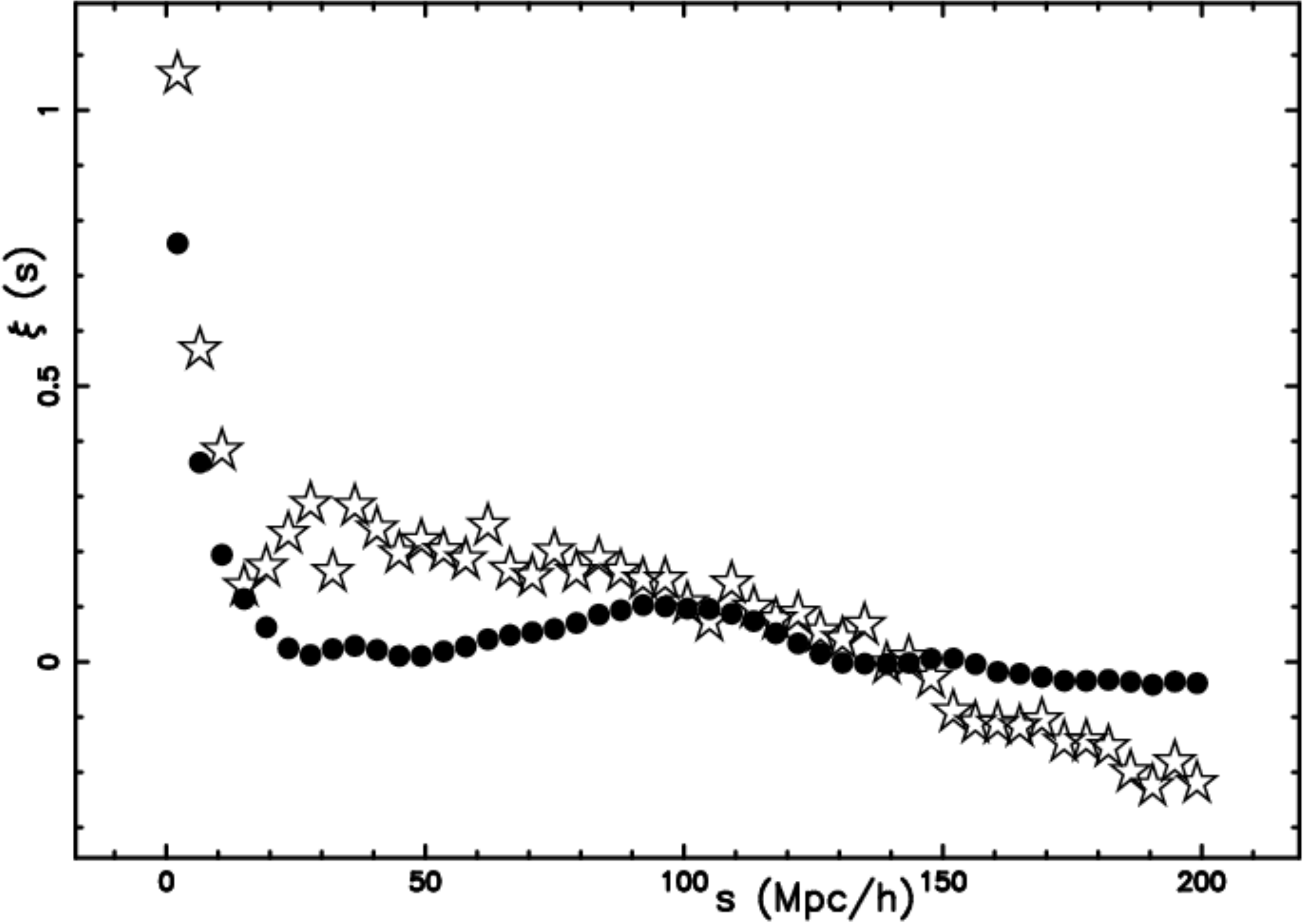}
\end{center}
\caption{
Redshift-space correlation function 
for  the 2dfGRS sample ( empty stars )
 and the Voronoi sample( full points ).  
The covered range is $[40-200] Mpc/h$.
        }
    \label{correlation_due}
    \end{figure}

The pair correlation function for the vertexes of the
Poissonian Voronoi Polyhedron
 presents a typical damped oscillation,
see Figure 5.4.11 in \cite{okabe}, Figure 2 in 
\cite{Martinez2009}
and Figure 3 in \cite{Heinrich2008}.
Here conversely : (a) we first consider a set of objects belonging
to the faces of the irregular Polyhedron
; (b) we extract from the previous set a subset which follows 
the photometric law and then  
we compute the pair correlation function. 
The difference between our model and the model in 
\cite{Martinez2009}  
for 2dFVL
can be due to the luminosity color segregation 
presents in 2dFVL but not in our Voronoi type model. 
A typical  result is  reported in Fig. \ref{correlation_due}
where it is possible to find the correlation function
of 2dfGRS   with astronomical 
data as reported in Fig. \ref{2df_all}
as well  as the correlation function of the Voronoi 
network with simulated data as reported
in Fig. \ref{voro_2df_cones}.

A careful analysis of Fig. \ref{correlation_due}
allows us to conclude that the behavior of the correlation
function is similar for the astronomical data as well
as the simulated Voronoi-data.
The oscillations 
after 100 $Mpc$ are classified as 
acoustic,  \cite{Eisenstein2005}.

\section{Conclusions}

{\bf Photometric maximum}

The  observed number of galaxies in a given  solid angle  
with a chosen   flux/magnitude  
versus the redshift presents 
a maximum  that is a function of the flux/magnitude.
From  a theoretical  point of view,   
the  photometric  properties of the galaxies 
depend on  the chosen law  for the $LF$.
The  three $LF$s   here adopted 
predict a maximum 
in the theoretical number of galaxies  as a  
function of the redshift
once the apparent flux/magnitude is fixed,
for the  Schechter             $LF$  see  formula(\ref{nfunctionz}),
for the ${\mathcal M}-L$       $LF$  see  formula(\ref{nfunctionz_mia})
and  for the generalized gamma $LF$  see  formula(\ref{nfunctionz_gammagene}).
  
The theoretical  fit  representing  the number 
of galaxies as a function 
of the redshift can be compared with the 
real number of galaxies
of the 2dFGRS which  is theory-independent.
The superposition of theoretical  and observed fit  
is   acceptable      ,
see Fig.~\ref{maximum_flux}.
Particular attention should be paid
to the  Malmquist bias and to equation~(\ref{range}) 
that  regulate  the upper value of the 
redshift that defines the complete sample.

\noindent
{\bf 3D Voronoi  Diagrams}
The intersection between a plane and the 3D Voronoi faces
is  well known as 
$V_p(2,3)$.
The intersection between a slice of a given opening angle,
for example $3^{\circ}$, and the 3D Voronoi faces
is less known  and has  been developed  in 
Section~\ref{faces}.
This intersection can be calibrated 
on the astronomical data  once the number of 
Poissonian 
seeds   is such that the largest observed void 
matches  the largest Voronoi volume.
Here  the largest observed void is 
2700 $Km/sec$ and  in order to simulate, for  example,
the 2dFGRS,   137998  Poissonian seeds  were inserted 
in a volume of
$(131908~Km/sec)^3$.
The intersection between a sphere 
and the 3D Voronoi faces
represents a new way to visualize 
the voids in the distribution of galaxies,
see Section~\ref{faces}.
In this  spherical  cut the intersection 
between a sphere
and the 3D Voronoi faces
is no longer represented 
by  straight lines but by curved lines 
presenting  in some cases a  cusp behavior at the 
intersection, see Fig.~\ref{aitof_sphere}.
In line of principle the spatial distribution
of  galaxies at a given redshift  should
follow   such  curved lines.

\noindent
{\bf Statistics of the voids}
The  statistical properties of the voids 
 between galaxies 
can be well described by the volume  distribution 
of the Voronoi Polyhedra.
Here two distributions of probability were carefully compared:
the old Kiang function here parametrized 
as a function of the dimension $d$  , see formula~(\ref{kiang}),
and the new distribution 
of Ferenc ~\&~ Neda~ , see formula~(\ref{rumeni}), 
which  is a function of the selected 
dimension $d$.
The analysis of the normalized
areas of $V_p(2,3)$ is a subject of research rather than a
well-established fact and we have fitted them with the Kiang
function and the exponential distribution. The $\chi^2$ value
indicates that the exponential distribution fits more closely the
normalized area distribution of $V_p(2,3)$ than does the Kiang
function, see Table~\ref{tablev23}. This fact follows from the
comparison between  the
exponential and Kiang distributions of the radius, see
Fig.~\ref{comparison_cut2}. Therefore, the one parameter survival
function of the radius of the exponential distribution for
$V_p(2,3)$, $S_{ER23} $, as represented by
(\ref{survival_expr23}), may model the voids between galaxies.

\noindent
{\bf Simulations of the catalogs of galaxies}
By combining  the photometric dependence 
in the number of galaxies 
as a function of the redshift with the intersection
between a slice and the Voronoi faces, it  is possible to simulate
the astronomical catalogs  
such as the  2dFGRS, see Section~\ref{cat2dFGRS}.
Other catalogs such as  
the RC3 which covers 
all the sky ( except the   Zone of Avoidance ) 
can be simulated through a given number of 
spherical cuts, for example 25,
with progressive increasing redshift.
This  simulation is visible 
in  Fig.~\ref{mix_rc3}  in which the 
theoretical influence   of the 
Zone of Avoidance  has been inserted,
and in Fig.~\ref{noavoid_rc3} 
in which the theoretical RC3 without the 
Zone of Avoidance has  been modeled.
Fig.~\ref{rc3_radio} reports the subset 
of the  galaxies which  are  radiogalaxies.

\noindent
{\bf Correlation function}
The standard behavior of the correlation function  for  
galaxies in the short range 
$[0-10~Mpc/h]$ 
can be simulated  once 12 Poissonian seeds are inserted 
in a  box of volume
$( 96.24~Mpc/h )^3$ .
In this case the model can be refined
by 
 introducing the concept
of galaxies generated in   a thick face 
belonging to the Voronoi  Polyhedron.
The behavior of the correlation function in the large range 
$[40-200~Mpc/h]$  of the Voronoi   simulations 
of the 2dFGRS  presents minimum variations from  
 the processed astronomical data, see 
Fig. ~\ref{correlation_due}.
We now extract a question from the conclusions of 
\cite {Martinez2009} ``Third, the minimum in 
the large-distance correlation functions
of some samples demands explanation: is it really the
signature of voids?''
Our answer is ``yes''. The  minimum in the large scale
correlation function is due to the combined effect 
of the large empty space between galaxies ( the voids )
and to  the photometric behavior of  the number of 
galaxies as  a function of the red-shift.

\noindent
{\bf Stereological approach}
In this review analytical  formulas have been provided  
which model the distributions of the lengths and areas
 of the planar sections of three-dimensional 
Poisson Voronoi diagrams:
in particular, it has been  shown that they are 
related to the Meier $G$ function.
This finding  is consistent with the analytical  results presented in
\cite{Springer1970}, where it is proved that 
nonlinear combinations of gamma variables, such as products or quotients,  
have distributions proportional, or closely related, to the Meijer $G$ distribution.
The analytical distributions $F_r$ and $F_A$ been compared 
with results of numerical simulations.

\noindent
{\bf Acknowledgements.} 
I thank the
2dF Galaxy Redshift Survey team for the use of
Fig. \ref{2df_cone}, which is
taken from the image gallery on the 2dFGRS website (see
http://www2.aao.gov.au/2dFGRS)
and the  
Stanford Encyclopedia of Philosophy    for the use of
Fig. \ref{descartes}.

\noindent
{\bf REFERENCES} 


\end{document}